\documentclass{jfm}

\usepackage{graphicx}
\usepackage{newtxtext}
\usepackage{newtxmath}
\usepackage{natbib}
\usepackage{hyperref}
\usepackage{comment}
\usepackage{color,soul}
\hypersetup{
    colorlinks = true,
    urlcolor   = blue,
    citecolor  = black,
}

\newcommand{\RomanNumeralCaps}[1]
\linenumbers

\title{Effects of surfactants on bubble-induced turbulence} 

\author{
Tian Ma\aff{1}\corresp{\email{tian.ma@hzdr.de}},
Hendrik Hessenkemper\aff{1}\corresp{\email{h.hessenkemper@hzdr.de}},
Dirk Lucas\aff{1}
\and Andrew D. Bragg\aff{2}\corresp{\email{andrew.bragg@duke.edu}}
}

\affiliation{\aff{1}Helmholtz-Zentrum Dresden -- Rossendorf, Institute of Fluid Dynamics, 01328 Dresden, Germany
\aff{2}Department of Civil and Environmental Engineering, Duke University, Durham, NC 27708, USA}

\begin{document}
\maketitle

\begin{abstract}

We use experiments to explore the effect of surfactants on bubble-induced turbulence (BIT) at different scales, considering how the bubbles affect the flow kinetic energy, anisotropy and extreme events. To this end, high-resolution Particle Shadow Velocimetry measurements are carried out in a bubble column in which the flow is generated by bubble swarms rising in water for two different bubble diameters ($3$ mm  $\&$ $4$ mm) and moderate gas volume fractions ($0.5\%\sim1.3\%$). To contaminate the flow, different amounts of 1-Pentanol were added to the flow, leading to different bubble shapes and surface boundary conditions. The results reveal that with increasing surfactant concentration, the BIT generated increases in strength, even though bubbles of a given size rise more slowly with surfactants. We also find that the level of anisotropy in the flow is enhanced with increasing surfactant concentration for bubbles of the same size, and that for the same surfactant concentration, smaller bubbles generate stronger anisotropy in the flow.  Concerning the intermittency quantified by the normalized probability density functions of the fluid velocity increments, our results indicate that extreme values in the velocity increments become more probable with decreasing surfactant concentration for cases with smaller bubbles and low gas void fraction, while the effect of the surfactant is much weaker for cases with larger bubble and higher void fractions. 

\end{abstract}

\begin{keywords}

\end{keywords}

\section{Introduction}\label{sec: introduction}

Surfactants are \textquoteleft surface-active\textquoteright\ molecules and/or particles that are easily adsorbed at surfaces and form monolayers \citep{2014_Langevin,2020_Manikantan}. They are present in most multiphase systems, being either naturally present or else introduced purposefully \citep{1962_Levich,1994_Stone,2019_Soligo,2022_Lohse}. In bubbly flows, it has been shown that even small amounts of surfactant can cause dramatic changes to the bubble shapes \citep{2002_Tomiyama,2014_Tagawa,2021_Hessenkemper}; bubble rise velocities \citep{1996_Bel,1997_Cuenot,2008_Takagi,2014_Tagawa}, lateral migration \citep{2017_Lu,2018_Hayashi,2020_Ahmed,2022_Atasi}, cluster formation \citep{2009_Takagi,2021_Maeda}, coalescence \citep{2016_Verschoof,2018_Lohse,2021_Neel} and mass transfer on the interfaces \citep{1997_Cuenot,2021_Schlueter}.

The mechanism by which surfactants influence the velocity field in the vicinity of a gas-liquid interface was first introduced by \cite{1947_Frumkin} and \cite{1962_Levich}, who showed that as a bubble rises in an aqueous surfactant solution, surfactant is swept off the front part of the bubble by surface convection and accumulates in the rear region. They lower the surface tension in the rear region relative to that at the front, leading to a gradient of surface tension along the interface. This gradient creates a tangential shear stress on the bubble surface (Marangoni effect) which opposes the surface flow, causes the interface to become more rigid, and increases the drag coefficient $C_D$. The free-slip boundary condition that occurs at a gas-liquid interface for an ideal purified system (e.g. \textquoteleft hyper clean\textquoteright\ water) breaks down, and the rise speed of the contaminated bubble decreases with increasing surfactant concentration, approaching the behaviour of a rigid body for sufficient surfactant concentration.

The effect of drag coefficient enhancement due to the adsorption of surfactants has been studied in great detail. Readers are referred to \cite{2000_Magnaudet}, \cite{2006_Palaparthi} and \cite{2011_Takagi} for detailed reviews from the perspective of hydrodynamics, and \cite{2020_Manikantan} from the perspective of surfactant dynamics. An early study \citep{1974_Bachhuber} measured the terminal velocity of small bubbles (bubble Reynolds number $Re_p\sim O(10)$) rising in distilled water at two heights in the flow and observed that the velocity at the lower height was in good agreement with that expected for a clean spherical bubble, while that measured at the greater height was consistent with a particle having a value of $C_D$ that corresponds to a rigid sphere. Their interesting observation reflects at least two well-known aspects of how surfactants impact bubble motion. First, water used in typical lab conditions is contaminated, possibly containing considerable amounts of surfactants that can influence the motion of bubbles. Second, it takes a finite time (that depends on the properties of the surfactants) for the surfactants to be adsorbed at the bubble surface before reaching an equilibrium state, with the implication that bubble rise velocities decrease with increasing travel distance until reaching a constant value \citep{1986_Durst,2014_Tagawa,2021_Hessenkemper}. 

The fact that the bubble rise velocity decreases in the presence of surfactants also leads to a smaller inertial force experienced by bubbles. As a result of this, for a fixed bubble size, the bubble shape is less flattened in a contaminated system than it is in a purified system, despite the fact that surface tension is reduced by surfactants.      

The aforementioned modifications in bubble rise velocity, surface boundary condition and shape due to surfactants have a strong impact on the wake structure and the path instability of a rising bubble. \cite{2014_Tagawa} used experiments to investigate the effect of surfactants on the path instability of an air bubble rising in quiescent water, and categorized the rising bubble trajectories as straight, helical, or zigzag based on the bubble Reynolds number and surface slip condition. They observed that the trajectory of the bubble was first helical and then transitioned to a zigzag motion in Triton X-100 solution -- a phenomenon never reported in purified water \citep{2001_Mougin,2006_Shew}. \cite{2018_Pesci} used numerical simulations to study the effects of soluble surfactants on the dynamics of a single bubble rising in a large spherical domain. They found that if the surface contamination is sufficiently high, a quasi-steady bubble velocity can be obtained over a wide range of surfactant concentrations, independent of the exact concentration value in the bulk. Furthermore, they also observed a transition from a helical to a
zig-zag rising bubble trajectory as reported experimentally by \cite{2014_Tagawa}, but also found that the nature of the trajectory depends on both the initial surface and bulk surfactant contamination. Pesci et al. also looked at the vorticity in the vicinity of the bubble and observed strong vorticity production very close to the bubble surface due to Marangoni forces, while in clean water this behaviour does not appear for path unstable bubbles \citep{2006_Mougin}. \cite{2009_Legendre} performed numerical simulations to study the two-dimensional flow past a cylinder and investigated the influence of a generic slip boundary condition on the wake dynamics. They showed that slip markedly decreases the vorticity intensity in the wake. \cite{1996_Mclaughlin} and \cite{1997_Cuenot} used a stagnant-cap approximation to compare the wake structure produced by contaminated bubbles and solid spheres. The former study revealed that the wake volume for contaminated bubbles is larger than for solid spheres moving at the same Reynolds number, and the latter study found that the wake length is larger for contaminated bubbles. They explained that the increase of the wake size is caused by the abrupt change to the dynamic boundary condition where the transition from a quasi-shear-free (the upper part of the bubble) to a quasi-no-slip (the rear region) condition generates strong vorticity at the interface. As a result, there is more vorticity injected into the flow for contaminated bubbles than for a rigid sphere with a uniform no-slip condition, resulting in a larger wake.

The observations summarized above were mainly for isolated bubbles in the absence of turbulence. Swarms of bubbles can however induce strong background turbulence, and the observations above could indicate that turbulence generated by wakes in bubble swarms could depend strongly on the degree of contamination in the fluid. Direct numerical simulations (DNS) have revealed the following properties of dilute dispersed bubbly flows: (i) the liquid velocity fluctuations are highly anisotropic, with fluctuations that are much larger in the direction of the mean bubble motion \citep{2013_Lu,2020_Ma,2022_Cluzeau}. It was recently demonstrated that this strong anisotropy also exists at the small-scales of BIT due to the energy being injected at the scale of the bubble \citep{2021_Ma}. (ii) The probability density functions (PDFs) of all fluctuating velocity components are non-Gaussian, with the PDF of the vertical velocity fluctuations being strongly positively skewed, while the other two directions have symmetric PDFs \citep{2011_Roghair,2013_Riboux}. (iii) There is strong enhancement of the turbulent kinetic energy dissipation rate in the vicinity of the bubble surface \citep{2016_Santarelli_b,2019_Cluzeau}. (iv) The energy spectra in either the frequency or wave number space exhibit a $-3$ power law scaling for a subrange for all the components of the fluctuating fluid velocity \citep{2020_Pandey,2021_Innocenti}. (v) \cite{2020_Pandey} and \cite{2021_Innocenti} showed that on average the energy transfer is from large to small scales in BIT, however, there is also evidence of an upscale transfer when considering the transfer of energy associated with particular components of the velocity field \citep{2021_Ma}. (vi) For a bubbly flow with background turbulence, the flow intermittency is significantly increased by the addition of bubbles to the flow when compared to the corresponding unladen turbulent flow with similar bulk Reynolds number \citep{2012_Biferale,2021_Ma}. Note that some of these DNS studies effectively considered contaminated bubbly flows since they used no-slip conditions on rigid bubble surfaces, while others used slip boundary conditions, and a clear understanding of exactly how the aforementioned properties of BIT depend on contaminants in the flow is not available.  For further details on the relevant DNS studies, readers are referred to the recent reviews of \cite{2018_Risso} and \cite{2020_Mathai}.  

Several experiments have also observed these properties for bubbly turbulent flows \citep{1991_Lance,2005_Rensen,2010_Riboux,2013_Mendez,2019_Lai,2021_Masuk}. However, in most of these experiments there will be some contamination in the liquid, and the impact of this on the results is not understood. Indeed, a systematic investigation into how contamination affects the properties of BIT is still missing, to the best of the authors’ knowledge.  Therefore, in the present paper we systematically explore the effect of surfactants on the properties of BIT produced by bubble swarms, considering both single-point and two-point turbulence statistics to characterize large and small scale flow properties.  The rest of this paper is organized as follows. In \S\,\ref{sec: Experimental method}, we describe the experimental set-up and the measurement techniques. We then give a brief overview regarding the effect of the surfactant on the single bubble behaviour in the chosen solution (\S\,\ref{sec: Surfactant & single bubble}), followed by a presentation of the single-point statistics for the bubble swarm in \S\,\ref{sec: one-point}. Finally, the multipoint results from the experiments that give insights into the properties of the flow at different scales are divided into two parts, namely, the flow anisotropy in \S\,\ref{sec: Anisotropy} and the extreme events in \S\,\ref{sec: Extreme velocity increment}.

\section{Experimental method}\label{sec: Experimental method}

\subsection{Experimental facility}\label{sec: Experimental facility}

\begin{figure}
	\centering
	\includegraphics[height=9.5cm]{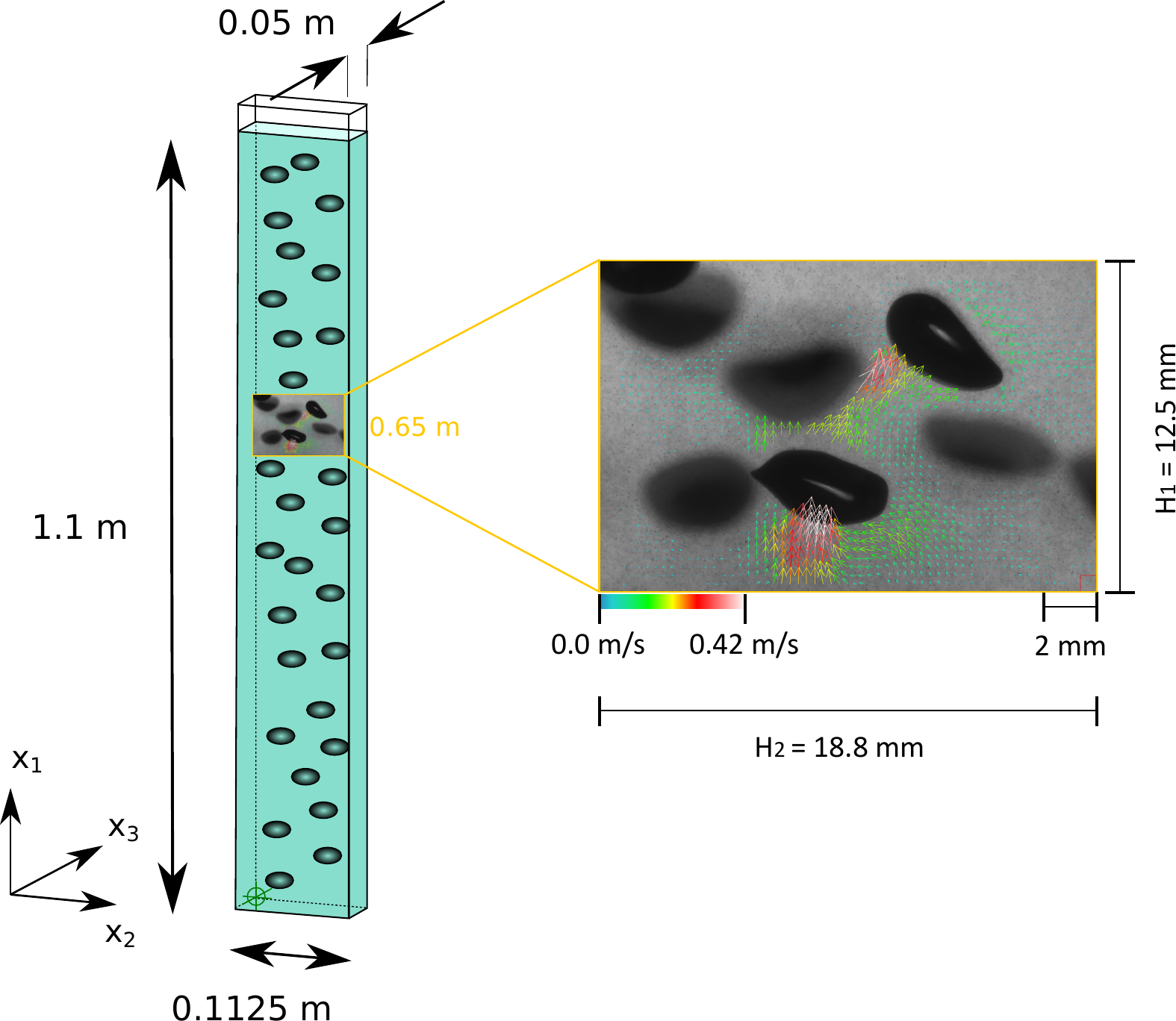}
	\caption{Sketch of the bubble column used in the experiments (note that in the actual experiment, the number of bubbles in the column is $O(10^3)$). The sketch is not to scale; the column depth is many times larger than the bubble diameter. Right shows an instantaneous realization of velocity vector over the FOV in the case \textit{LaTap} with two in-focus bubbles recognizable by their sharp interfaces and associated wakes.} \label{fig: Bubble_column}
\end{figure}

The experimental apparatus is identical to that in \cite{2022_Ma}, and we therefore refer the reader to that paper for additional details; here, we summarize. Figure \ref{fig: Bubble_column} shows a sketch of the experimental set-up, consisting of a rectangular column (depth $50\,\mathrm{mm}$ and width $112.5\,\mathrm{mm}$) made of acrylic glass with a water fill height of $1,100\,\mathrm{mm}$. Air bubbles are injected through $11$ spargers which are homogeneously distributed at the bottom of the column. 

We use tap water in the present work as the base liquid and add 1-Pentanol as an additional surfactant with varying bulk concentration $C_\infty$ of $0$ ppm, $333$ ppm and $1,000$ ppm. The surfactants properties and the single-bubble behaviour in these solutions will be discussed in detail in \S\,\ref{sec: single bubble}. Note that the tap water will already be sightly contaminated prior to adding the surfactants, and the bubbles can behave differently in this tap water without surfactants compared to that in a pure water system \citep[no surfactants, see e.g.][]{2008_Veldhuis,2011_Takagi}. For the bubble sizes considered in the present study (with $d_p>2$ mm), however, the slight contamination in the tap water does not have a significant effect on the bubble motion \citep{2001_Ellingsen}. This point was confirmed as well by our previous study \citep{2021a_Hessenkemper} for the bubble sizes considered here, with both the bubble rise velocity and bubble aspect ratio showing little difference between tap water and purified water systems.

We consider two different bubble sizes by using spargers with different inner diameters. For each bubble size, we maintain the same gas inlet velocity and ensure that all cases are not in the heterogeneous regime of dispersed bubbly flows. In total, we have six mono-dispersed cases labelled as \textit{SmTap}, \textit{SmPen}, \textit{SmPen+}, \textit{LaTap}, \textit{LaPen}, and \textit{LaPen+} in table \ref{tab: bubble para}, including some basic characteristic dimensionless numbers for the bubbles. Here, \textquotedblleft\textit{Sm}/\textit{La}\textquotedblright\ stand for smaller/larger bubbles and \textquotedblleft\textit{Tap}/\textit{Pen}/\textit{Pen+}\textquotedblright\ stand for corresponding cases, having 1-Pentanol concentration with $C_\infty=$ $0$ ppm, $333$ ppm and $1,000$ ppm, respectively. It should be noted that the three cases with larger bubble sizes have higher gas void fraction than the three cases with smaller bubbles. This is because in our setup it is not possible to have the same flow rate for two different spargers while also maintaining a homogeneous gas distribution for mono-dispersed bubbles.  

\begin{table}
	\begin{center}
		\def~{\hphantom{0}}
		\begin{tabular}{ccccccc}
			Parameter  
			&\textsl{SmTap}&\textsl{SmPen}&\textsl{SmPen+}&\textsl{LaTap}&\textsl{LaPen}&\textsl{LaPen+}\\	
			\hline
			$C_\infty$ (ppm) &0&333&1000&0&333&1000\\
			$\alpha$   &0.46\%&0.47\%&0.54\%&1.36\%&1.33\%&1.33\%\\
			$d_p \;(\mathrm{mm})$    &3&2.5&2.6&4.1&3.8&3.8\\
			$\chi$     &1.8&1.3&1.2&1.8&1.5&1.3\\
			$Ga$       &504&394&420&815&729&719\\
			$Eo$       &1.27&0.91&0.99&2.4&2.07&2.03\\
			$Re_p$     &798&587&528&986&866&817\\
			$C_D$      &0.531&0.599&0.841&0.900&0.933&1.018\\
		\end{tabular}
		\caption{Selected characteristics of the six bubble swarm cases investigated. Here, $C_\infty$ is the bulk concentration of 1-Pentanol, $\alpha$ the averaged gas void fraction, $d_p$ the equivalent bubble diameter, $\chi$ the aspect ratio, $Ga\equiv\sqrt{\left|\pi_\rho-1\right|gd_{p}^{3}}/\nu$ the Galileo number, $Eo\equiv\Delta\rho gd_p^2/\sigma$, the E\"{o}tv\"{o}s number. The values of $Re_p$, the bubble Reynolds number and $C_D$, the drag coefficient are based on $d_p$ and the bubble to fluid relative velocity obtained from the experiment.} \label{tab: bubble para}
	\end{center}
\end{table}

\subsection{Flow imaging}\label{sec: Flow imaging}

\subsubsection*{Liquid phase}
For all measurements stated below, we use a $2.5$ megapixel CMOS camera (Imaging Solutions) equipped with a $100$ mm focal length macro lense (Samyang). To measure the liquid velocity we use Particle Shadow Velocimetry (PSV), which is similar to planar Particle Image Velocimetry (PIV) with the only difference being that backlight illumination together with a shallow Depth of Field (DoF) is used. The velocity is then determined by correlating the displacement of sharp tracer particles inside a narrow sharpness region. A detailed description of the image processing procedure used can be found in \cite{2018_Hessenkemper}. The measurement setup and data acquisition are similar to our previous work (Ma et al., 2022), so that only the key aspects for the liquid velocity measurements are stated in what follows. 

The liquid velocity measurements take place along the $x_1-x_2$ symmetry plane in the centre of the depth ($x_3$). The measurement height with $x_1 = 0.65$ m is based on the centre of the field of view (FOV) (figure \ref{fig: Bubble_column}). We use $10\,\mathrm{\mu m}$ hollow glass spheres (Dantec) as tracer particles with estimated Stokes number $O(10^{-3})$, and they are hydrophilic and so are not adsorbed by the bubble surface. The flow section is illuminated with a $200$ W LED-lamp and an f-stop of $2.8$ is set at the camera lens to provide an effective shallow DoF $\sim 370\,\mathrm{\mu m}$. The captured images cover a FOV of $18.8\,\mathrm{mm}\,(H_2) \times 12.5\,\mathrm{mm}\,(H_1)$, which results in a pixel size of  $9.8\,\mathrm{\mu m}$. For each case, $15000$ image pairs are recorded, with a time delay of about $0.5$ sec before the next image pair is acquired, providing us $15000$ uncorrelated velocity fields. The comparatively large bubble shadows in the images are masked and the corresponding areas are not considered in the following correlation step. The final interrogation window is $64 \times 64$ pixels with $50\%$ overlap, resulting in a vector spacing of $0.627$ mm. Here, we use standard PIV-processing steps to determine the velocity, including  multi-pass/window refinement steps and universal outlier detection. A representative transient FOV for the case \textit{LaTap} is shown in figure \ref{fig: Bubble_column}, overlaid with the resulting liquid velocity vector field. In this figure, two in-focus bubbles can be identified by their sharp interfaces and the associated wakes in the velocity vector field.

\subsubsection*{Gas phase}

\begin{figure}
	\centering
	\includegraphics[height=6.8cm]{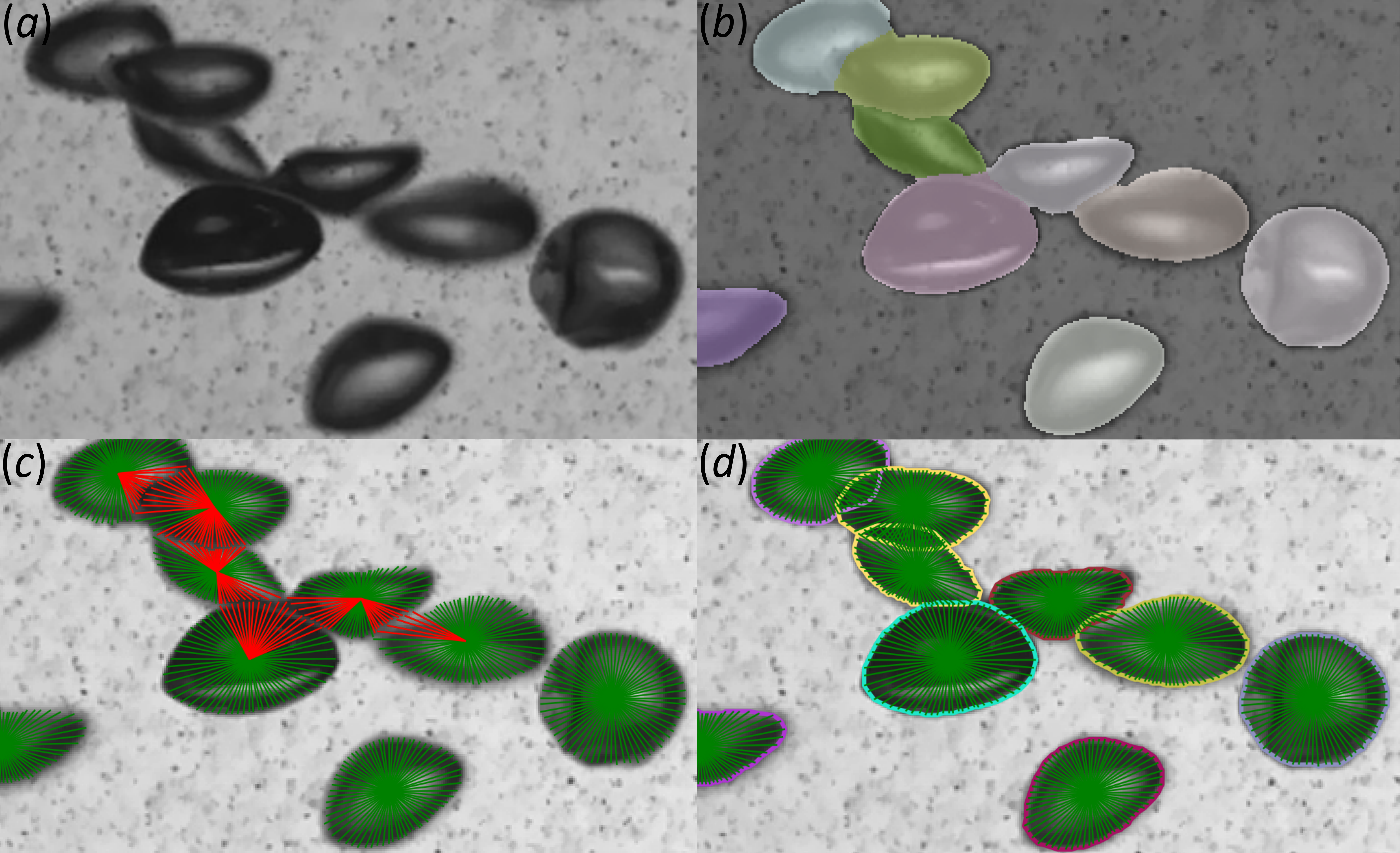}
	\caption{(\textit{a}) Example image of the bubbles from a fragment in the middle of figure \ref{fig: real bubbles}(\textit{b}), identified by the area marked with white dashed line there for the \textit{LaTap} case. (\textit{b} - \textit{d}) Three steps to reconstruct hidden bubble parts for an irregular shaped bubble: (\textit{b}) segmentation mask, (\textit{c}) radial distances, and (\textit{d}) ground truth radial distances.} \label{fig: RDC_Principel}
\end{figure}

\begin{figure}	
	\begin{minipage}[b]{1.0\linewidth}
		\begin{minipage}[b]{0.5\linewidth}
			\centering
			\makebox[2em][l]{\raisebox{-\height}{(\textit{a})}}%
			\raisebox{-\height}{\includegraphics[height=4.8cm]{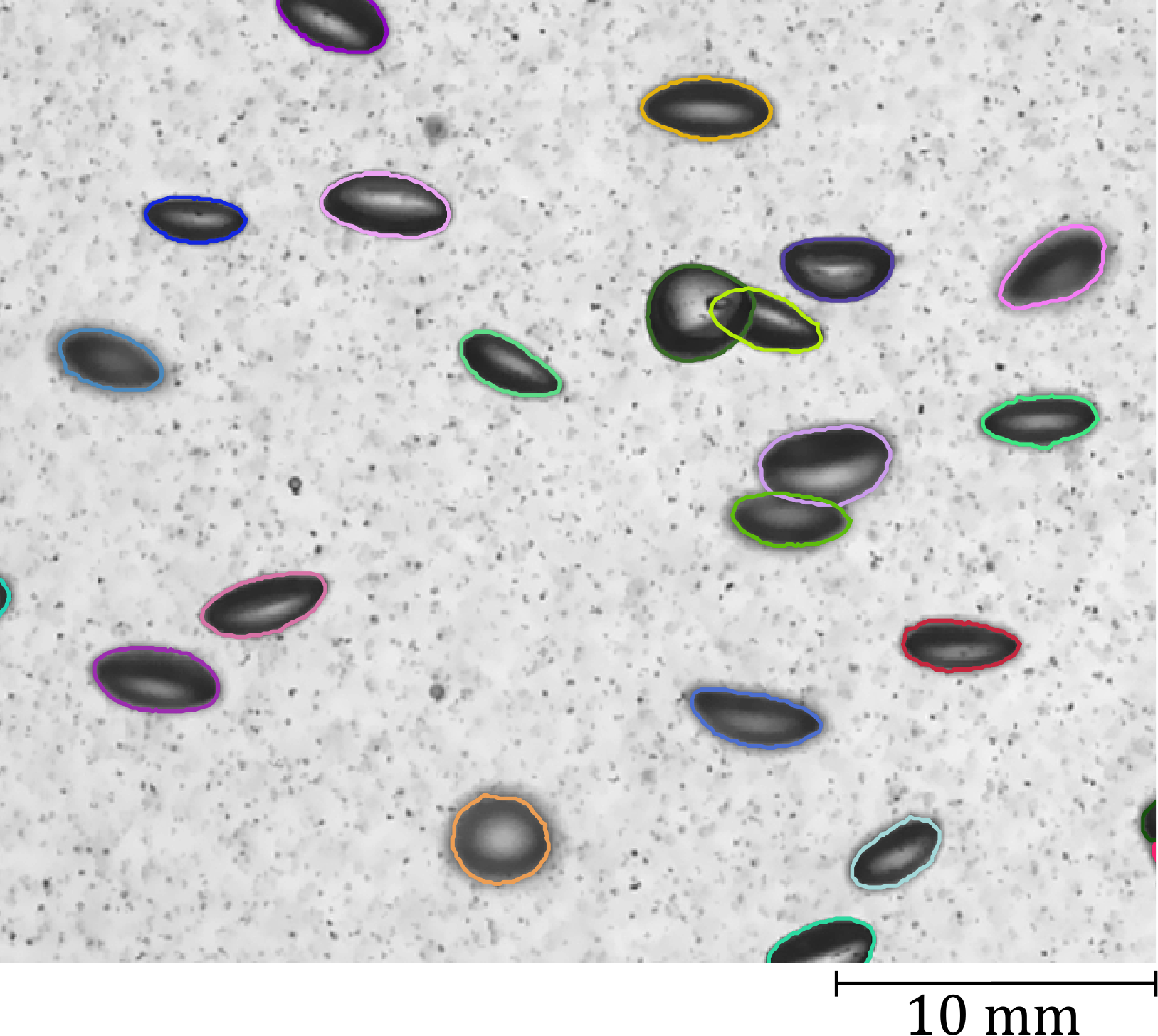}}
		\end{minipage}
		\begin{minipage}[b]{0.5\linewidth}
			\centering
			\makebox[2em][l]{\raisebox{-\height}{(\textit{b})}}%
			\raisebox{-\height}{\includegraphics[height=4.8cm]{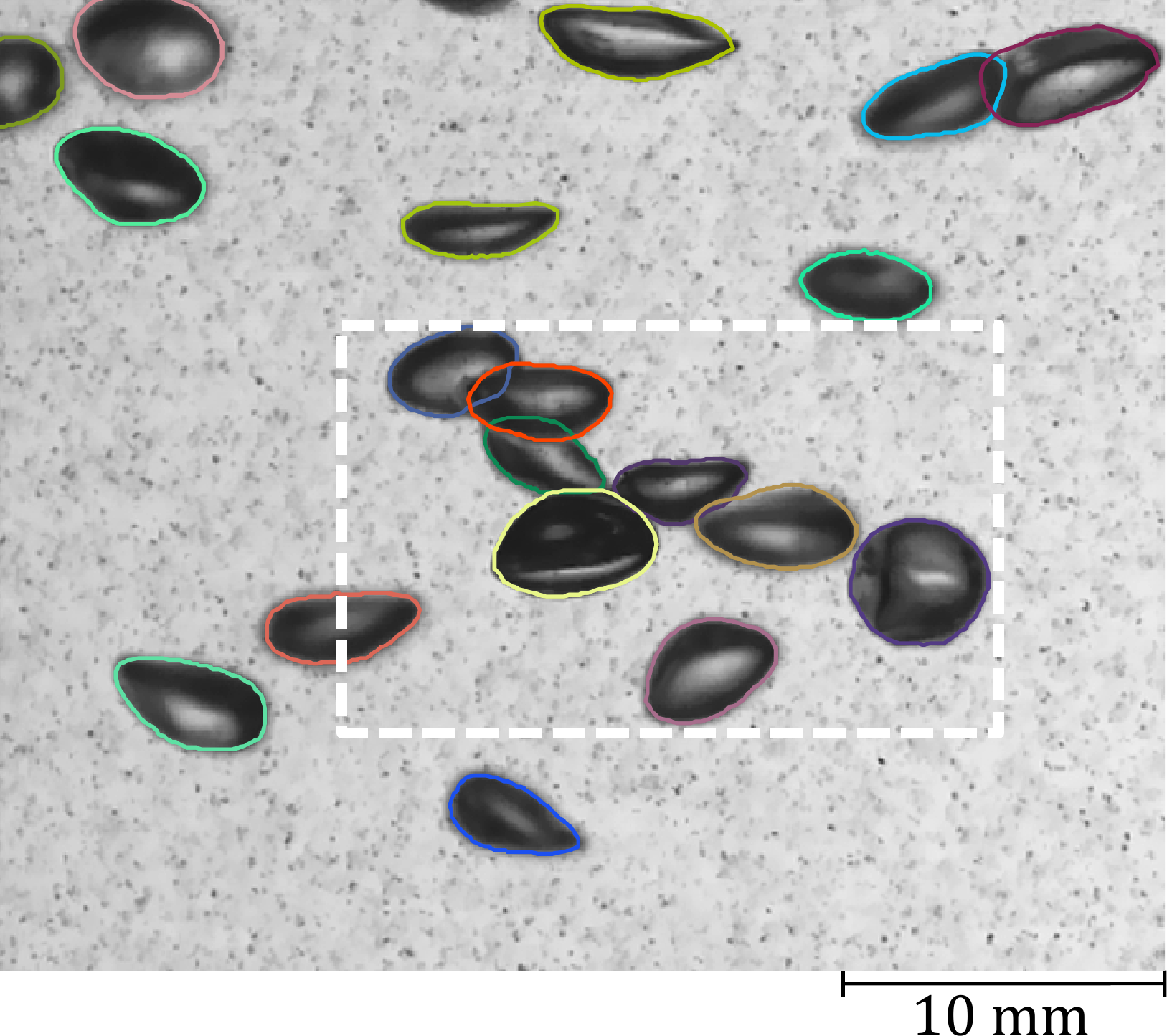}}
		\end{minipage}
	\end{minipage}	
	\begin{minipage}[b]{1.0\linewidth}
		\vspace{3mm}
		\begin{minipage}[b]{0.5\linewidth}
			\centering
			\makebox[2em][l]{\raisebox{-\height}{(\textit{c})}}%
			\raisebox{-\height}{\includegraphics[height=4.8cm]{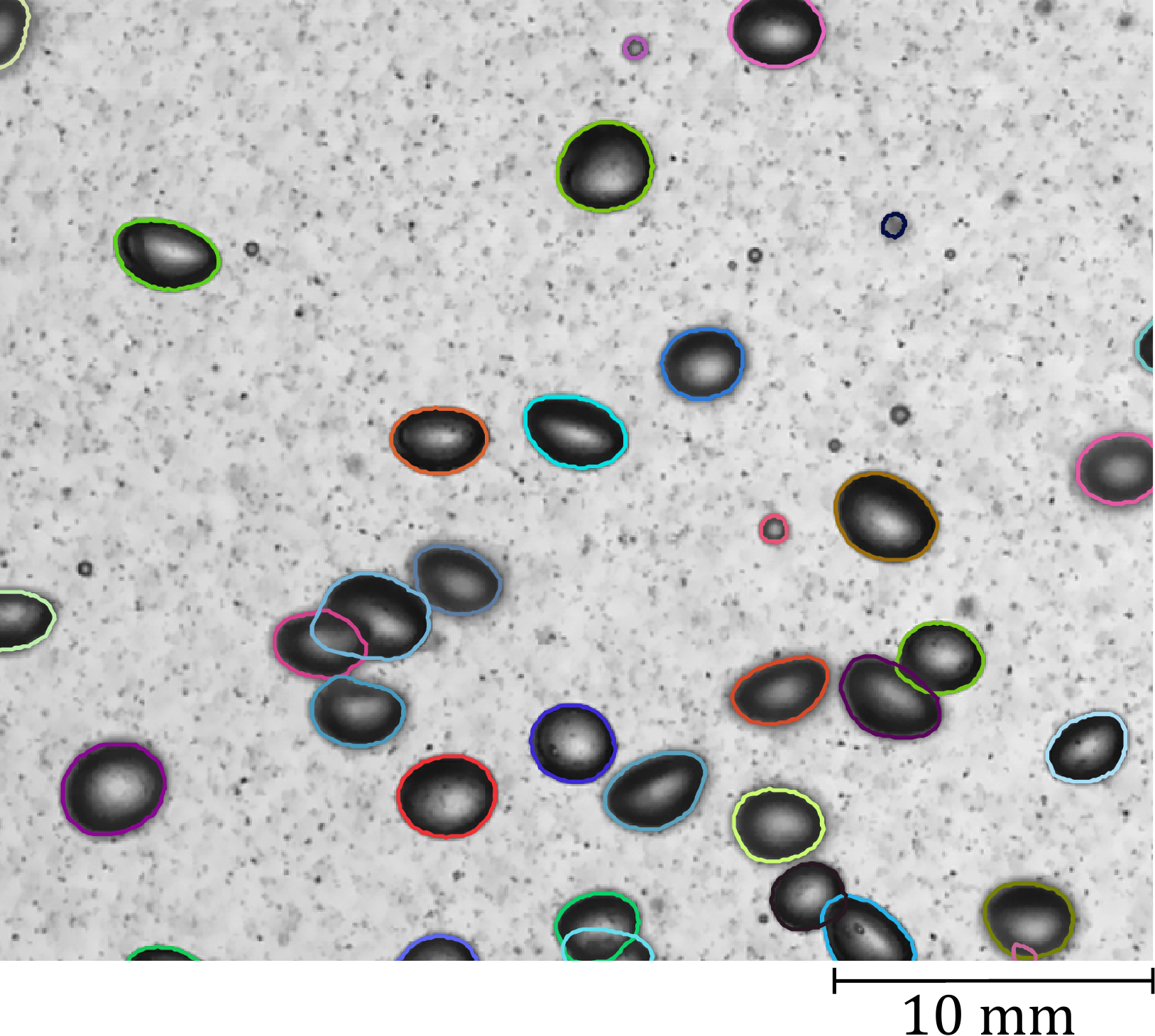}}
		\end{minipage}
		\begin{minipage}[b]{0.5\linewidth}
			\centering
			\makebox[2em][l]{\raisebox{-\height}{(\textit{d})}}%
			\raisebox{-\height}{\includegraphics[height=4.8cm]{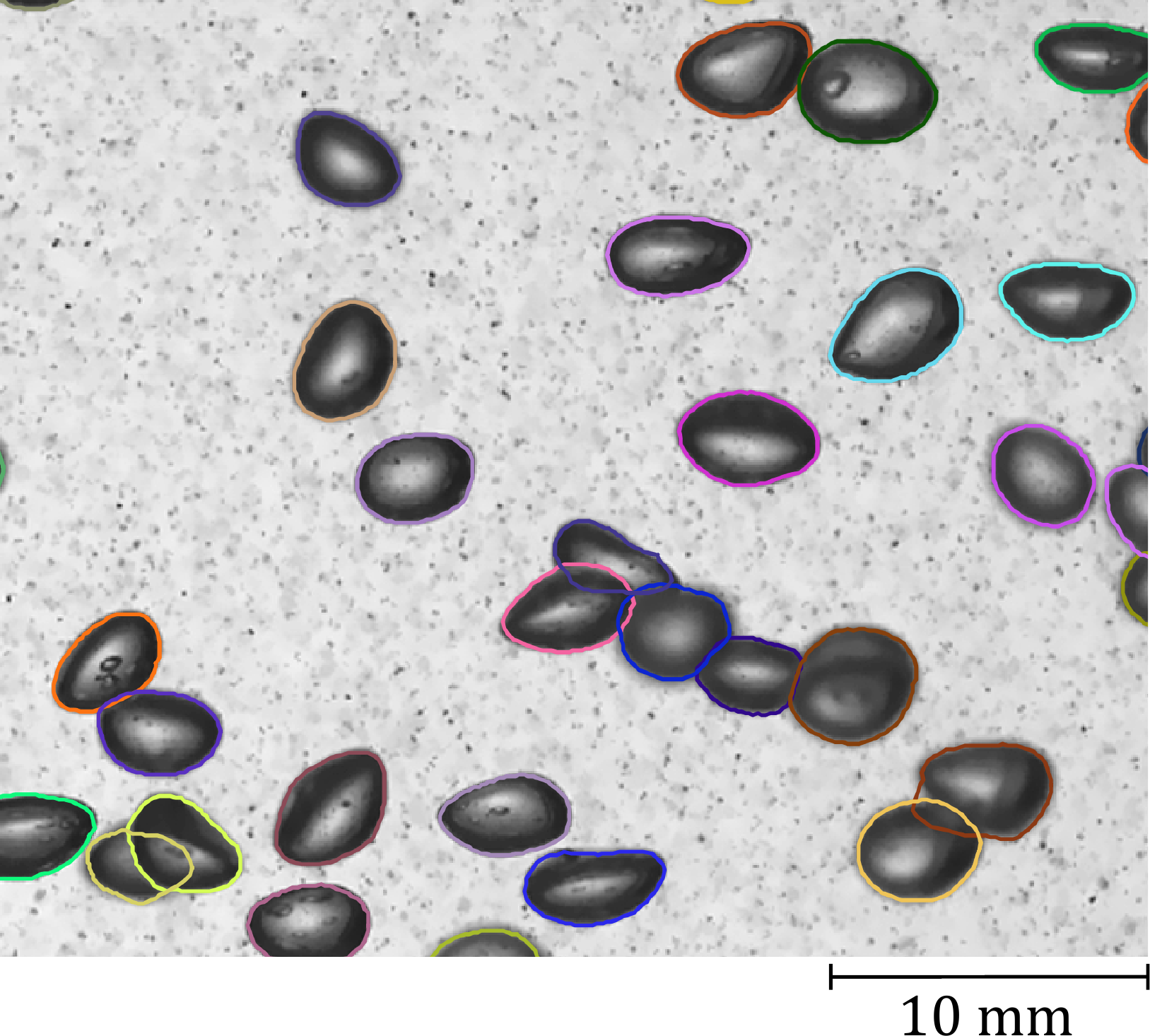}}
		\end{minipage}
	\end{minipage}	
	\begin{minipage}[b]{1.0\linewidth}
		\vspace{3mm}
		\begin{minipage}[b]{0.5\linewidth}
			\centering
			\makebox[2em][l]{\raisebox{-\height}{(\textit{e})}}%
			\raisebox{-\height}{\includegraphics[height=4.8cm]{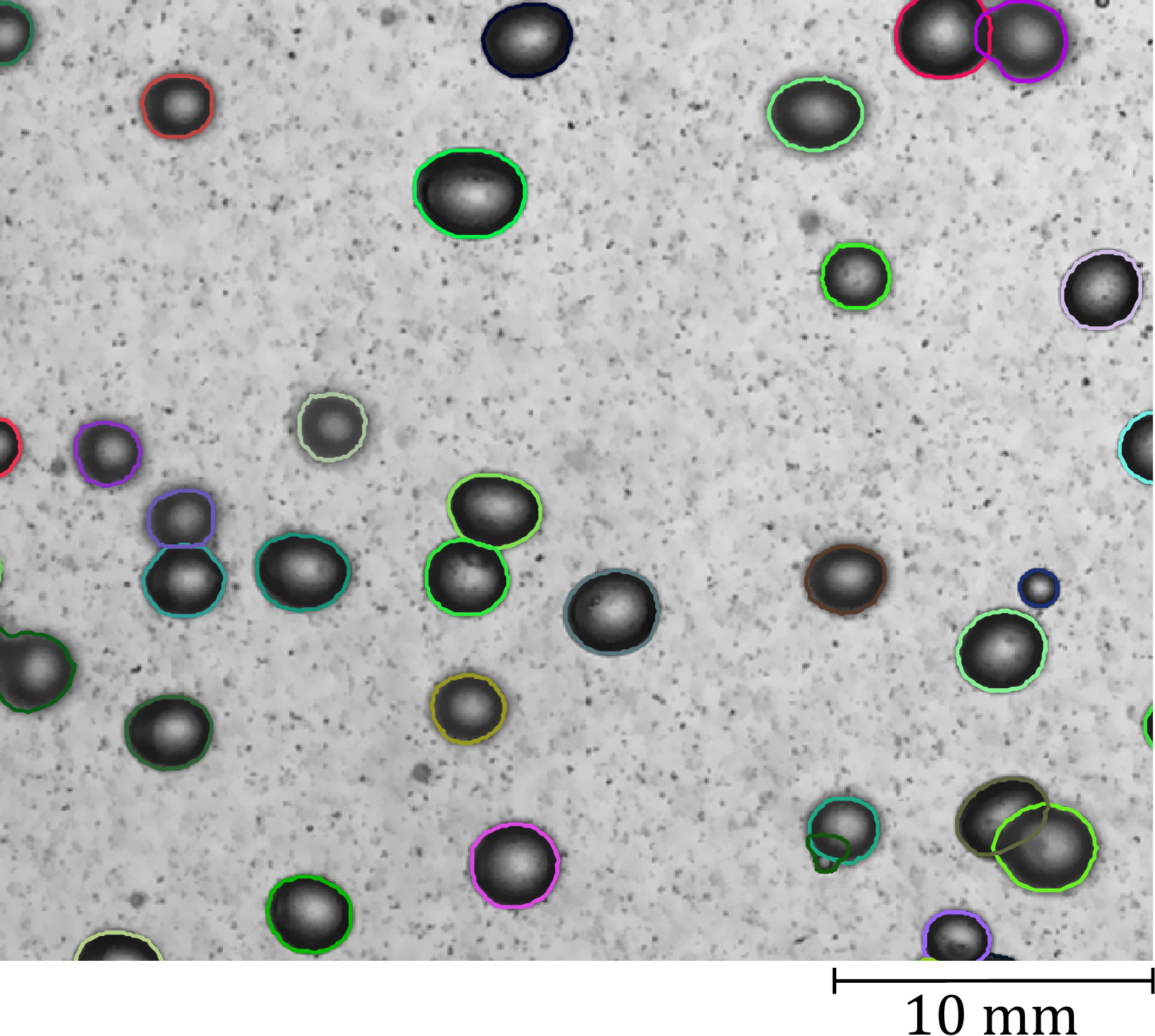}}
		\end{minipage}
		\begin{minipage}[b]{0.5\linewidth}
			\centering
			\makebox[2em][l]{\raisebox{-\height}{(\textit{f})}}%
			\raisebox{-\height}{\includegraphics[height=4.8cm]{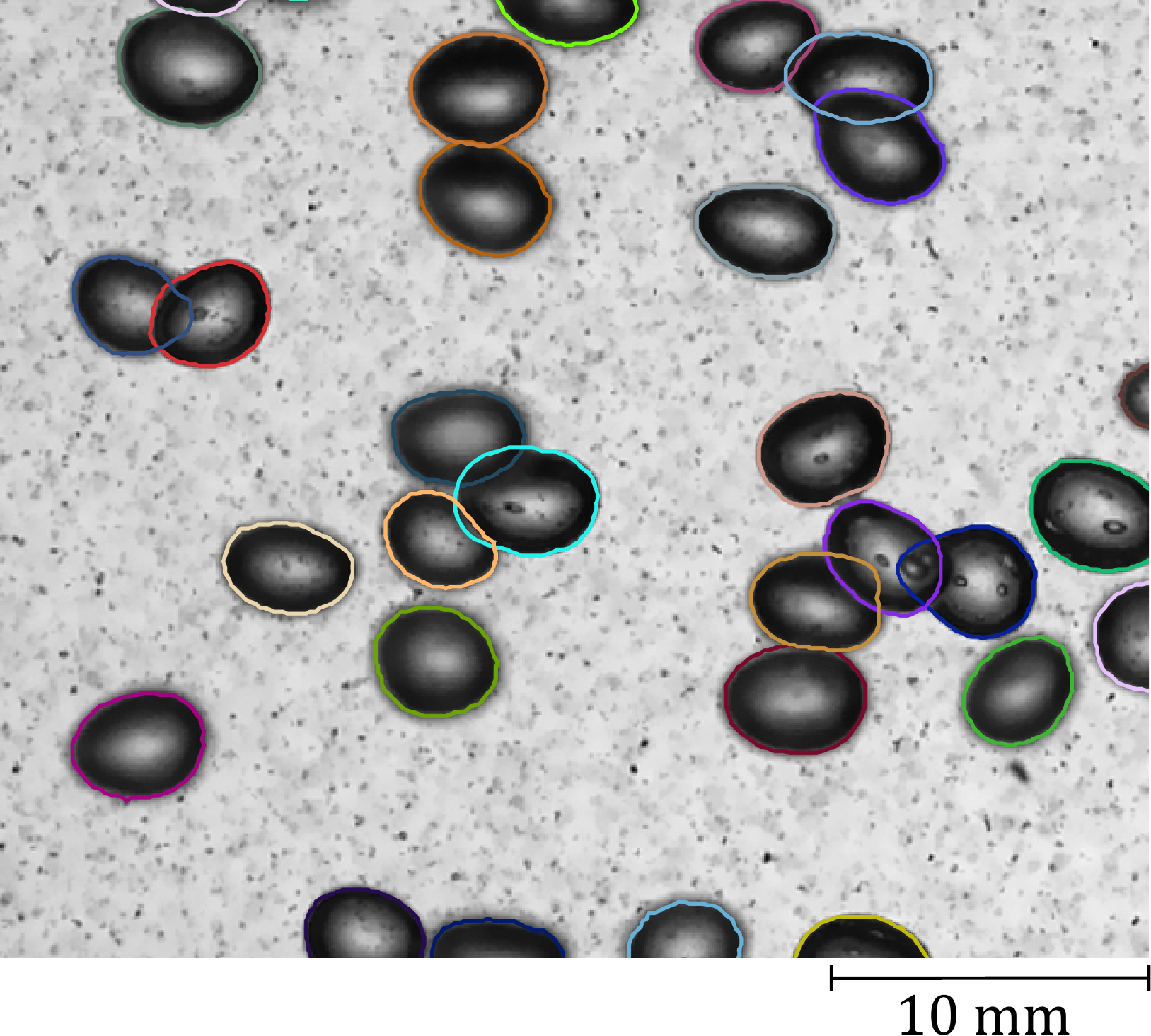}}
		\end{minipage}
	\end{minipage}	
	\caption{Example images of the bubbles with fitted ellipses for an arbitrary instant. Left: smaller bubbles with (\textit{a}) Case \textit{SmTap}, (\textit{c}) Case \textit{SmPen} and (\textit{e}) Case \textit{SmPen+}. Right: larger bubbles with (\textit{b}) Case \textit{LaTap}, (\textit{d}) Case \textit{LaPen} and (\textit{f}) Case \textit{LaPen+}.} \label{fig: real bubbles}
\end{figure}

Similar to our previous work \citep{2022_Ma}, compared to the liquid phase we use a larger FOV $(40\,\mathrm{mm} \times 80\,\mathrm{mm})$ for the gas phase to capture the bubble statistics in all the cases considered. For sufficient statistics, $1500$ image pairs are evaluated at a frame rate of $500$ f.p.s.. The bubbles are detected with a novel convolutional neural network (CNN) based bubble identification algorithm, which is trained to segment bubbles in crowded situations (coloured patches in figure \ref{fig: RDC_Principel}\textit{b}). Furthermore, the hidden part of a partly occluded bubble is estimated with an additionally trained neural network. For this, an equal number of radial rays are generated for each identified segment, starting from the segment centre to its boundary (green and red lines in figure \ref{fig: RDC_Principel}\textit{c}). Afterwards, radial rays that touch neighbouring segments (red lines in figure \ref{fig: RDC_Principel}\textit{c}) are corrected with the neural network in order to account for the occluded part of the bubble. These steps are illustrated
in figure \ref{fig: RDC_Principel}, and we refer the reader to the paper \citep{2022_Hessenkemper} for more technical detail. Here, we show in figure \ref{fig: real bubbles} the example images of detected bubbles with corrected outline for all the six cases. While the $SmTap$, $LaTap$ and $LaPen$ cases have mostly irregular bubble shapes associating with shape oscillations (see supplementary movie), the cases $SmPen$, $SmPen+$ and $LaPen+$ have fixed shapes. The bubble size is calculated using the volume-equivalent bubble diameter of a spheroid as $d_p=(d_{maj}^2d_{min})^{1/3}$, where $d_{maj}$ and $d_{min}$ are the lengths of the major and minor axes of the fitted ellipse, respectively. We observe that the surfactant changes the bubble shape dramatically. For the smaller bubbles with $C_\infty=1000$, the shape is close to a sphere with a small aspect ratio $\chi=d_{maj}/d_{min}=1.2$, while for tap water case, i.e. $C_\infty=0$ we obtain $\chi=1.8$. For the larger bubbles we also have aspect ratios in a similar range, with $\chi$ decreasing from $1.8$ to $1.3$ when going from higher to lower surfactant concentrations.  

After the bubble detection, the centroids are tracked in each image pair to obtain corresponding bubble velocities. Again a shallow DoF is used, which allows to select only sharp bubbles in the column centre by evaluating the grey value derivative along their contour. We choose a grey value derivative threshold that provides a centre region of about $15$ mm thickness $(3d_p\sim4d_p)$ in which the bubbles are evaluated. Some of the important bubble properties are listed in table \ref{tab: bubble para}.

It should be noted here that in the case of ellipsoidal bubbles, i.e. the tap water cases, errors in the determined size are larger due to the bubbles being tilted with respect to the camera. This error can result in a bubble size overprediction of up to $25\%$ for a strongly tilted spheroidal bubble with fixed shape, which is, however, on average much lower due to the normal distribution of the orientation angle with a maximum around zero \citep{2007_Broder}. Using the single bubble data with two cameras (separate experiments reported in \S\,\ref{sec: single bubble}), we estimate the overprediction of the average volume-equivalent diameter for a fixed spheroidal shaped bubble (corresponding to Case \textit{SmTap}) to be about $4\%$. Due to the irregular (wobbling) bubble surfaces, the error for the larger bubbles in the \textit{LaTap} case are unknown, but we expect it to be of the same order of magnitude. The error for the void fraction is about $5-8\%$ \citep{2022_Hessenkemper} and the error of the bubble velocity is around $4\%$.   

\section{Surfactant properties and single bubble behaviour} \label{sec: Surfactant & single bubble}

\subsection{Surfactant properties}\label{sec: Surfactant}

The effect of surfactants on the bubbles arises fundamentally due to their impact on the shear stress at the gas-liquid interface, and depends upon the species of surfactants along with their concentration. The variety of surfactants is huge and our quantitative understanding of their effect on the base-fluid is still in its infancy. There are also sometimes \textquotedblleft untypical\textquotedblright\ scenarios reported. \cite{1998_Ybert} showed the effect of surfactant desorption, leading to a remobilization of the interface by considering bubbles contaminated with sodium dodecyl sulfate (SDS). This surfactant has a significant desorption velocity, and they found that due to the progressive remobilization of the interface, the rise velocity of the SDS-preloaded bubble increases over a large portion of its trajectory (see their figure 12), until reaching a constant value. Similar phenomena was also reported by the same group \citep{2000_Ybert} for short-chain alcohols, where they illustrated bubble rise velocities in ethanol-water solution that were almost indistinguishable from those in pure water, because of its fast desorption kinetics. 

In the present study, we focus on the limit of high bubble P\'{e}clet number, $Pe$ ($=d_p\left\|\boldsymbol{U}^G\right\|/D$, where $d_p$ is the bubble diameter, $\boldsymbol{U}^G$ is the averaged bubble velocity and $D$ is the diffusion coefficient of the surfactant in the liquid), relatively low surfactant concentration compared to the critical micelle concentration (CMC) and low rate of desorption. Under these conditions, surface convection is dominant compared to the adsorption–desorption kinetics and diffusive transport of surfactants on the bubble surface. Surfactants collect in a stagnant cap at the back end of the bubble while the front end is stress free and mobile. \cite{2006_Palaparthi} call this the \textquotedblleft stagnant cap regime\textquotedblright\ -- that most commonly realized in typical cases of bubbles moving in a contaminated solution, and, in this regime the bubbles are significantly affected by the surfactant, as shown by numerical and experimental results of the rise velocity and wake structure.

In our experiments we use 1-Pentanol as the surfactant and use bulk concentrations $C_\infty$ of 0 ppm, 333 ppm and 1000 ppm for the six bubble swarm cases. Tests for a single bubble rising in the column were also conducted with concentrations up to 2000 ppm  (discussed in \S\,\ref{sec: single bubble}). Here, the main reason for choosing 1-Pentanol is that with this surfactant (and these concentrations) the bubbles adapt very quickly to the surfactants and no transitions in their motion type appear within the FOV \citep{2014_Tagawa}.

We measure the surface tension separately with a bubble pressure tensiometer, allowing a dynamic determination of the surface tension. For 1-Pentanol concentration of 1000 ppm, using profile analysis tensiometry \citep{2021_Eftekhari} we observe a surface tension reduction $\sim 7\%$ compared to tap water. The P\'{e}clet number is of order of $10^5$, indicting the convection time scale is very small compared to the time scale of the diffusion on the bubble surface. The solubility for 1-Pentanol is 2500 $\mathrm{mol/m}^3$, which is much higher than the present cases (1000 ppm $\sim9.19\,\mathrm{mol/m}^3$). Micelles do not form in the bulk aqueous phase as indicted by our experimental results to be reported later for a single bubble.

\subsection{Preliminary test for single bubble}\label{sec: single bubble}

To provide reference cases later for the bubble swarms analysis, we first examine the effect of the surfactants on an isolated bubble in the same setup by varying the concentration of 1-Pentanol $C_\infty$ with levels 0 ppm, 333 ppm, 666 ppm, 1000 ppm, 1500 ppm and 2000 ppm in tap water. This is a wider range than that used for the bubble swarms to check if any interface remobilization occurs at high concentration. We use a single sparger in the column centre, but vary two types of spargers sizes for each $C_\infty$, which are also used for the bubble swarm results to generate (approximately) two bubble sizes. Here, we continuously generate single bubbles by setting a small gas flow rate as we also did in our previous investigation \citep{2021_Hessenkemper}. The low generation frequency of ~1 Hz ensures a large enough distance between successive bubbles to avoid any influence from a leading bubble, but allows to study multiple same sized bubbles under the same conditions for better statistics. 

For the evaluation of the single bubble rising trajectory in three-dimensions (3D), we conducted stereoscopic measurements with an additional second camera of the same type placed perpendicular with respect to the imaging direction of the first camera. The single bubble images are also captured with a frame rate of $500$ f.p.s., but with a larger f-stop of $11$, since sharp bubble outlines at all depth positions are required. As no overlapping bubbles have to be distinguished in the single bubble experiments, we use a conventional image processing approach here instead of the CNN-based one that is used for bubble swarms. At first, sharp edges marking the outline of the bubbles are detected with a Canny-edge detector. The solid of revolution is then used to determine the volume of the bubble. Further geometric properties like the bubble major axis, defined as the largest extent in a $2$D projection, and the bubble minor axis, defined as the largest extent perpendicular to the bubble major axis, are extracted as well from the bubble projections \citep{2017_Ziegenhein}. In comparison to the usual approach of fitting ellipses around the contour of detected bubbles, almost the same volume and semi-axis length are obtained, with only minor deviations for irregular shaped bubbles \citep{2019_Ziegenhein}. Afterwards, the centroid of the bubble projections is tracked through successive images to obtain time-resolved instantaneous bubble velocities $\tilde{\boldsymbol{u}}^G$, which can be decomposed into a ensemble-averaged part $\boldsymbol{U}^G$ and a fluctuating part $\boldsymbol{u}^G$. The same decomposition is also used for the liquid phase with $\tilde{\boldsymbol{u}}^L=\boldsymbol{U}^L+\boldsymbol{u}^L$ in the later sections.

\begin{figure}	
	\begin{minipage}[b]{1.0\linewidth}
		\begin{minipage}[b]{0.5\linewidth}
			\centering
			\makebox[0.5em][l]{\raisebox{-\height}{(\textit{a})}}%
			\raisebox{-\height}{\includegraphics[height=3.6cm]{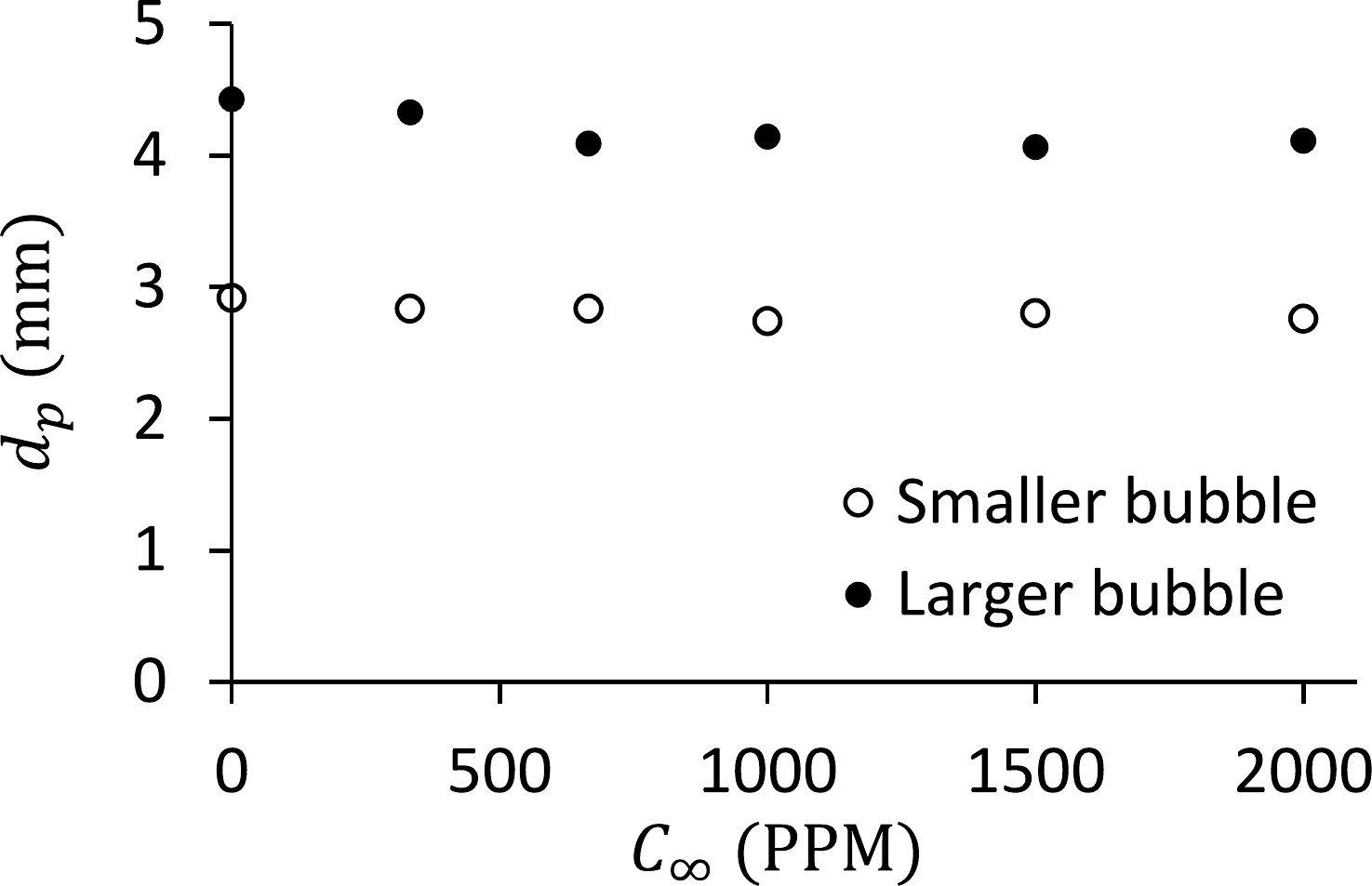}}
		\end{minipage}
		\begin{minipage}[b]{0.5\linewidth}
			\centering
			\makebox[0.5em][l]{\raisebox{-\height}{(\textit{b})}}%
			\raisebox{-\height}{\includegraphics[height=3.6cm]{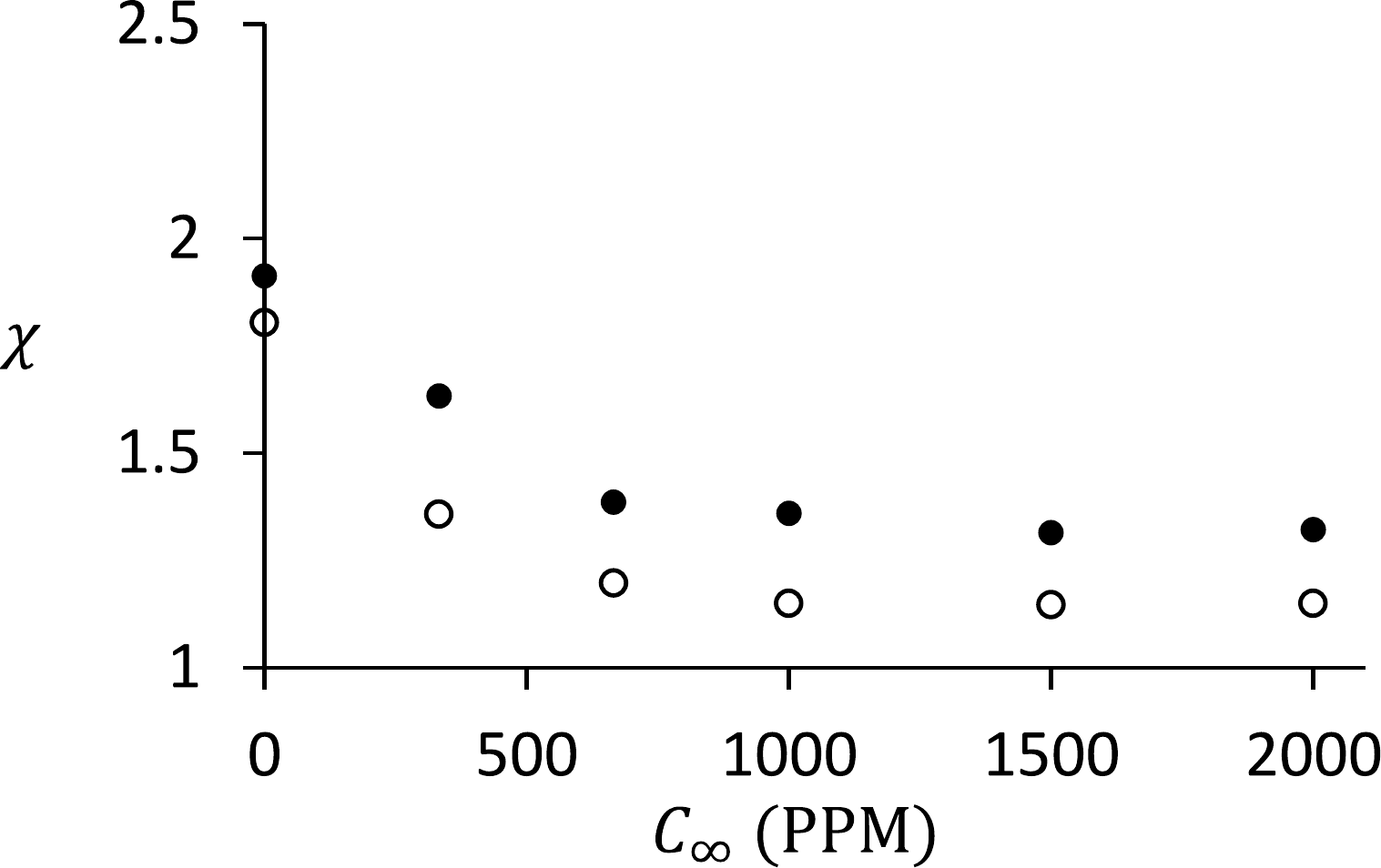}}
		\end{minipage}
	\end{minipage}	
	\begin{minipage}[b]{1.0\linewidth}
		\centering
		\vspace{4mm}
		\makebox[0.5em][l]{\raisebox{-\height}{(\textit{c})}}%
		\raisebox{-\height}{\includegraphics[height=3.6cm]{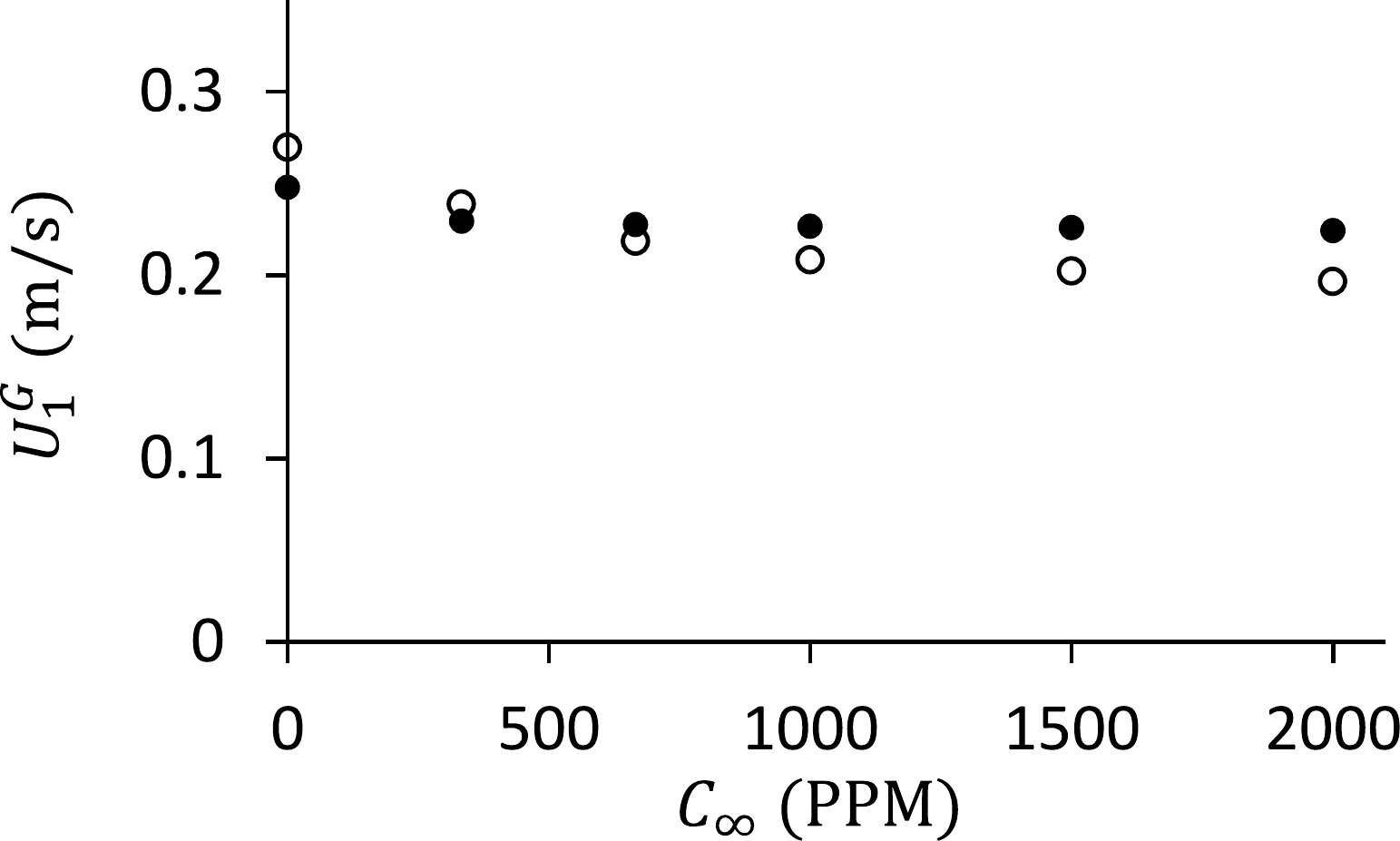}}
	\end{minipage}
	\caption{Measured single bubble size (\textit{a}), aspect ratio (\textit{b}) and rising velocity (\textit{c}) in different 1-Pentanol concentration.} \label{fig: ave single-bub statistic}
\end{figure}

Figures \ref{fig: ave single-bub statistic}(\textit{a}) shows the bubble diameters using two types of spargers size as a function of $C_\infty$. The bubble size is slightly reduced when adding 1-Pentanol for both types of sparger, generating smaller bubbles from 2.9 mm ($C_\infty=0$ ppm) to 2.7  mm ($C_\infty=1000$ ppm) and larger bubbles from 4.4 mm ($C_\infty=0$ ppm) to 4.1  mm ($C_\infty=1000$ ppm). This is due to the influence of the surfactants that reduce the surface tension and hence affect the bubble formation at the rigid orifice \citep{2004_Loubiere,2015_Drenckhan}. A similar trend can also be found in table \ref{tab: bubble para} for the corresponding bubble swarm cases. In contrast to bubble size, the bubble shape changes dramatically (figure \ref{fig: ave single-bub statistic}\textit{b}), with the aspect ratio reducing from $1.8$ at $0$ ppm to $1.15$ at $1000$ ppm for the smaller bubble and $1.9\,(0\,\mathrm{ppm}) \rightarrow1.36\,(1000\,\mathrm{ppm})$ for the larger bubble. Further increasing $C_\infty$ does not change the bubble shapes for the present smaller and larger bubbles. A similar trend can be found for the averaged rise velocity $U^G_1$ for the both bubble sizes, namely, increasing 1-Pentanol concentration above 1000 ppm does not affect $U^G_1$. For both bubble sizes and for all concentrations considered we found no increase of the rise velocity, nor any change in terms of aspect ratio above $C_\infty=1000$ ppm, hence, no surface remobilization has occurred. Based on these we assume that for both bubble sizes a saturated contamination state (at least from the hydrodynamic perspective) is reached at the threshold $C_\infty\approx1000$ ppm. 

\begin{figure}	
	\begin{minipage}[b]{1.0\linewidth}
		\begin{minipage}[b]{0.5\linewidth}
			\centering
			\makebox[1.2em][l]{\raisebox{-\height}{(\textit{a})}}%
			\raisebox{-\height}{\includegraphics[height=4.3cm]{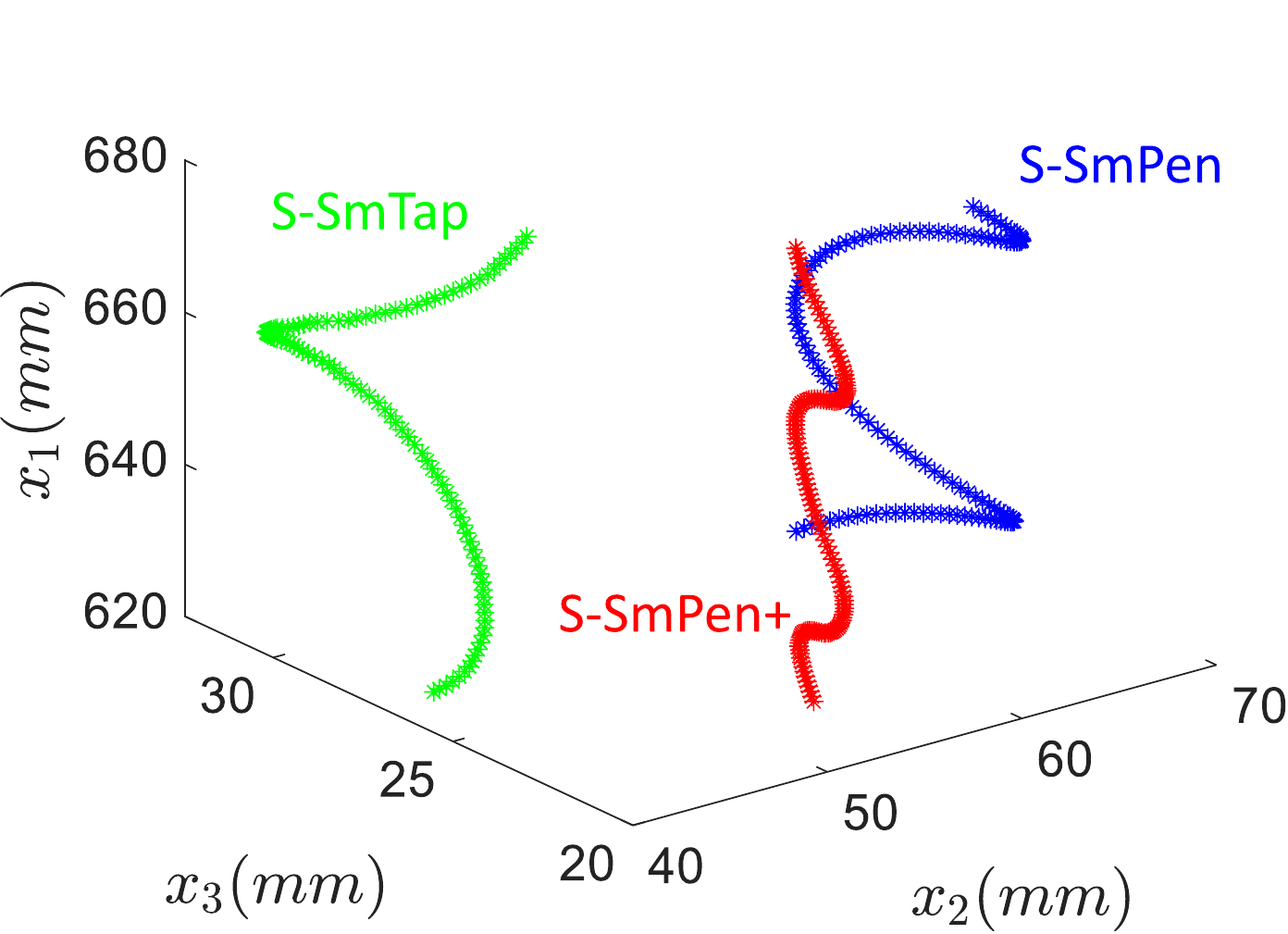}}
		\end{minipage}
		\begin{minipage}[b]{0.5\linewidth}
			\centering
			\makebox[1.2em][l]{\raisebox{-\height}{(\textit{b})}}%
			\raisebox{-\height}{\includegraphics[height=4.3cm]{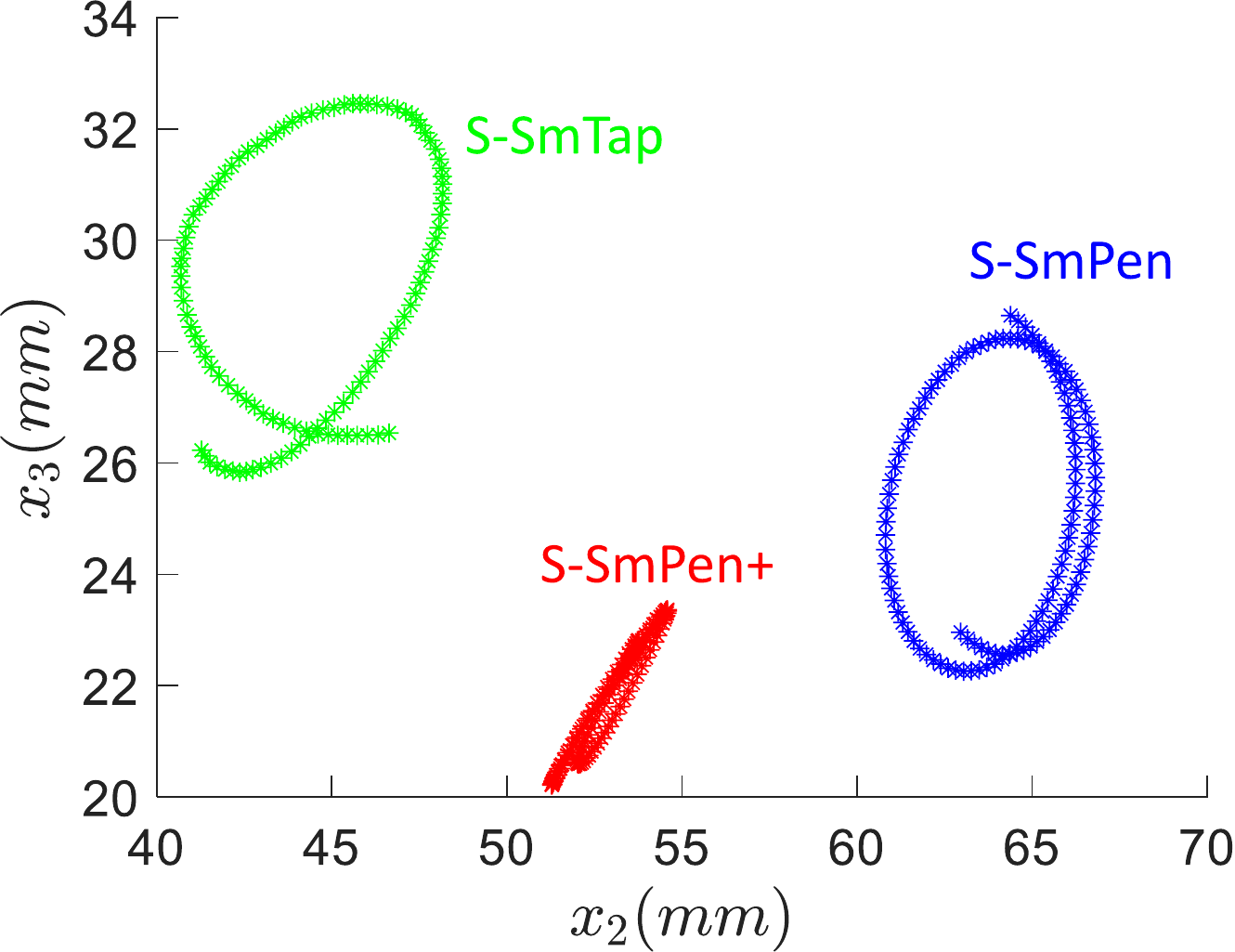}}
		\end{minipage}
	\end{minipage}
	\begin{minipage}[b]{1.0\linewidth}
		\vspace{4mm}
		\begin{minipage}[b]{0.33\linewidth}
			\centering
			\makebox[0.5em][l]{\raisebox{-\height}{(\textit{c})}}%
			\raisebox{-\height}{\includegraphics[height=2.7cm]{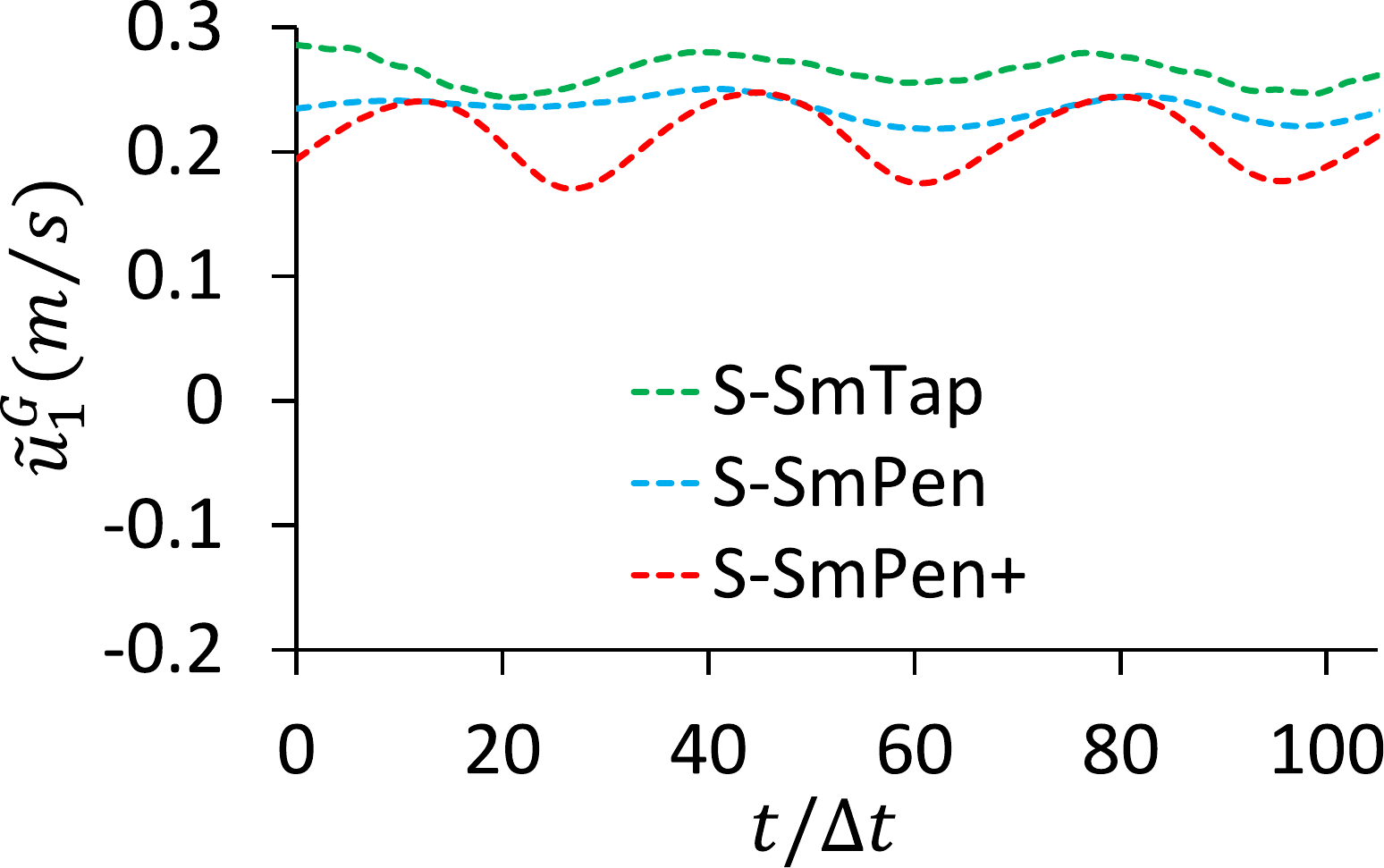}}
		\end{minipage}
		\begin{minipage}[b]{0.33\linewidth}
			\centering
			\makebox[0.5em][l]{\raisebox{-\height}{(\textit{d})}}%
			\raisebox{-\height}{\includegraphics[height=2.7cm]{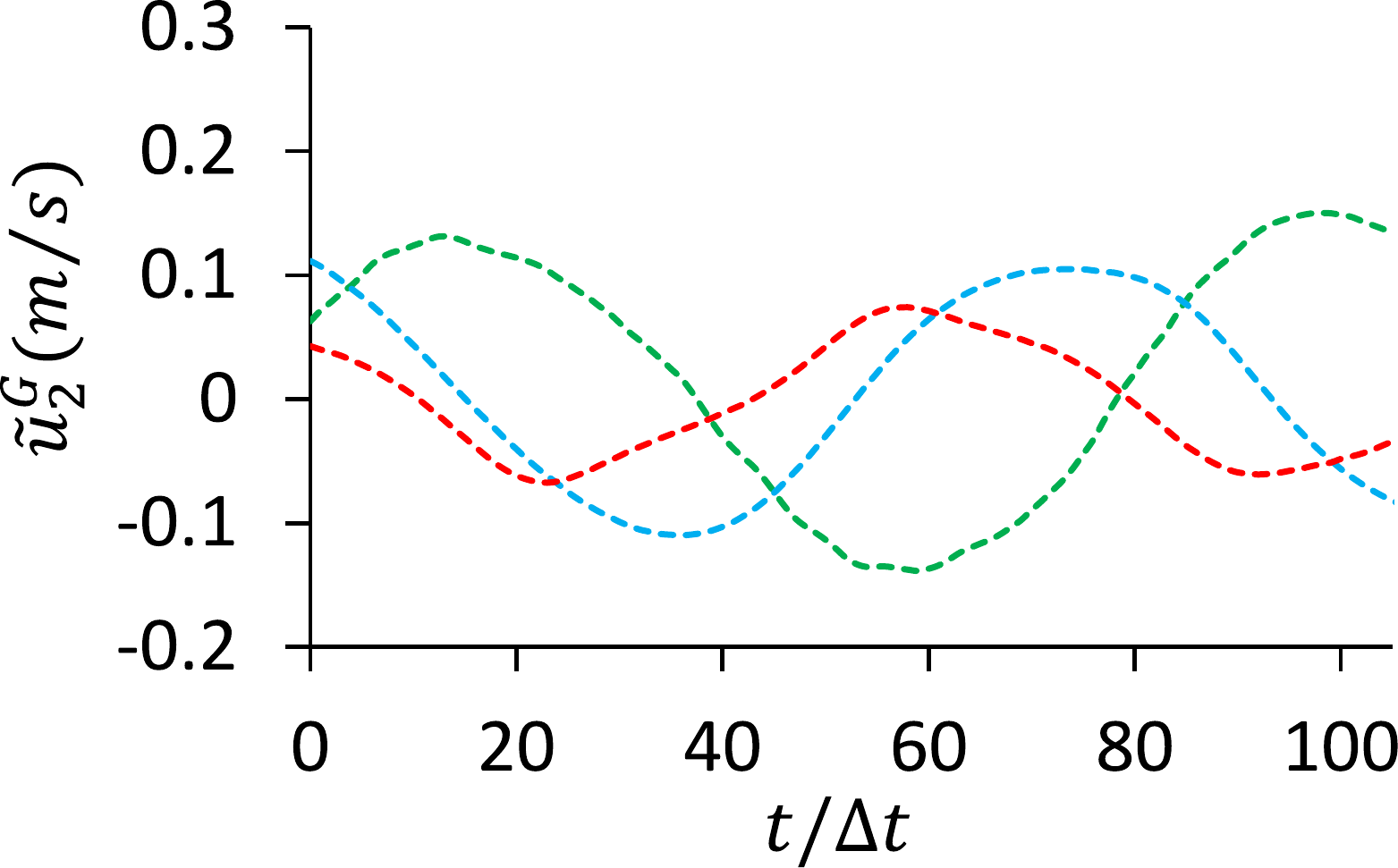}}
		\end{minipage}
		\begin{minipage}[b]{0.33\linewidth}
			\centering
			\makebox[0.5em][l]{\raisebox{-\height}{(\textit{e})}}%
			\raisebox{-\height}{\includegraphics[height=2.7cm]{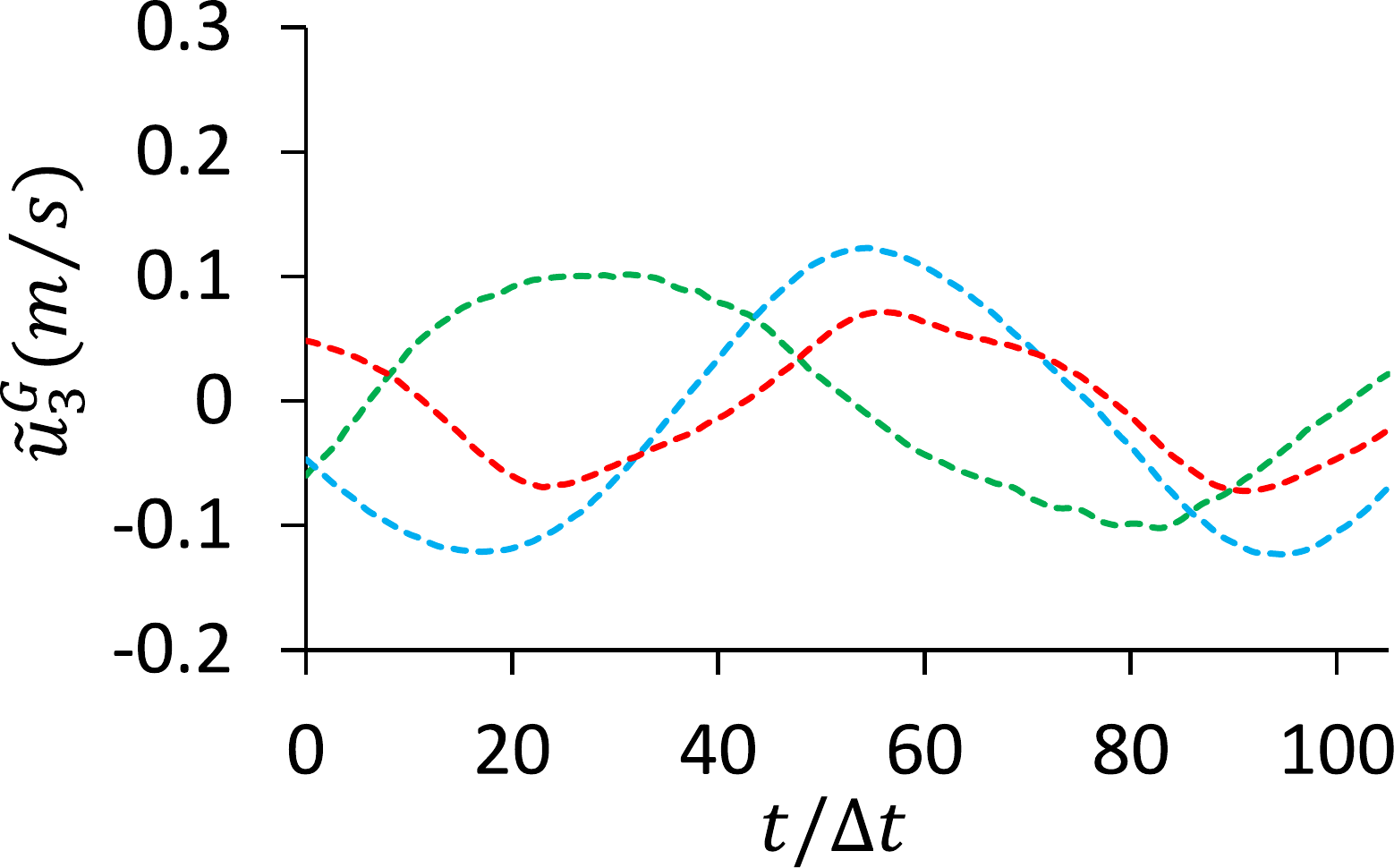}}
		\end{minipage}	
	\end{minipage}	
	\caption{3D trajectories (\textit{a}), top view of those (\textit{b}) and corresponding instantaneous bubble velocities with components $\tilde{u}^G_1$ (\textit{c}), $\tilde{u}^G_2$ (\textit{d}), $\tilde{u}^G_3$ (\textit{e}) over time (normalized by $\Delta t=2$ ms) of smaller single bubbles at $C_\infty=0$ ppm (\textit{S-SmTap}), $C_\infty=333$ ppm (\textit{S-SmPen}) and $C_\infty=1000$ ppm (\textit{S-SmPen+}). All figures are from the same track of the particular case.} \label{fig: velo & trajectory Sin-Sm}
\end{figure}

We now look more in detail at the single bubble cases with $C_\infty=0$ ppm, $C_\infty=333$ ppm and $C_\infty=1000$ ppm, since they have the same set of $C_\infty$ values as the swarm cases we will discuss later. We label these cases \textit{S-SmTap}, \textit{S-SmPen} and \textit{S-SmPen+} for the smaller bubble and  \textit{S-LaTap}, \textit{S-LaPen} and \textit{S-LaPen+} for the larger bubble, respectively (the first letter \textit{S} denotes a single bubble). Figure \ref{fig: velo & trajectory Sin-Sm}(\textit{a}) shows a 3D view of the bubble trajectories for the three smaller bubble cases, while figure \ref{fig: velo & trajectory Sin-Sm}(\textit{b}) shows a view of these trajectories from the top. The trajectories are similar to those found by \cite{2014_Tagawa} for 2 mm bubbles, with helical trajectories at lower 1-Pentanol concentration (see \textit{S-SmTap} and \textit{S-SmPen}) and zigzag rising paths for higher concentration (\textit{S-SmPen+}). These zigzag and helical motions are accompanied by oscillations in the vertical velocity $\tilde{u}^G_1$ plot of figure \ref{fig: velo & trajectory Sin-Sm}(\textit{c}) (corresponding to the paths in figure \ref{fig: velo & trajectory Sin-Sm}\textit{a}), i.e. the bubbles alternately speed up and slow down as they rise. As expected, the \textit{S-SmPen+} case has much higher oscillation in $\tilde{u}^G_1$, compared to the \textit{S-SmTap} and \textit{S-SmPen} cases. This is caused by the relatively unstable wake structure that is associated with zigzag paths, rather than the more stable one associated with the helical trajectories \citep{2016_Cano}. Moreover, we observe the helical trajectories of \textit{S-SmTap} and \textit{S-SmPen} are not as regular as those in \cite{2014_Tagawa} due to our larger bubble size $\sim3$ mm. This is also indicated by the vertical velocities (figure \ref{fig: velo & trajectory Sin-Sm}\textit{c}), showing oscillations in \textit{S-SmTap} and \textit{S-SmPen} compared to the constant $\tilde{u}^G_1$ for the cases with relatively perfect helical rising paths in \cite{2014_Tagawa}.   

\begin{figure}	
	\begin{minipage}[b]{1.0\linewidth}
		\vspace{4mm}
		\begin{minipage}[b]{0.5\linewidth}
			\centering
			\makebox[1.2em][l]{\raisebox{-\height}{(\textit{a})}}%
			\raisebox{-\height}{\includegraphics[height=4.3cm]{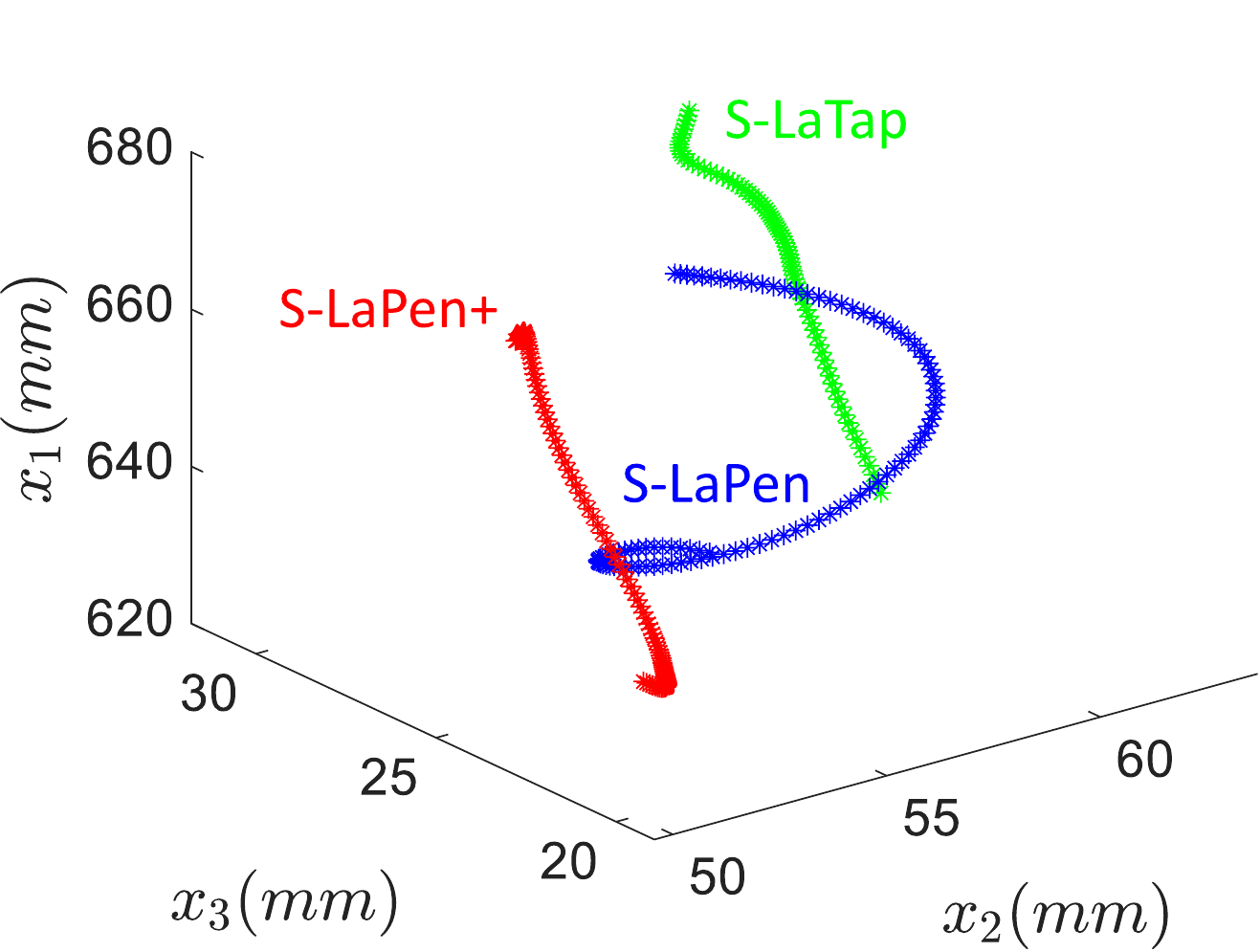}}
		\end{minipage}
		\begin{minipage}[b]{0.5\linewidth}
			\centering
			\makebox[1.2em][l]{\raisebox{-\height}{(\textit{b})}}%
			\raisebox{-\height}{\includegraphics[height=4.3cm]{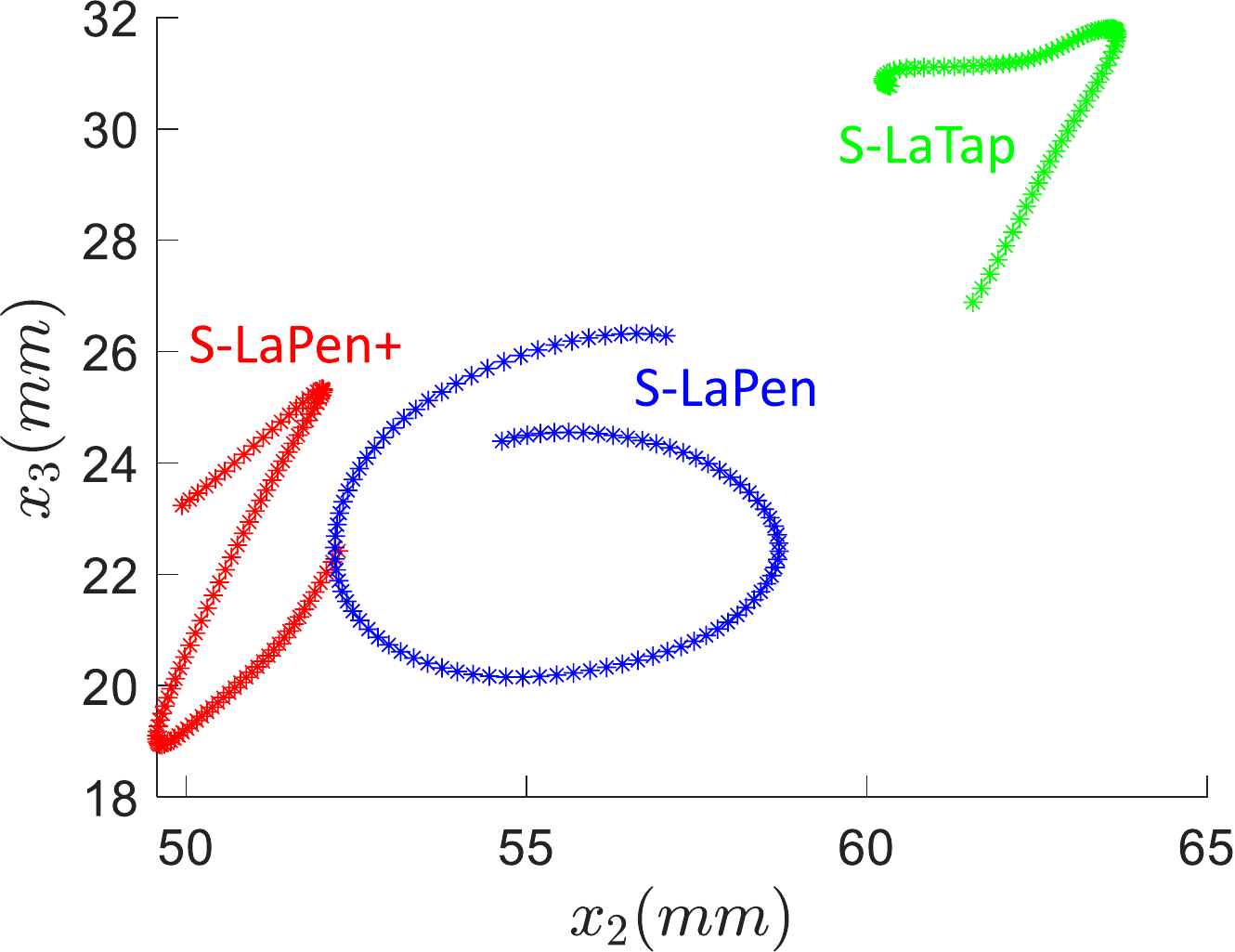}}
		\end{minipage}
	\end{minipage}
	\begin{minipage}[b]{1.0\linewidth}
		\vspace{4mm}
		\begin{minipage}[b]{0.33\linewidth}
			\centering
			\makebox[0.5em][l]{\raisebox{-\height}{(\textit{c})}}%
			\raisebox{-\height}{\includegraphics[height=2.7cm]{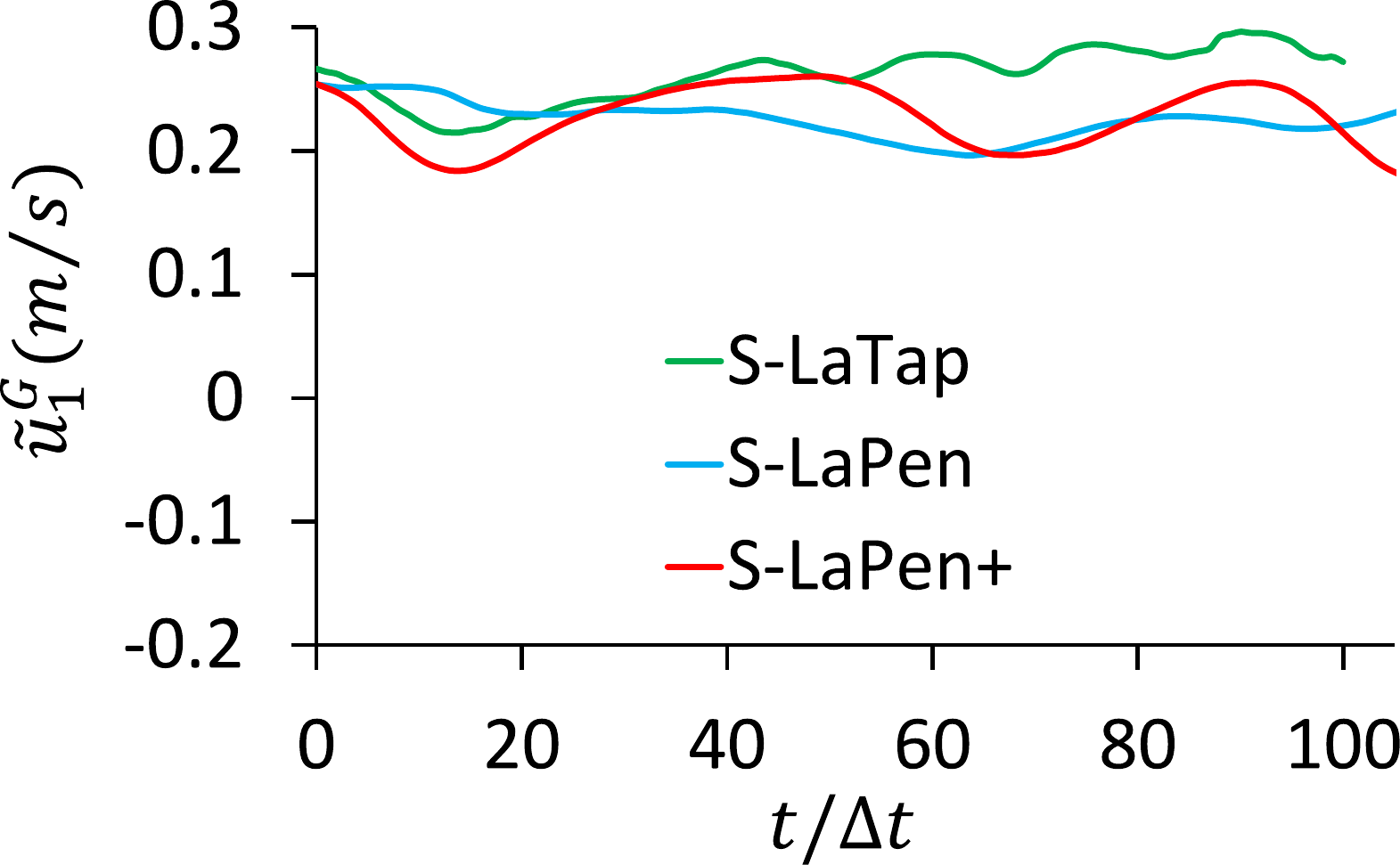}}
		\end{minipage}
		\begin{minipage}[b]{0.33\linewidth}
			\centering
			\makebox[0.5em][l]{\raisebox{-\height}{(\textit{d})}}%
			\raisebox{-\height}{\includegraphics[height=2.7cm]{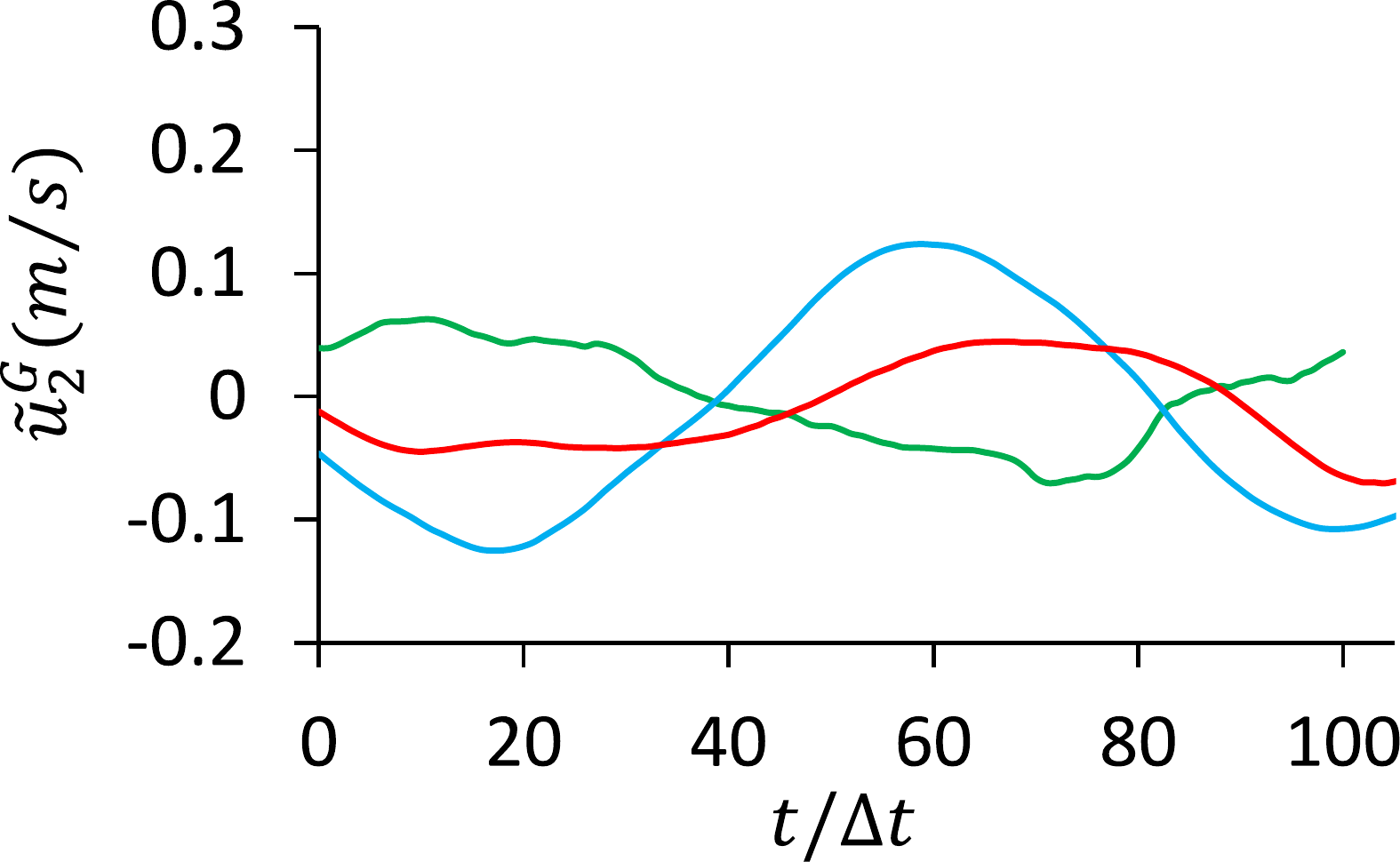}}
		\end{minipage}
		\begin{minipage}[b]{0.33\linewidth}
			\centering
			\makebox[0.5em][l]{\raisebox{-\height}{(\textit{e})}}%
			\raisebox{-\height}{\includegraphics[height=2.7cm]{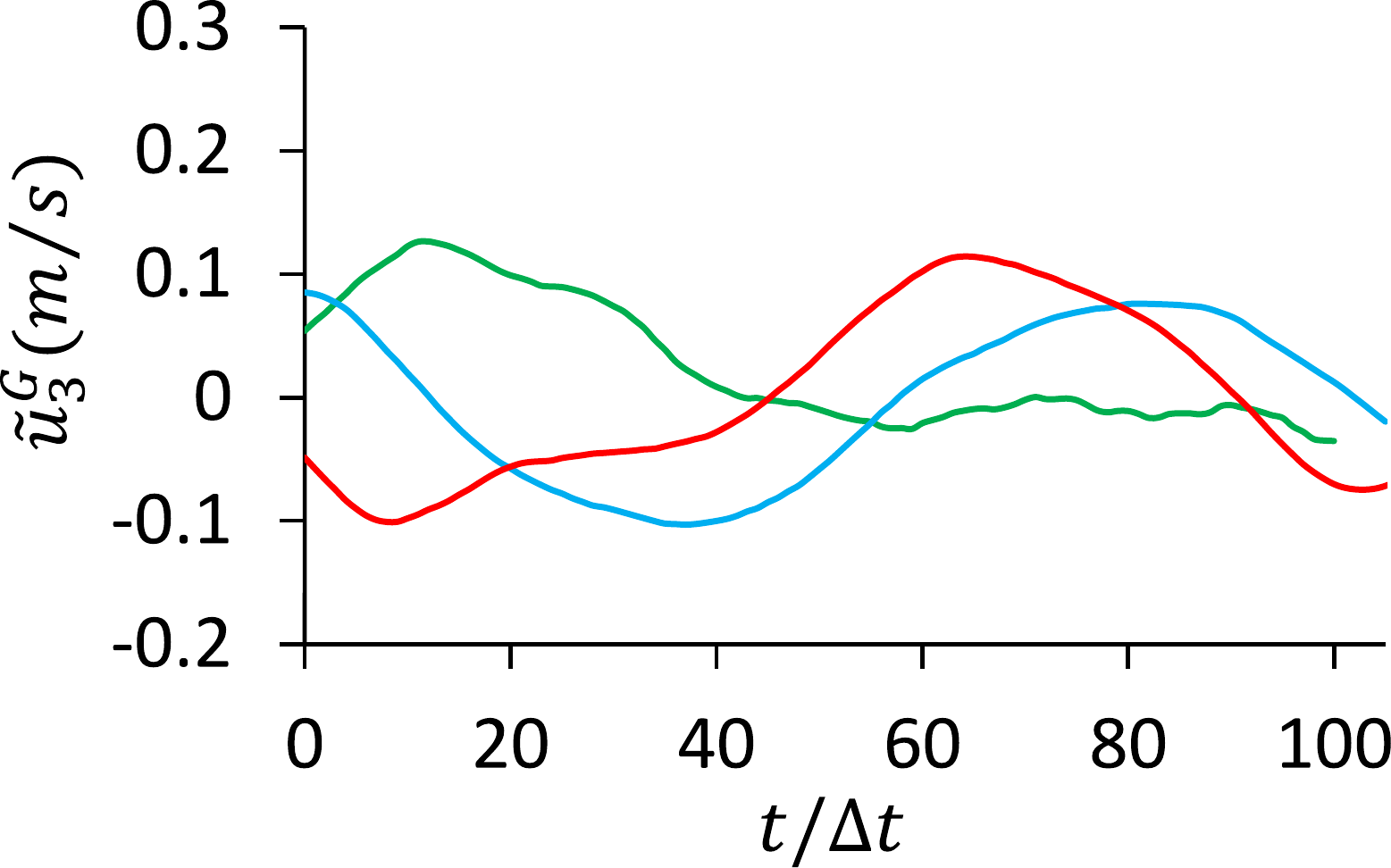}}
		\end{minipage}	
	\end{minipage}	
	\caption{3D trajectories (\textit{a}), top view of those (\textit{b}) and corresponding instantaneous bubble velocities with components $\tilde{u}^G_1$ (\textit{c}), $\tilde{u}^G_2$ (\textit{d}), $\tilde{u}^G_3$ (\textit{e}) over time (normalized by $\Delta t=2$ ms) of larger single bubbles at $C_\infty=0$ ppm (\textit{S-LaTap}), $C_\infty=333$ ppm (\textit{S-LaPen}) and $C_\infty=1000$ ppm (\textit{S-LaPen+}). All figures are from the same track of the particular case.} \label{fig: velo & trajectory Sin-La}
\end{figure}

Another important observation from figures \ref{fig: velo & trajectory Sin-Sm}(\textit{c-e}) is that the frequency of the vertical velocity is approximately twice as large as that of the horizontal velocities due to the frequency of the force oscillations in the corresponding directions \citep{2006_Mougin}. Moreover, there is a trend that with increasing $C_\infty$, the frequency of the oscillation in velocities increases. This is in line with the observation reported in \cite{2014_Tagawa} for their smaller bubbles.

In figure \ref{fig: velo & trajectory Sin-La}, we plot the same quantities for the three larger bubble cases. A fixed path type cannot be found for Case \textit{S-LaTap}. Although in figure \ref{fig: velo & trajectory Sin-La}(e) it shows a zigzagging trend, there are many other snapshots, showing a flattened helix (not shown here). Case \textit{S-LaTap} belongs to the chaotic regime due to the very large $Re_p$ and $Eo$, while \textit{S-LaPen} exhibits a flattened helical motion, and \textit{S-LaPen+} converges toward a zigzag path. Moreover, compared to the smaller bubbles, the vertical velocities (figure \ref{fig: velo & trajectory Sin-La}\textit{a}) in all three larger bubble cases display irregular oscillations.

\section{Flow characterization for bubble swarm} \label{sec: one-point}

\begin{figure}	
	\begin{minipage}[b]{1.0\linewidth}
		\begin{minipage}[b]{0.5\linewidth}
			\centering
			\makebox[0.5em][l]{\raisebox{-\height}{(\textit{a})}}%
			\raisebox{-\height}{\includegraphics[height=4cm]{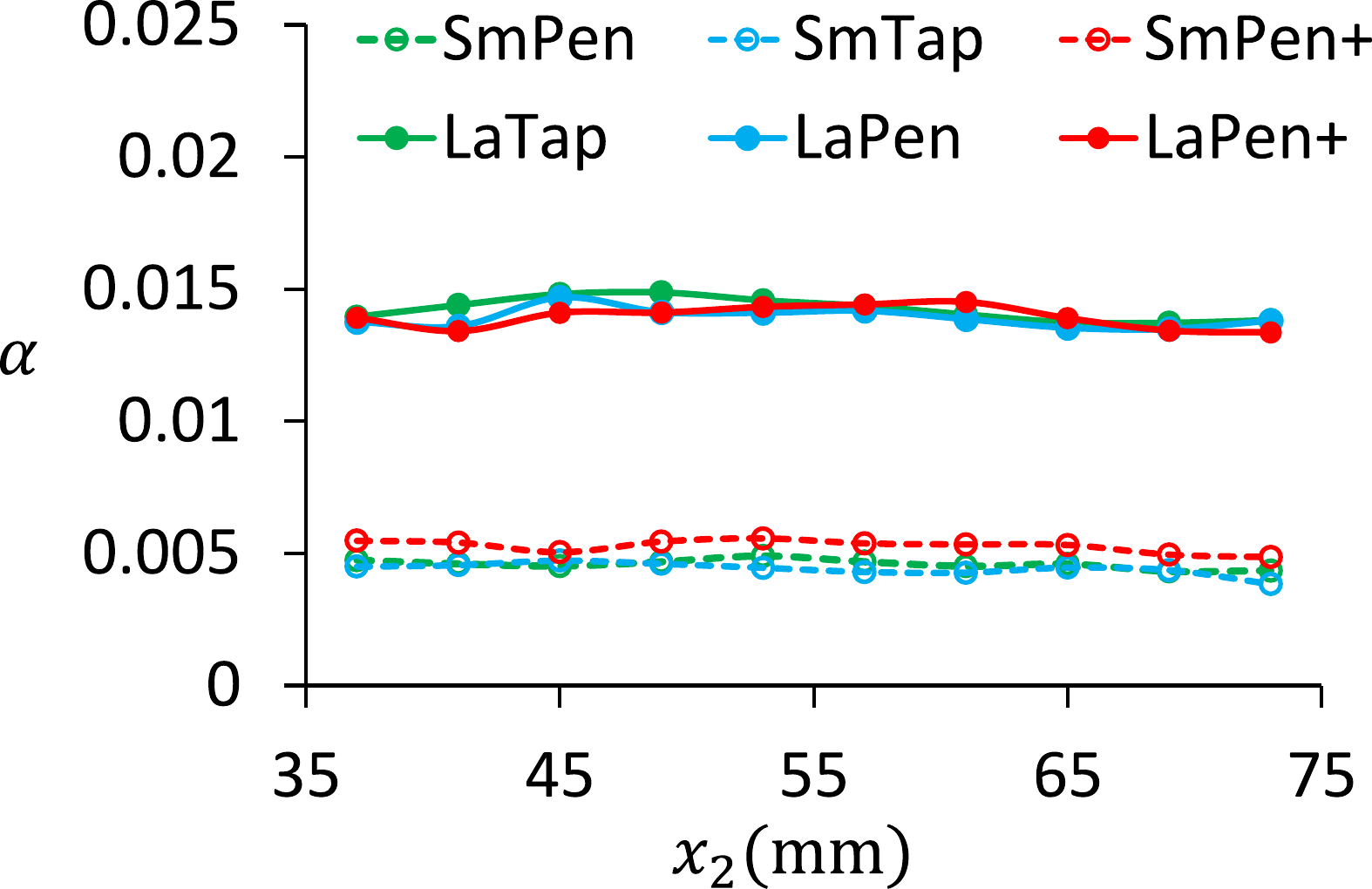}}
		\end{minipage}
		\begin{minipage}[b]{0.5\linewidth}
			\centering
			\makebox[0.5em][l]{\raisebox{-\height}{(\textit{b})}}%
			\raisebox{-\height}{\includegraphics[height=4cm]{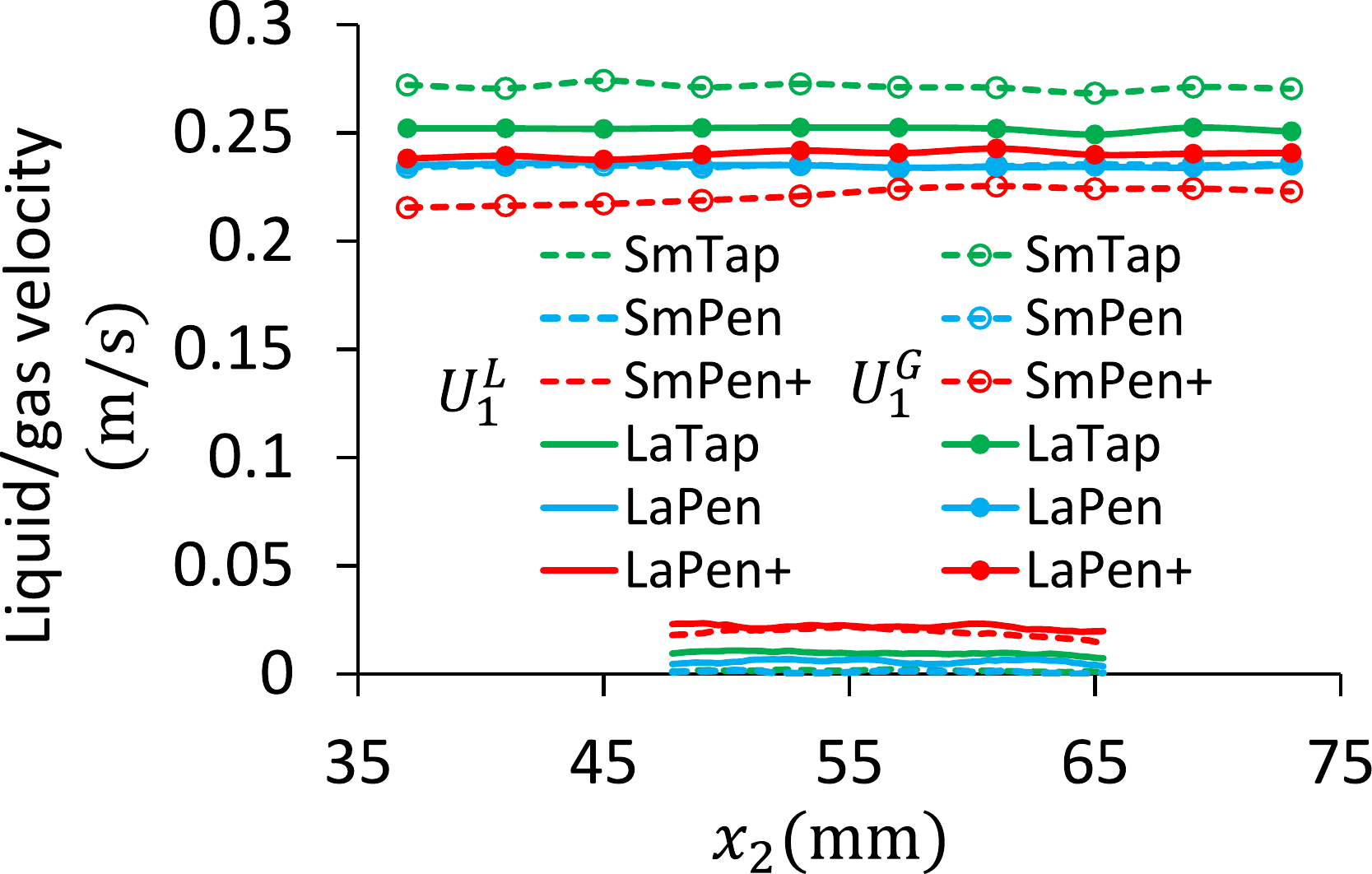}}
		\end{minipage}
	\end{minipage}	
	\caption{Gas void fraction (\textit{a}) and liquid/gas vertical velocity (\textit{b}) along the horizontal axis of FOV.} \label{fig: void & mean flow}
\end{figure}

Basic statistics of both phases for the considered cases are plotted in figure \ref{fig: void & mean flow}. Both the void fraction (figure \ref{fig: void & mean flow}\textit{a}) and the vertical gas/liquid velocity (figure \ref{fig: void & mean flow}\textit{b}) have very flat profiles, indicating statistical homogeneity of the current flow in the FOV. Even with zero-bulk-flow (averaged over the entire flow cross section), we observe for almost all the cases $U^L_1\neq0$ along the horizontal axis of the FOV (especially for the cases \textit{SmPen+} and \textit{LaPen+}), so that the relative velocity in the FOV is in general not equal to the bubble terminal rise velocity. Furthermore, due to the Marangoni effect we find that the bubble swarms rise more slowly with increasing $C_\infty$ for both smaller and larger bubbles, consistent with our results for the corresponding single bubble in \S\,\ref{sec: single bubble}. It is worth noting although \textit{LaPen} and \textit{LaPen+} have a similar $U^G_1$, \textit{LaPen+} has a smaller relative velocity due to the higher $U^L_1$, as indicated by its smaller bubble Reynolds number (table \ref{tab: bubble para}). 

\begin{figure}	
	\begin{minipage}[b]{1.0\linewidth}
		\begin{minipage}[b]{0.5\linewidth}
			\centering
			\makebox[0.5em][l]{\raisebox{-\height}{(\textit{a})}}%
			\raisebox{-\height}{\includegraphics[height=4cm]{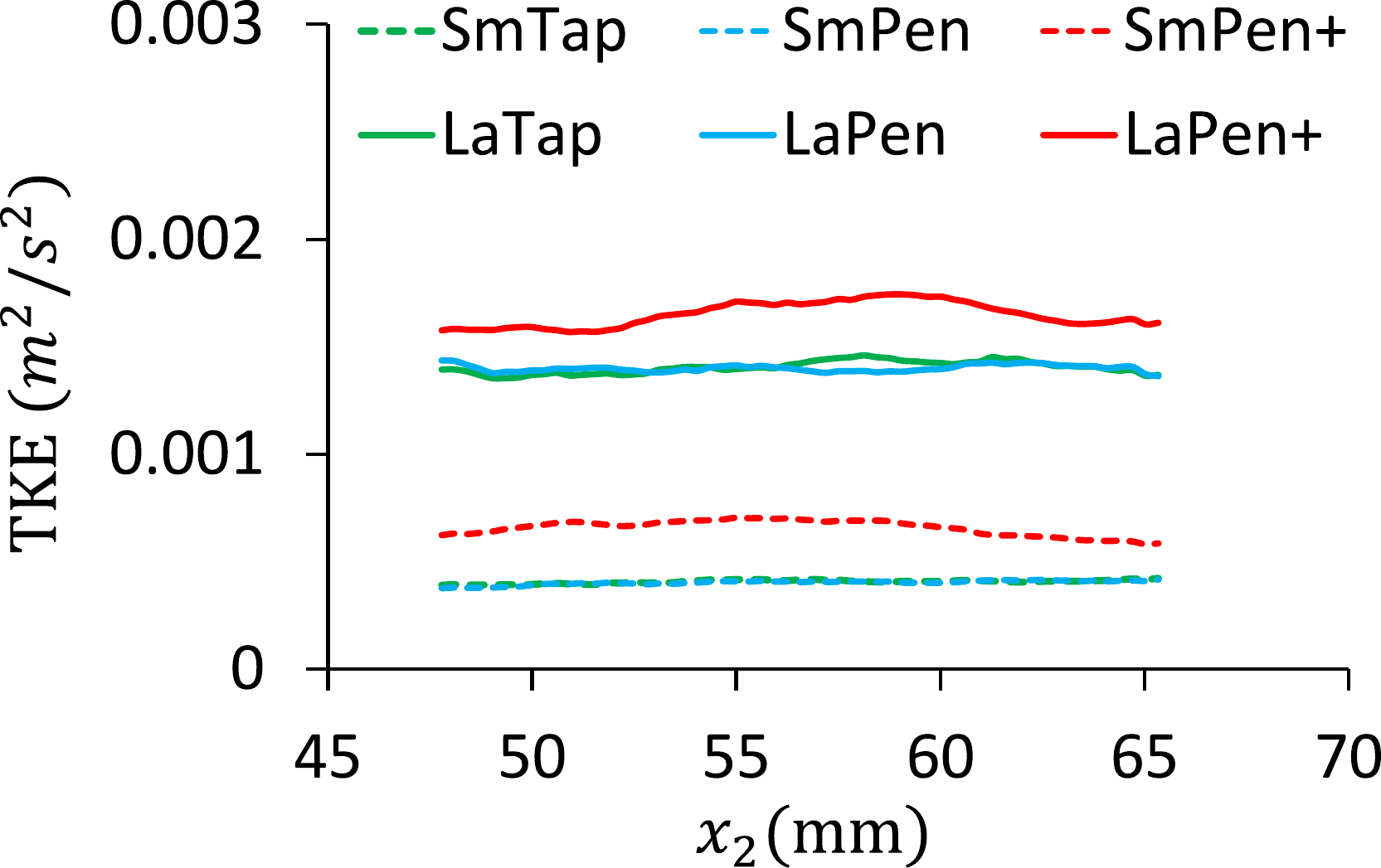}}
		\end{minipage}
		\begin{minipage}[b]{0.5\linewidth}
			\centering
			\makebox[0.5em][l]{\raisebox{-\height}{(\textit{b})}}%
			\raisebox{-\height}{\includegraphics[height=4.2cm]{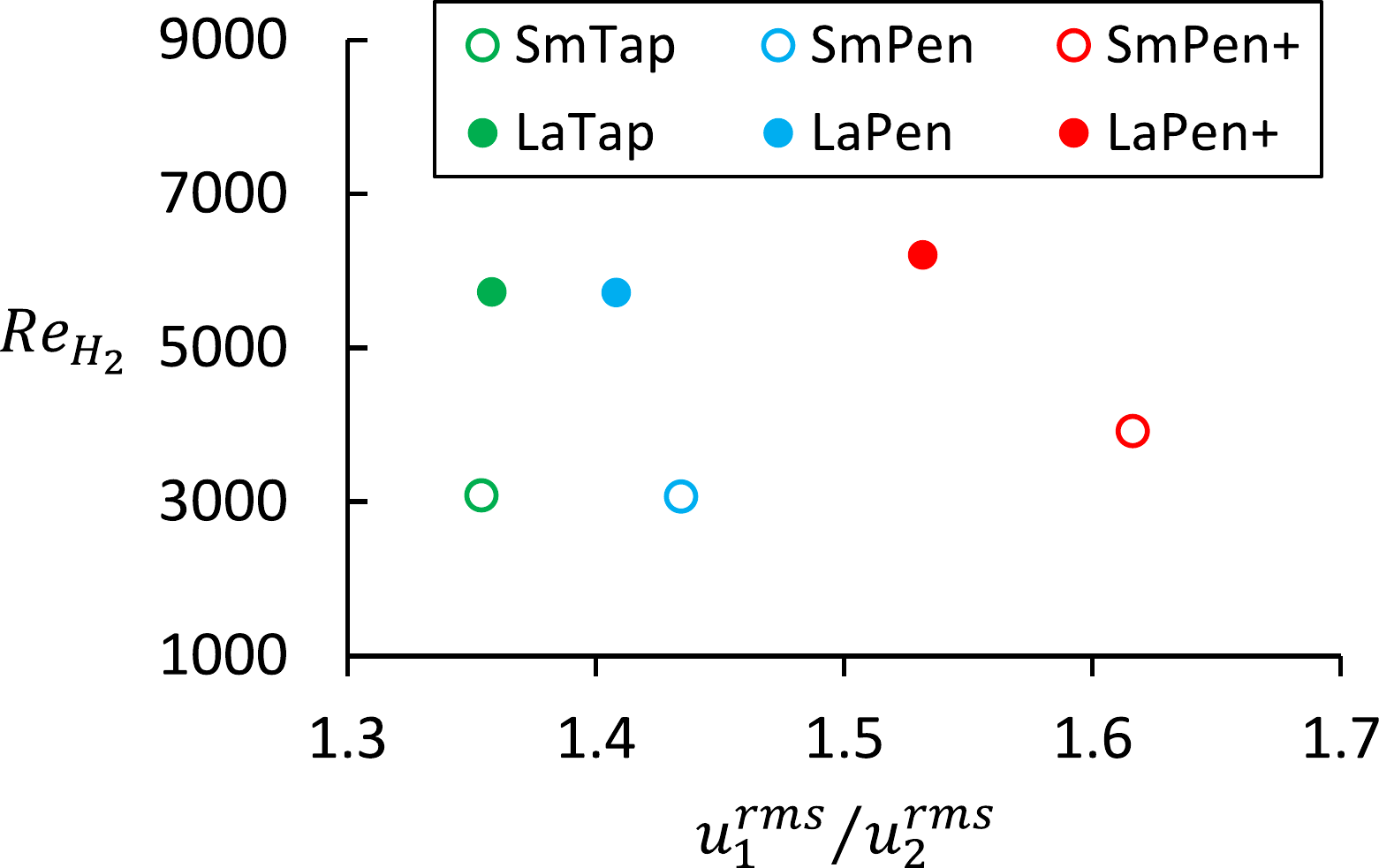}}
		\end{minipage}
	\end{minipage}	
	\caption{(\textit{a}) Turbulent kinetic energy along the horizontal axis of FOV. (\textit{b}) Reynolds number, $Re_{H_2}$ plotted versus large-scale anisotropy ratio, $u_1^{rms}/u_2^{rms}$.} \label{fig: TKE & Re_H2}
\end{figure}

We now turn to consider the role played by the surfactant in generating BIT. Hereafter, all average quantities refer to the liquid, so that the upper index $L$ is dropped for simplicity. In figure \ref{fig: TKE & Re_H2}(\textit{a}) we plot the turbulent kinetic energy (TKE), $k$, calculated as
\begin{equation}
k=\frac{1}{2}\left ( (u_1^{rms})^2+2(u_2^{rms})^2\right ) \;,\label{eq: F}
\end{equation}
assuming the out-of-plane velocity variance is equal to the measured horizontal component. This approximation -- axisymmetry about the vertical direction is expected for the BIT dominated flows far from the wall. Here, $u_1^{rms}$ and $u_2^{rms}$ are the root-mean-square values of the vertical and the horizontal velocity fluctuation, respectively. For both smaller and larger bubbles, surprisingly, the TKE is highest for the cases with the highest $C_\infty$, although \textit{SmPen+} and \textit{LaPen+} have the lowest $Re_p$ in the cases of smaller and larger bubbles, respectively. It is also interesting to note that TKE in the cases \textit{SmPen} and \textit{LaPen} do not change much compared to \textit{SmTap} and \textit{LaTap}, respectively, indicating that the amount of 1-Pentanol added ($C_\infty=333$ ppm) is not enough to modify the BIT already initiated in the tap water system. However, we should keep in mind that the bubble sizes in \textit{SmPen} and \textit{LaPen} are slightly smaller than their corresponding tap water cases (see table \ref{tab: bubble para}), so that if the TKE is almost the same for the $C_\infty=0$ ppm and $C_\infty=333$ ppm cases, then the surfactant is in fact leading to a positive contribution to the BIT generation. 

In table \ref{tab: factors change BIT} we summarise the effect of the surfactant on three aspects that influence BIT. Aside from reducing the bubble Reynolds number, the bubble surface instability (e.g. deformation and wobbling) also reduces when increasing $C_\infty$, and both of these reductions have a negative impact on BIT production. This suggests that the change of boundary condition induced by increasing $C_\infty$ plays the key role in causing the TKE to be enhanced by the addition of surfactants. 
\begin{table}
	\begin{center}
		\def~{\hphantom{0}}
		\begin{tabular}{ccc}
			Factor influencing BIT \; &Change by increasing $C_\infty$ \; &Contribution to BIT\\	
			\hline
			Bubble Reynolds number &reduced&negative\\
			Boundary condition     &free-slip $\rightarrow$ no-slip&positive\\
			Surface instability    &reduced&negative\\
		\end{tabular}
		\caption{Surfactant effect on multiple factors that influence bubble-induced turbulence.}
		\label{tab: factors change BIT}
	\end{center}
\end{table}
Following our previous study \citep{2021_Ma}, we define a Reynolds number $Re_{H_2}\equiv u^\ast H_2/\nu$, indicating the range of scales in the turbulent bubbly flows. Here,
$u^\ast\equiv \sqrt{(2/3)k_{\mathrm{FOV}}}$, and $k_{\mathrm{FOV}}$ is the TKE averaged over the FOV of the liquid phase. In figure \ref{fig: TKE & Re_H2}(\textit{b}) we depict $Re_{H_2}$ versus the large-scale anisotropy ratio $u_1^{rms}/u_2^{rms}$ (also averaged over the FOV). The figure reflects the same behaviour as the TKE, namely the case with the highest $C_\infty$ also has the largest $Re_{H_2}$ for the corresponding bubble size group. 
Furthermore, we find that for large scales whose fluctuating velocities are characterized by $u_1^{rms}$ and $u_2^{rms}$, the smaller bubbles produce more anisotropy in the flow than the larger bubbles, as reflected by a larger ratio of $u_1^{rms}/u_2^{rms}$ for the cases \textit{SmTap}, \textit{SmPen} and \textit{SmPen+}. This is in very close agreement with our previous study based on DNS data of bubble-laden turbulent channel flow driven by a vertical pressure gradient and with no-slip boundary conditions on the bubble surfaces \citep{2021_Ma}. 

The more important point prompted by figure \ref{fig: TKE & Re_H2}(\textit{b}) is that the large-scale anisotropy increases significantly with increasing surfactant concentration for both smaller and larger bubbles. This behaviour is due to the effect of the surfactant on both the wake structure and trajectory type of the bubbles as $C_\infty$ changes, as was already seen for different single bubbles in figures \ref{fig: velo & trajectory Sin-Sm} and \ref{fig: velo & trajectory Sin-La}.

\section{Turbulence modification across scales}\label{sec: Two-point}

\subsection{Turbulence anisotropy}\label{sec: Anisotropy}

The components of the second-order velocity structure function are defined as
\begin{equation}
	D_2^{ij}(\boldsymbol{x},\boldsymbol{r},t)\equiv\langle \Delta u_i(\boldsymbol{x},\boldsymbol{r},t)\Delta u_j(\boldsymbol{x},\boldsymbol{r},t) \rangle  \;, \label{eq: Dij}
\end{equation}
where $\Delta u_i(\boldsymbol{x},\boldsymbol{r},t)$ denotes the difference in the velocity at positions $\boldsymbol{x}$ and $\boldsymbol{x}+\boldsymbol{r}$ at time $t$. Hereafter, we suppress the space $\boldsymbol{x}$ and time $t$ arguments since we are considering a flow which is statistically homogeneous and stationary over the FOV. The PSV measurement provides access to data associated with separations along two directions, namely, the vertical separation $\boldsymbol{r}=r_1\boldsymbol{e}_1$ ($r_1=\|\boldsymbol{r}\|$) and the horizontal separation $\boldsymbol{r}=r_2\boldsymbol{e}_2$ ($r_2=\|\boldsymbol{r}\|$). Hence, we are able to compute the four contributions $D^L_2(r_1)=D_2^{11}(r_1)$, $ D^L_2(r_2)=D_2^{22}(r_2)$, $D^T_2(r_1)=D_2^{22}(r_1)$, and $D^T_2(r_2)=D_2^{11}(r_2)$ based on the Cartesian coordinate system depicted in figure \ref{fig: Bubble_column}.

\begin{figure}	
	\begin{minipage}[b]{1.0\linewidth}
		\begin{minipage}[b]{0.5\linewidth}
			\centering
			\makebox[0.5em][l]{\raisebox{-\height}{(\textit{a})}}%
			\raisebox{-\height}{\includegraphics[height=3.8cm]{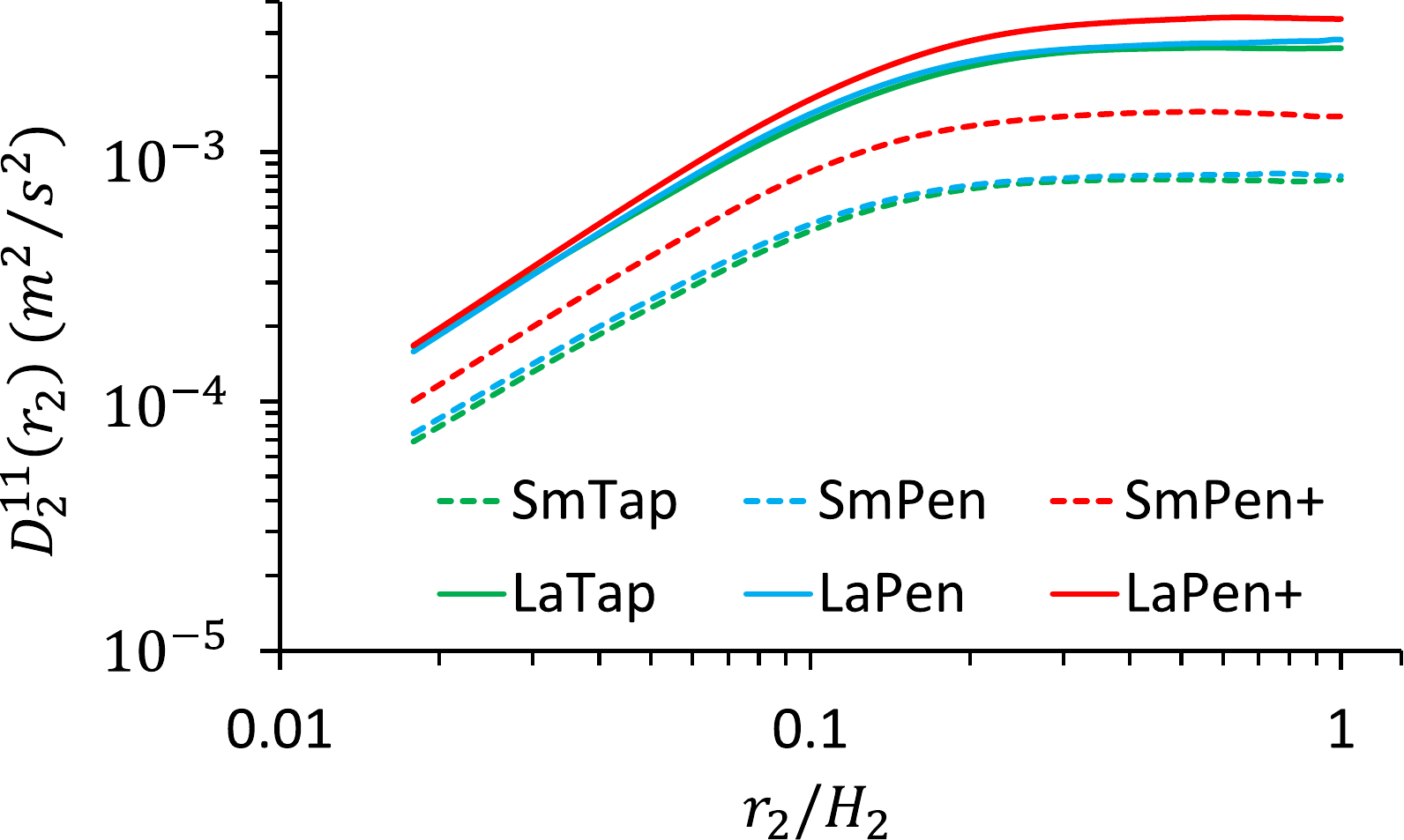}}
		\end{minipage}
		\begin{minipage}[b]{0.5\linewidth}
			\centering
			\makebox[0.5em][l]{\raisebox{-\height}{(\textit{b})}}%
			\raisebox{-\height}{\includegraphics[height=3.8cm]{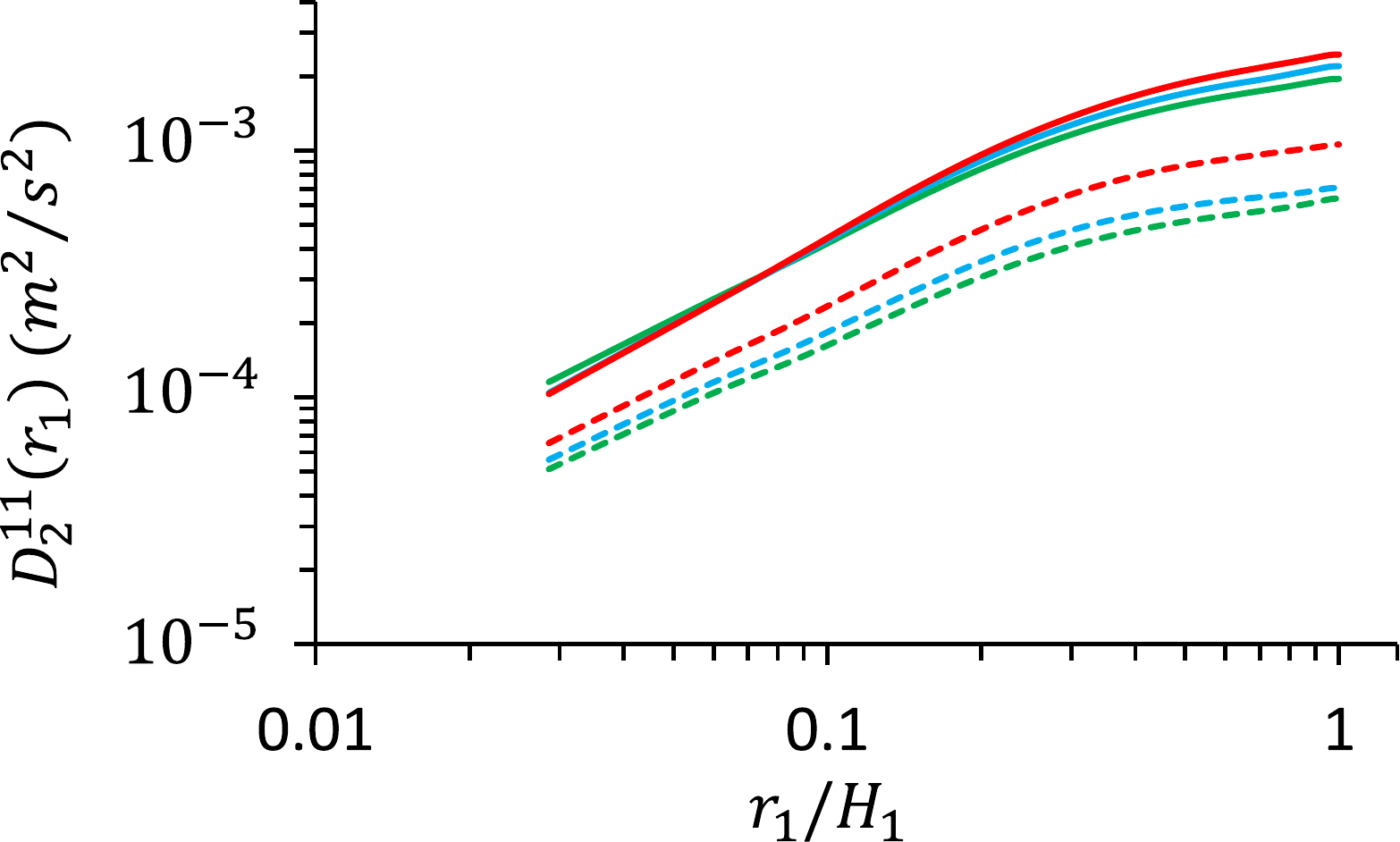}}
		\end{minipage}
	\end{minipage}	
	\caption{Transverse (\textit{a}) and longitudinal (\textit{b}) second-order structure function of the $u_1$ component, with separations
		along the horizontal (\textit{a}) and the vertical (\textit{b}) directions.} \label{fig: D11}
\end{figure}

Figures \ref{fig: D11}(\textit{a}) and (\textit{b}) show the measured transverse and longitudinal second-order structure functions of the $u_1$ component, respectively. The results show that the values of the structure functions increase in the order \textit{SmTap}, \textit{SmPen}, \textit{SmPen+}, \textit{LaTap}, \textit{LaPen}, \textit{LaPen+}, which corresponds to larger bubble size and higher surfactant concentration. This ordering also holds for the $u_2$ component computed (not shown). Similar to the results for the TKE, while the difference between \textit{Sm(La)Tap} and \textit{Sm(La)Pen} is small, $D_2^{\gamma\gamma}$ (no index summation is implied for $\gamma$) for \textit{Sm(La)Pen+} have noticeably higher values across most scales for both smaller and larger bubble sizes.

\begin{figure}
	\begin{minipage}[b]{1.0\linewidth}
		\begin{minipage}[b]{0.5\linewidth}
			\centering
			\makebox[1em][l]{\raisebox{-\height}{(\textit{a})}}%
			\raisebox{-\height}{\includegraphics[height=3.8cm]{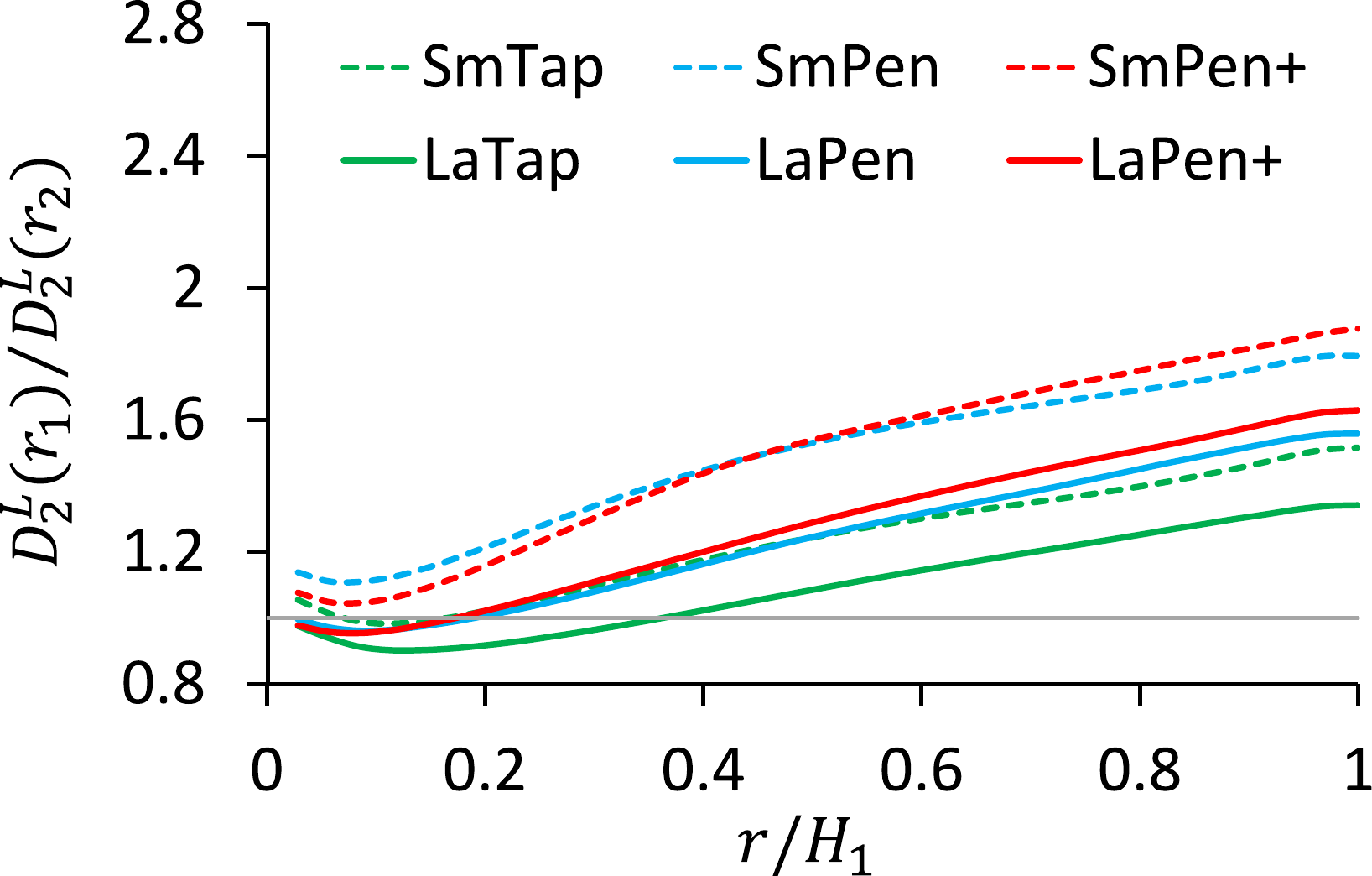}}
		\end{minipage}
		\begin{minipage}[b]{0.5\linewidth}
			\centering
			\makebox[1em][l]{\raisebox{-\height}{(\textit{b})}}%
			\raisebox{-\height}{\includegraphics[height=3.8cm]{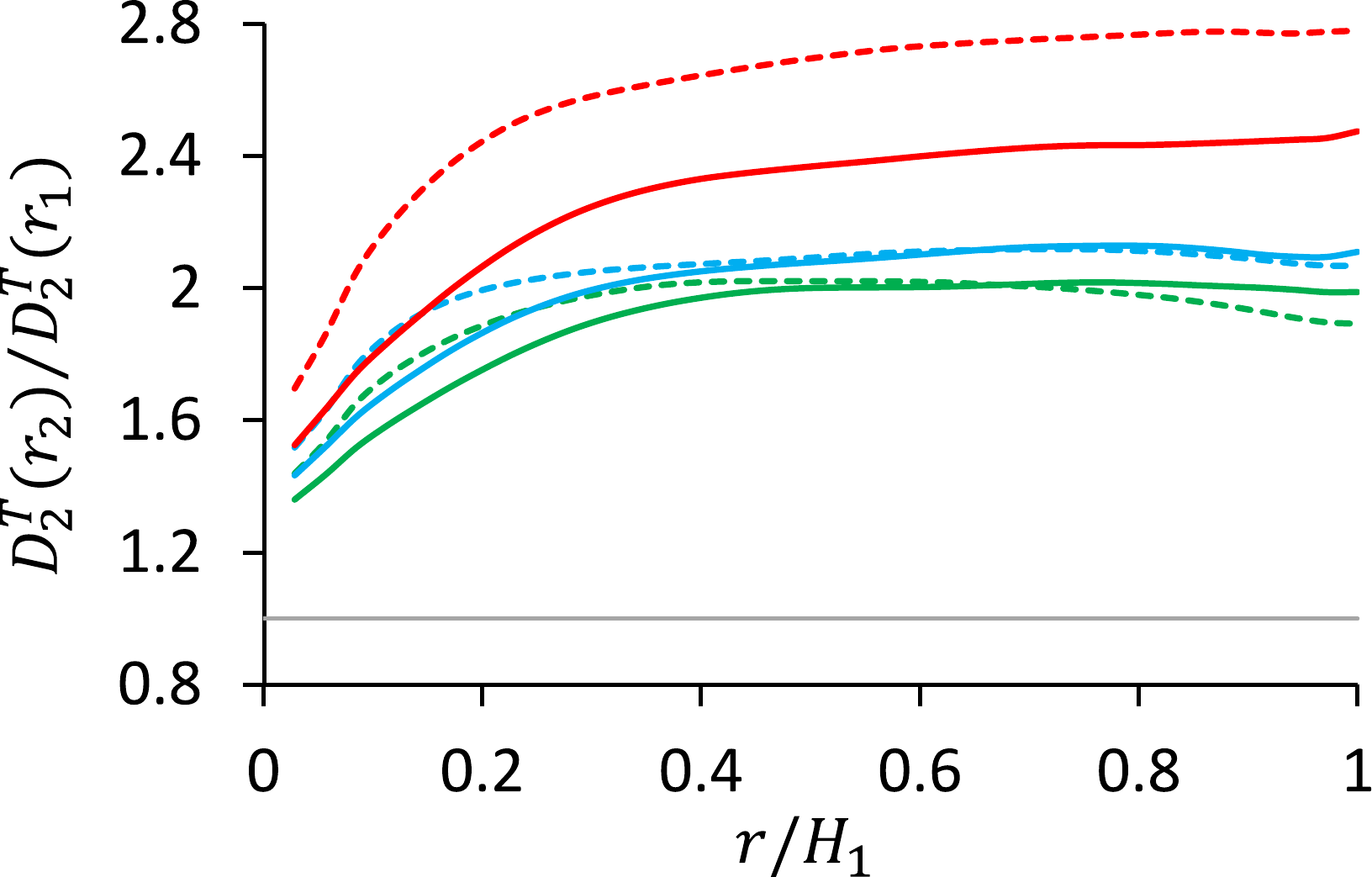}}
		\end{minipage}
	\end{minipage}
	\caption{Ratio of longitudinal (\textit{a}) and transverse (\textit{b}) structure functions in different separation directions for all the cases. In (\textit{a,b}) the horizontal lines indicate the value of unity.} \label{fig: D_LT_ratio}
\end{figure}

To characterizing the multiscale anisotropy associated with $D_2^{ij}$, we plot the ratios of components $D^L_2(r_1)/D^L_2(r_2)$ and $D^T_2(r_2)/D^T_2(r_1)$ in figure \ref{fig: D_LT_ratio}, which would be equal to unity for an isotropic flow \citep{2017_Carter}. Generally, the results show that both ratios decrease monotonically for decreasing separation. Moreover, the ratio of the transverse structure functions departs more strongly from unity than the longitudinal ones. By comparing the cases with the same bubble size, we find that generally the cases with higher surfactant concentration generate stronger anisotropy in the flow across the scales, and this trend is most obvious in the plot for $D^T_2(r_2)/D^T_2(r_1)$. This is in agreement with the results of the large-scale anisotropy in \S\,\ref{sec: one-point}.

Furthermore, considering the cases with the same $C_\infty$, our results indicate that smaller bubbles generate stronger anisotropy in the flow, consistent with our previous studies \citep{2021_Ma,2022_Ma} that only considered fully-contaminated bubbles.  However, the results in figure \ref{fig: D_LT_ratio} show that it is not true in general that smaller bubbles generate stronger anisotropy in the flow, because depends on the surfactant concentration. For example, figure \ref{fig: D_LT_ratio} shows that $LaPen+$ can be more anisotropic than $SmTap$.

\subsection{Extreme fluctuations in the flow}\label{sec: Extreme velocity increment}

\begin{figure}	
	\begin{minipage}[b]{1.0\linewidth}
		\begin{minipage}[b]{0.5\linewidth}
			\centering
			\makebox[1.5em][l]{\raisebox{-\height}{(\textit{a})}}%
			\raisebox{-\height}{\includegraphics[height=4.2cm]{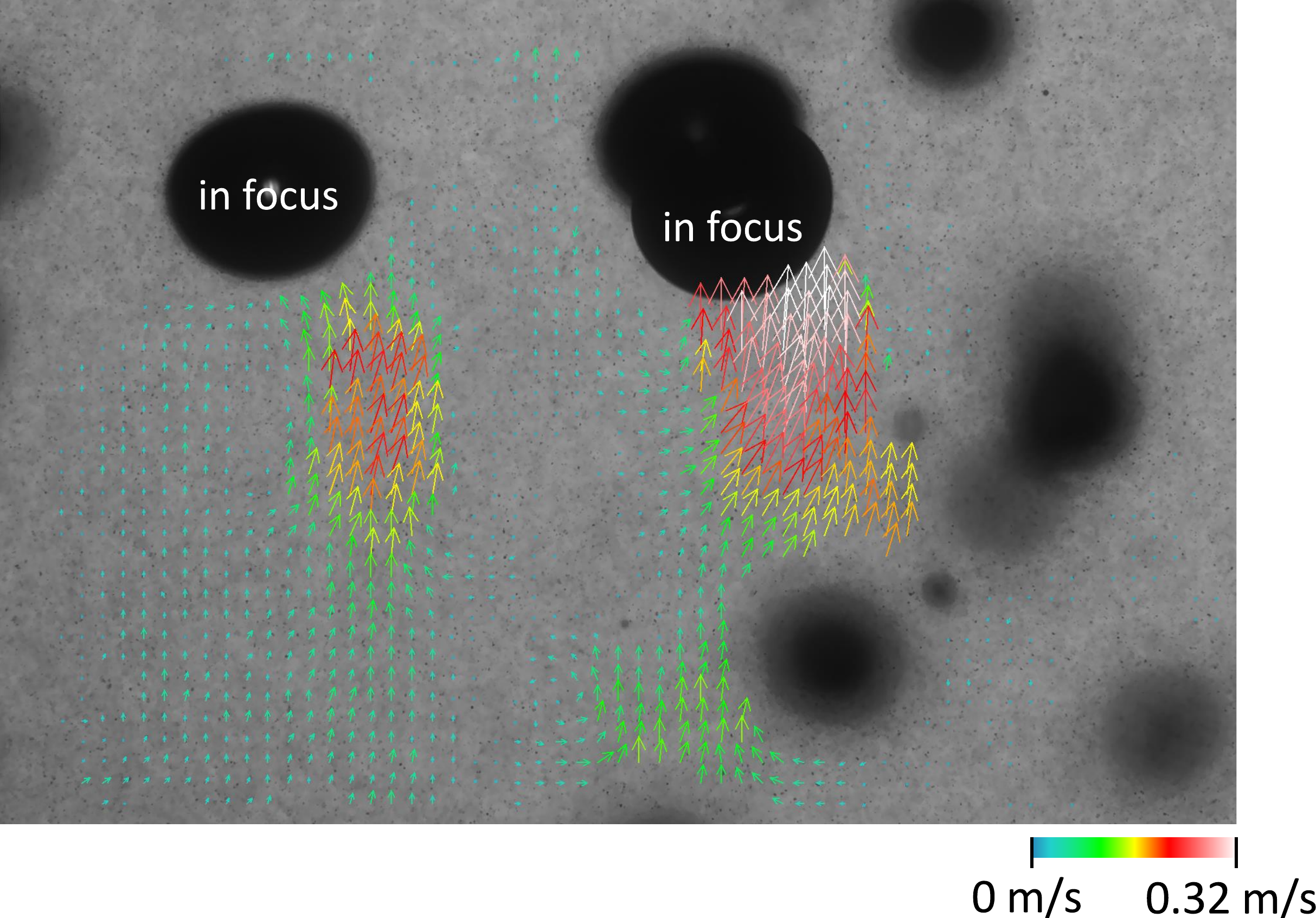}}
		\end{minipage}
		\begin{minipage}[b]{0.5\linewidth}
			\centering
			\makebox[1.5em][l]{\raisebox{-\height}{(\textit{b})}}%
			\raisebox{-\height}{\includegraphics[height=3.8cm]{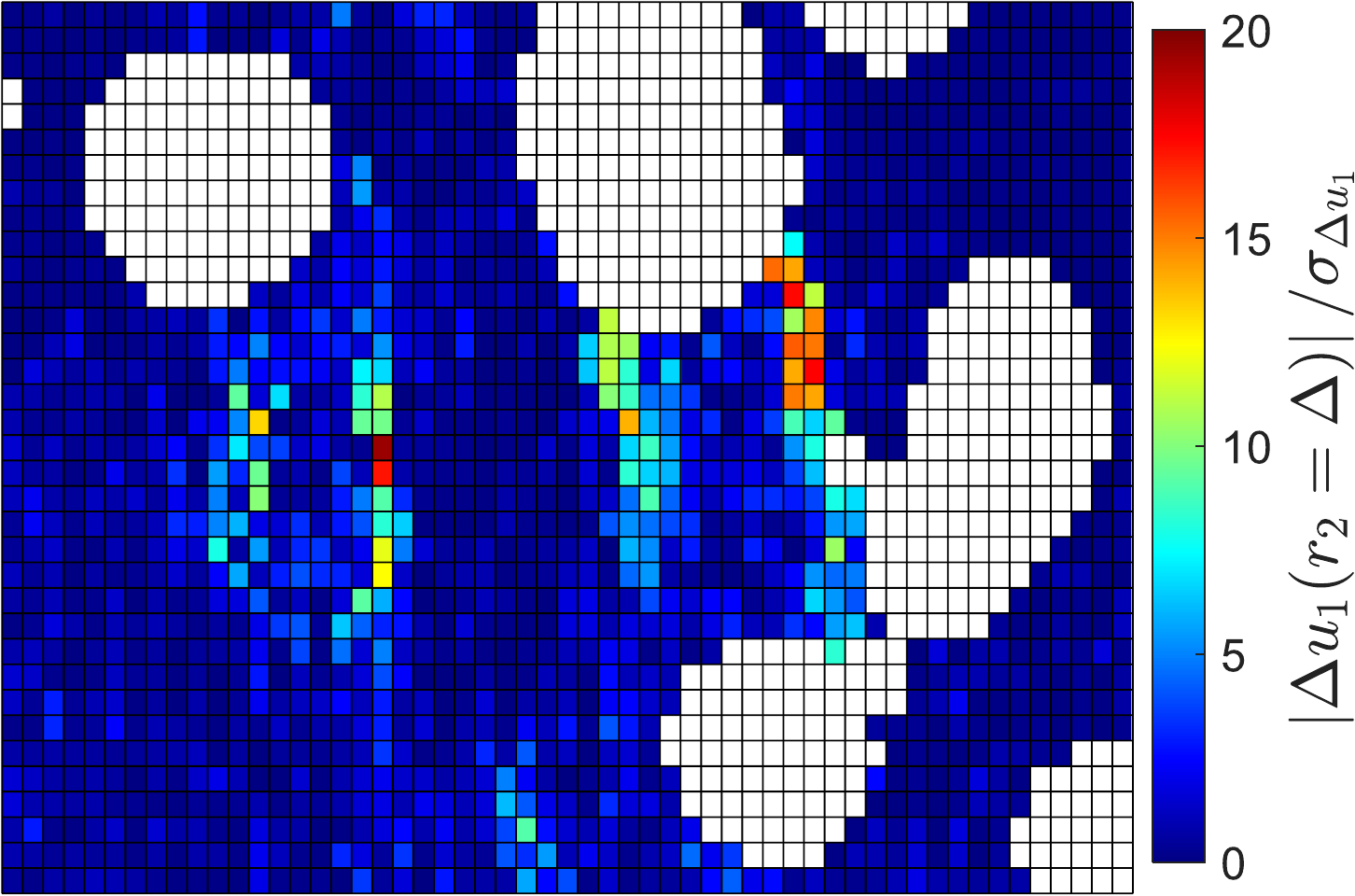}}
		\end{minipage}
	\end{minipage}	
	\caption{Snapshot of the original velocity vector (\textit{a}) and the intensity distributions of the normalized velocity increment $\left |\Delta u_1(r_2=\Delta)\right |/\sigma_{\Delta u_1}$ (\textit{b}) from the same instant based on the \textit{SmPen+} case. The in-focus bubbles are denoted in (\textit{a}).} \label{fig: snapshot_SmPen+}
\end{figure}

Having explored the role of surfactants on the flow anisotropy, we now turn to consider the effect of surfactants on extreme fluctuations of the velocity increments -- a phenomenon associated with internal intermittency in single-phase turbulence \citep{1995_Frisch,1997_Sreenivasan}. In the present bubbly flows, extreme events in the liquid phase result from the boundary of the wakes produced by the bubbles \citep{2022_Ma}. These contributions can be seen in figure \ref{fig: snapshot_SmPen+}, with the normalized velocity increments showing large values at the edge of the wakes, while the velocities are largest inside the wakes. Another possible contribution to extreme events would come from the flow in the boundary layer at the top of the rising bubble where the flow may abruptly change from being equal to the bubble surface velocity to the background bulk velocity over a thin boundary layer.  For spherical bubbles with large $Re_p$ rising in a quiescent flow the thickness of this boundary layer is $O(d_pRe_p^{-1/2})$ \citep{1963_Moore,1967_Batchelor}. However, we cannot resolve the velocity fluctuations in this thin boundary layer with our current experimental method (and the results discussed below should be interpreted accordingly), and further work is needed to understand how these regions might contribute to extreme fluctuations in the flow. 

\begin{figure}
	\begin{minipage}[b]{1.0\linewidth}
		\begin{minipage}[b]{0.5\linewidth}
			\centering
			\makebox[0.5em][l]{\raisebox{-\height}{(\textit{a})}}%
			\raisebox{-\height}{\includegraphics[height=4cm]{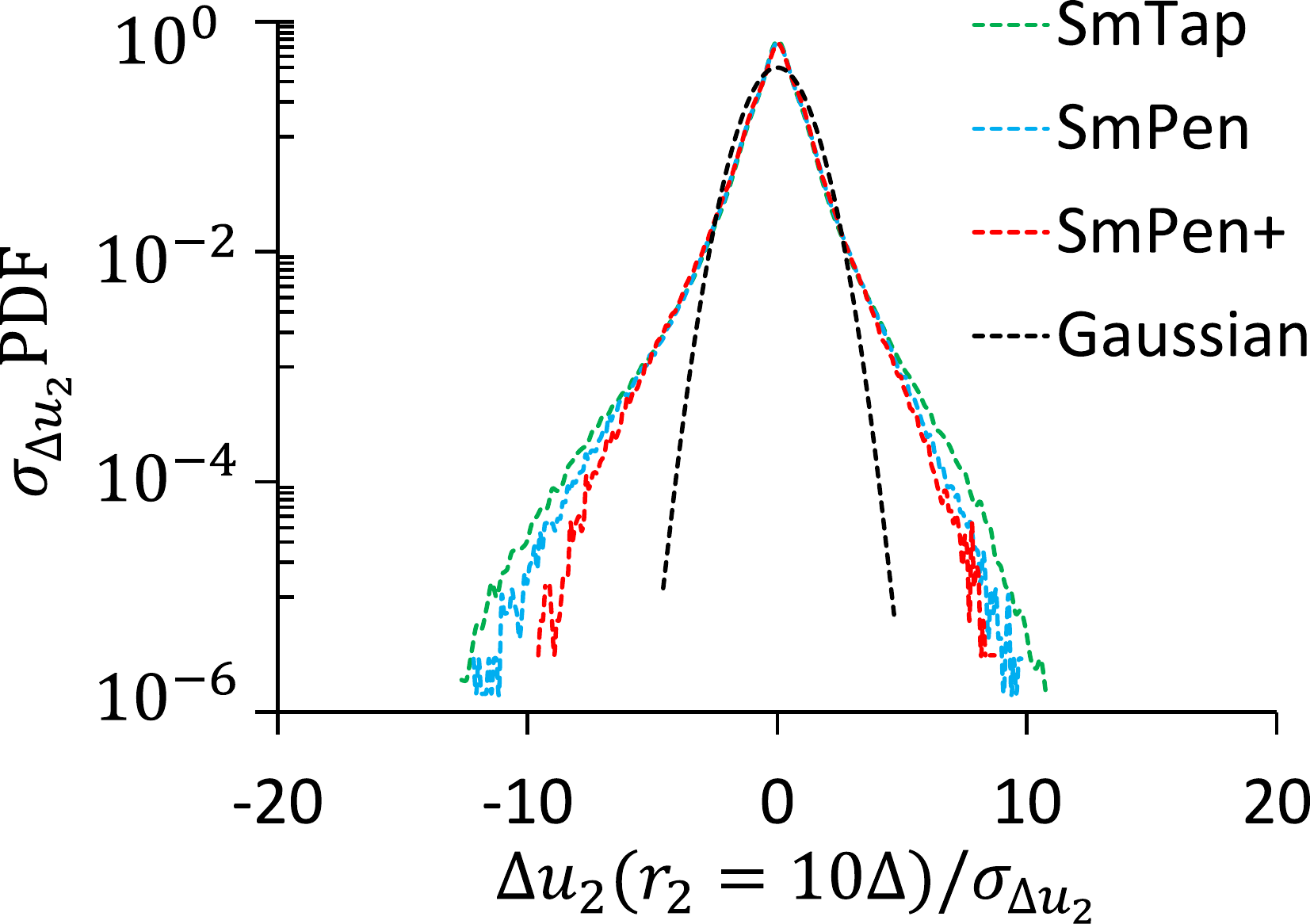}}
		\end{minipage}
		\begin{minipage}[b]{0.5\linewidth}
			\centering
			\makebox[0.5em][l]{\raisebox{-\height}{(\textit{b})}}%
			\raisebox{-\height}{\includegraphics[height=4cm]{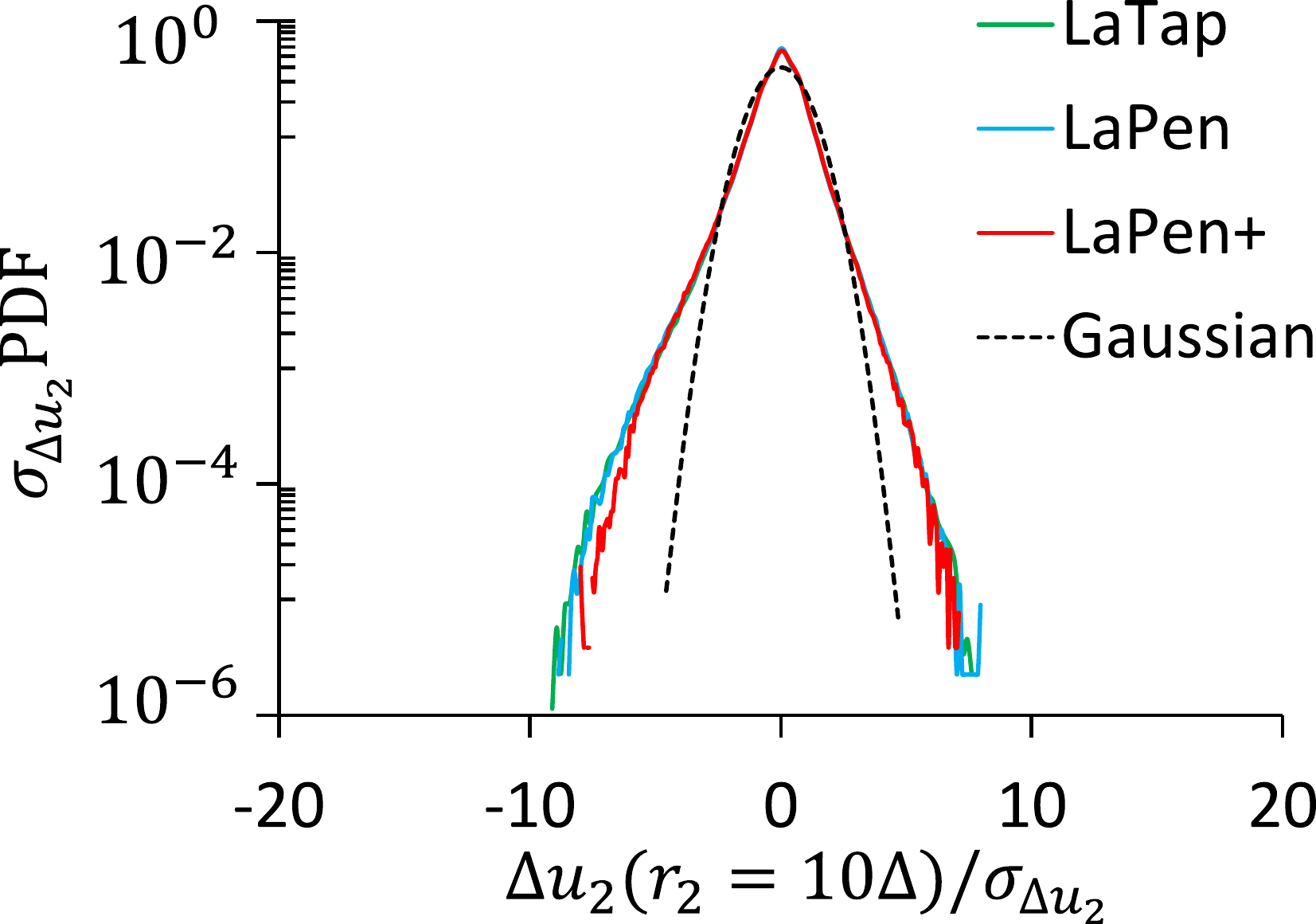}}
		\end{minipage}
	\end{minipage}
	\begin{minipage}[b]{1.0\linewidth}
		\vspace{3mm}
		\begin{minipage}[b]{0.5\linewidth}
			\centering
			\makebox[0.5em][l]{\raisebox{-\height}{(\textit{c})}}%
			\raisebox{-\height}{\includegraphics[height=4cm]{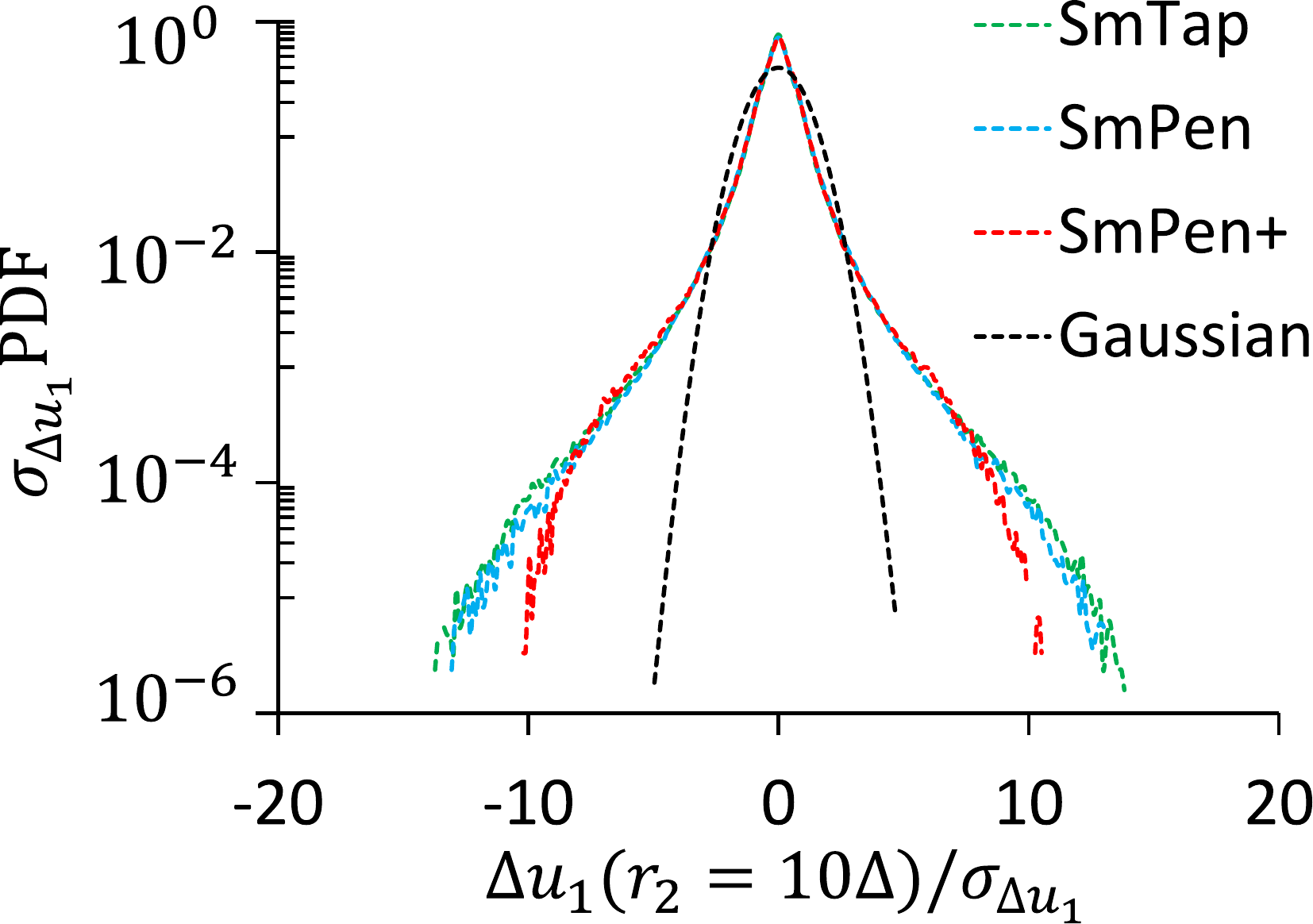}}
		\end{minipage}
		\begin{minipage}[b]{0.5\linewidth}
			\centering
			\makebox[0.5em][l]{\raisebox{-\height}{(\textit{d})}}%
			\raisebox{-\height}{\includegraphics[height=4cm]{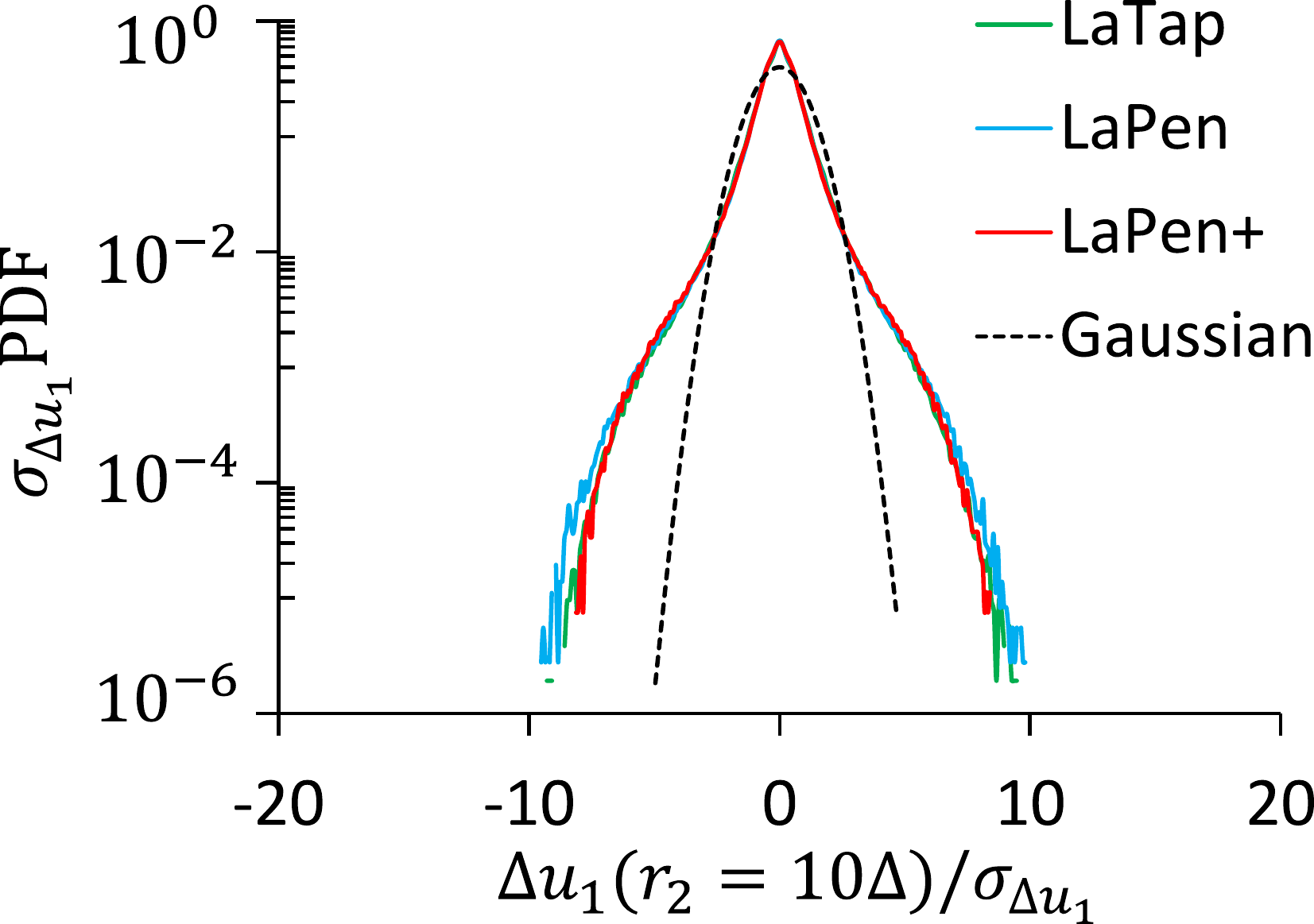}}
		\end{minipage}
	\end{minipage}
	\caption{Normalized probability density functions of the longitudinal  (\textit{a,b}) and transverse (\textit{c,d}) velocity increments for the three smaller bubble cases (\textit{a,c}) and three larger bubble cases (\textit{b,d}) with the separations along the horizontal direction ($r_2= 10\Delta$, where $\Delta$ is one PSV grid).} \label{fig: pdf 10 grids}
\end{figure}

In figure \ref{fig: pdf 10 grids} we plot the PDFs of the velocity increments for separations equal to ten PSV grids $(r=10\Delta=0.18H_2)$, corresponding to an \textquoteleft eddy size\textquoteright\ with $O(d_p)$. (The results for other separations show qualitatively similar trends and so are not shown). First, all of the six cases show that the velocity increments have strongly non-Gaussian PDFs, just as in single-phase turbulence. Second, while the PDFs of the transverse velocity increments in figure \ref{fig: pdf 10 grids}(\textit{c,d}) are almost symmetric, the PDFs of the longitudinal velocity increments in plots (\textit{a,b}) are negatively skewed for the horizontal separations $r_2$. Finally, with respect to the effect of bubble size and surfactant concentration, the results show very interesting behaviours: for the smaller bubbles, both the longitudinal and the transverse PDFs (figure \ref{fig: pdf 10 grids}\textit{a,c}) become increasingly non-Gaussian in the order of \textit{SmPen+, SmPen, SmTap}, which corresponds to the order of decreasing $C_\infty$ of 1-Pentanol. This is consistent with our previous finding \citep{2022_Ma} that the non-Gaussianity of the PDFs of the velocity increments becomes stronger as the Reynolds number (here, $Re_{H_2}$) is decreased in bubble-laden turbulent flow considered here, while the opposite occurs for single-phase turbulence \citep{1995_Frisch}. An explanation for why the intermittency is largest for the cases with lower surfactants concentration can be given as follows. For these three cases the volume fraction is not large, and there are relatively few regions in the flow where turbulence is produced due to the bubble wakes, meaning that turbulence in the flow is very patchy and therefore intermittent. For the cases with higher $C_\infty$, they have larger flow Reynolds numbers accompanied by longer and wider bubble wakes. As a result the wake regions are more space filling and hence the flow is less intermittent than the cases with lower surfactant concentration which have shorter and narrower bubble wakes. Again, this difference compared to intermittency in single-phase turbulence can be traced back to the fact that the regions of intense small-scale velocity increments occur near the turbulent/non-turbulent interface at the boundary of the bubble wake, which is more similar to so-called external intermittency \citep{1949_Townsend} than internal intermittency.

The $C_\infty$-dependency of the small-scale intermittency just discussed for the smaller bubbles is however much weaker for the larger bubbles (figure \ref{fig: pdf 10 grids}\textit{b,d}). A potential reason for this difference is that the three larger bubble cases have volume fractions that are almost three times larger than those for the smaller bubbles, and the bubble wake volume is also much larger, with much higher turbulence intensity generated by the wakes. As a result of these properties, while there are significsant regions of the flow that are almost quiescent for the smaller bubble cases the same is not true for the larger bubbles, with turbulent activity filling a significant fraction of the flow. Due to this, the velocity difference in and outside the wake is smaller for the larger bubble cases than the smaller bubble cases, and the intermittency is less dependent on $C_\infty$ since even for the largest $C_\infty$ case, these velocity differences across the wak boundaries are already smaller. This then explains why \textit{LaTap}, \textit{LaPen} and \textit{LaPen+} do not show the same $C_\infty$-dependence in the PDFs as the smaller bubble cases. 

\begin{figure}	
	\begin{minipage}[b]{1.0\linewidth}
		\begin{minipage}[b]{0.5\linewidth}
			\centering
			\makebox[1.5em][l]{\raisebox{-\height}{(\textit{a})}}%
			\raisebox{-\height}{\includegraphics[height=3.9cm]{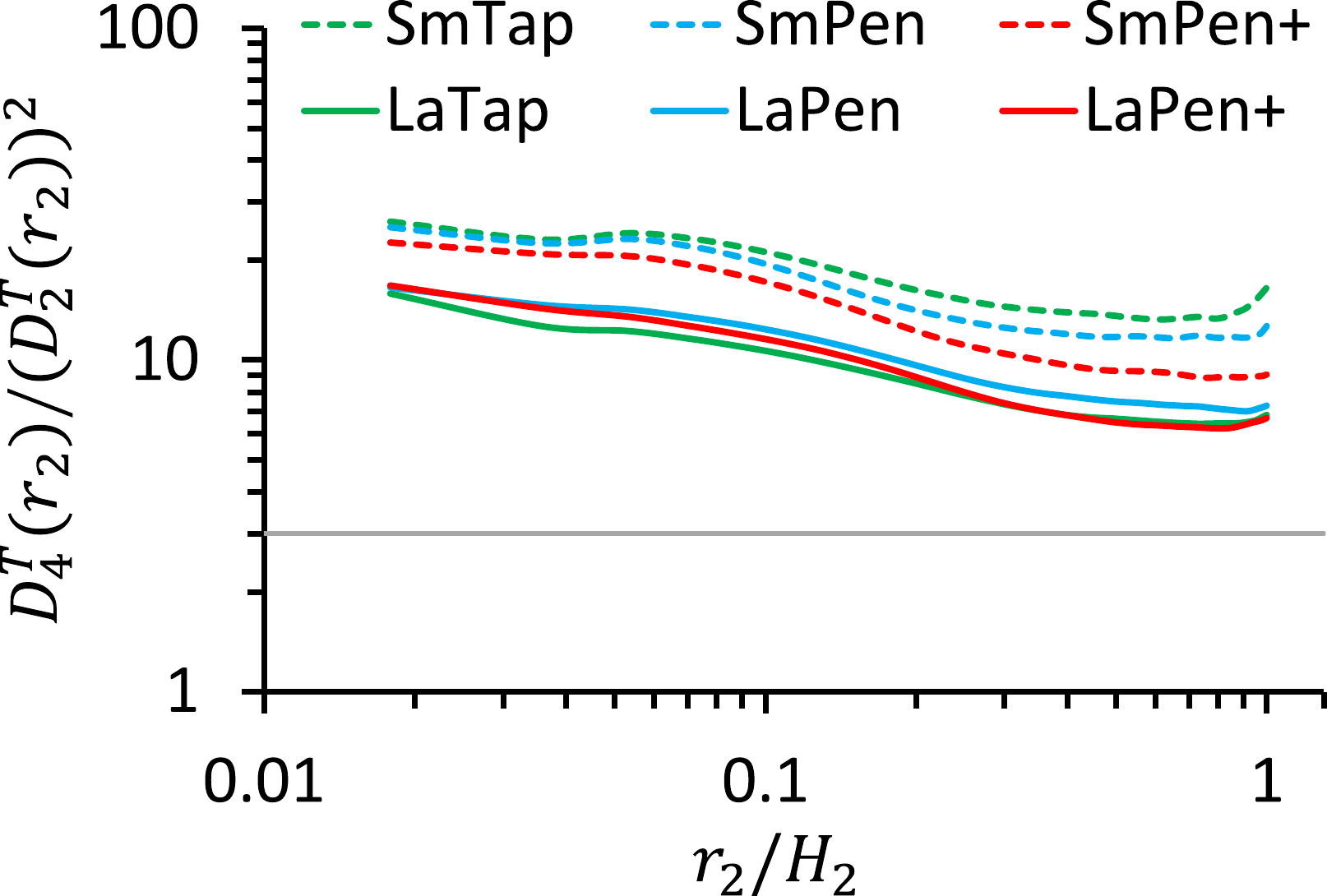}}
		\end{minipage}
		\begin{minipage}[b]{0.5\linewidth}
			\centering
			\makebox[1.5em][l]{\raisebox{-\height}{(\textit{b})}}%
			\raisebox{-\height}{\includegraphics[height=3.9cm]{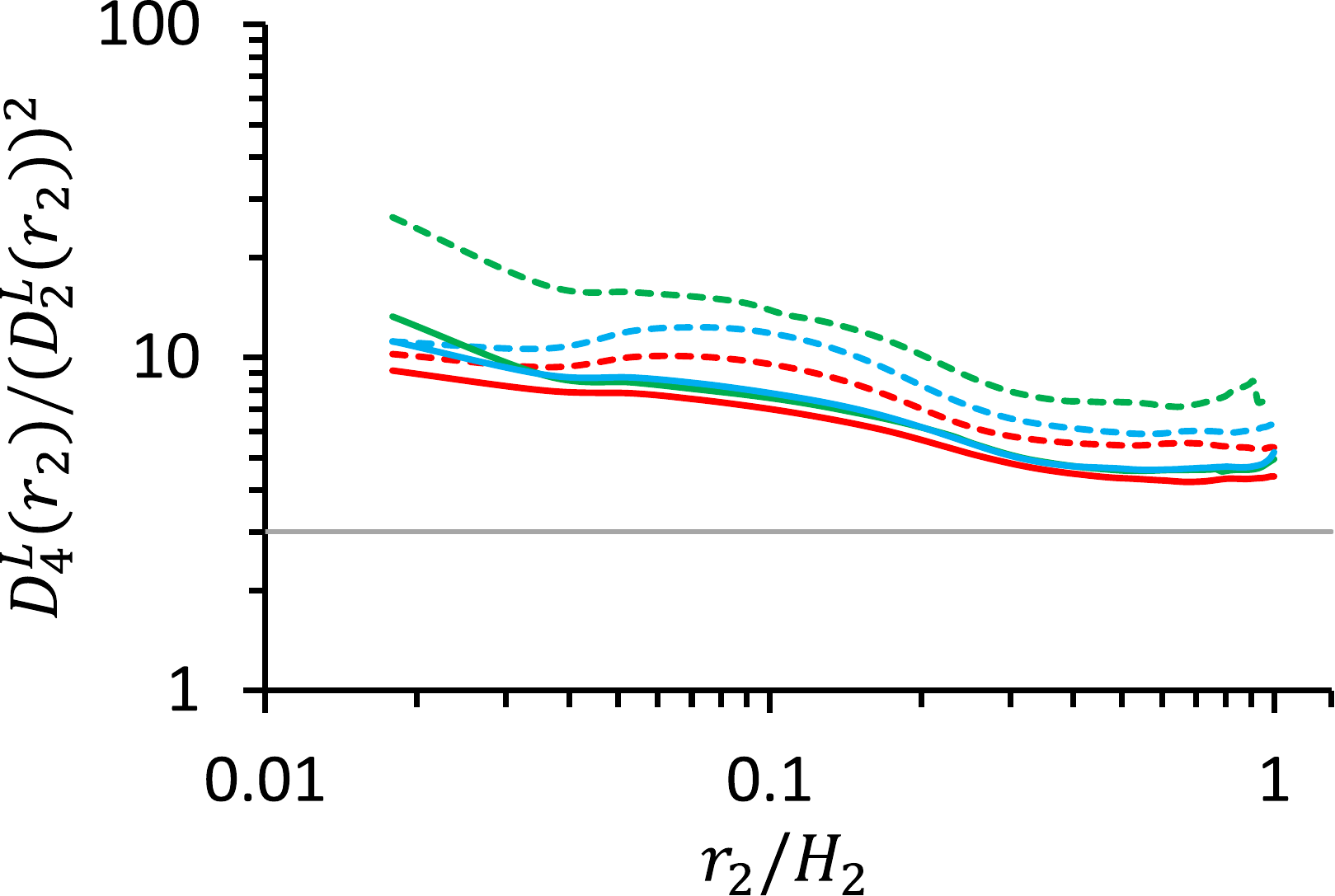}}
		\end{minipage}
	\end{minipage}	
	\caption{Normalized fourth-order transverse (\textit{a}) and longitudinal (\textit{b}) structure functions, corresponding to the flatness of the velocity increments along the horizontal direction. The horizontal lines in (\textit{a,b}) indicate to the Gaussian value of $3$ for the flatness.} \label{fig: flatness}
\end{figure}

While the PDFs results present the effect of the surfactants on the extreme events at a fixed separation, we look at now this property across all the scales quantified by the flatness $D^T_4(r_2)/(D^T_2(r_2))^2$ and $D^L_4(r_2)/(D^L_2(r_2))^2$ (figure \ref{fig: flatness}). The first observation from the results in figure \ref{fig: flatness} is that all six bubble-laden cases show a similar behaviour as $r$ increases, with values of up to 30 gradually reducing as $r$ increases. However, there is still considerable deviation from the Gaussian value of 3 even at the largest scale $r_2/H_2=1$. For both the transverse or longitudinal component, the flatness is larger for the cases with smaller bubbles, reflecting the same trend as the PDFs of the velocity increments at the single scale $r_2=10\Delta$ (see figures \ref{fig: pdf 10 grids}\textit{a,b} or figures \ref{fig: pdf 10 grids}\textit{c,d}). The qualitative effect of the surfactant concentration on the extreme events as quantified by the PDF results for $\Delta\boldsymbol{u}(r_2=10\Delta)$ (figures \ref{fig: pdf 10 grids}) can be extended to almost all the scales. While the flatness decreases in the sequence \textit{SmTap, SmPen} to \textit{SmPen+} across the scales of the flow, the values of the three cases with larger bubbles are very similar.

\section{Conclusions}

In this work we have used experiments to investigate the effect of surfactants on BIT using recently developed PSV and bubble identification techniques for bubble/particle-laden flows. The experiments consider two bubble sizes, 3 mm and 4 mm, and three different surfactant (1-Pentanol) concentrations for each bubble size. The addition of surfactants changes the bubble shape, as well the interface boundary condition from a quasi-free-slip to a no-slip condition. 

To provide some reference cases, we first investigated how 1-Pentanol influences single bubbles in terms of the influence on the size of the bubbles formed in the system, the bubble aspect ratios, the bubble rise velocity and the nature of their trajectories. We find that for the two bubble sizes considered, an approximately saturated contamination state is reached at $C_\infty\approx1000$ ppm. The effect of the surfactant on the bubble trajectories is similar to that observed in \cite{2014_Tagawa}, with helical trajectories at lower 1-Pentanol concentrations and zigzag rising trajectories for higher concentration for smaller bubble cases. The effect of the surfactant was similar for the larger bubbles, although in the absence of the surfactants, the trajectories of the larger bubbles appear chaotic.

Our results for bubble swarms show that for BIT-dominated flows the level of anisotropy is strong in general, and not negligible even at small scales in the flow, consistent with previous results. However, our results reveal that for the same bubble size the flow anisotropy can be strongly enhanced by increasing the surfactant concentration. We also investigated extreme events in the flow by considering the normalized probability density functions of the velocity increments at the scale of the bubble size. For the smaller bubbles, the PDFs become increasingly non-Gaussian with when the surfactant concentration is decreased. This is consistent with our previous finding that the non-Gaussianity of the PDFs of the velocity increments becomes stronger as the Reynolds number is decreased in the bubble-laden turbulent flows considered here, the opposite of what occurs for internal intermittency for single-phase turbulence which increases with increasing Reynolds number. However, the dependency of the non-Gaussianity of the velocity increments on the surfactant concentration is much weaker for the cases with larger bubbles. An explanation for this difference is that the larger bubble cases have much larger volume fractions compared to those for the smaller bubbles, and the bubble wake volume is also much larger, with much higher turbulence intensity generated by the wakes. Due to this, the velocity difference in and outside the wake is smaller for the larger bubble cases than the smaller bubble cases, and the intermittency is less dependent on the surfactants concentration.

While this study has focused on the impact of surfactants on the properties of the liquid turbulence in BIT, another important topic is to investigate how the surfactants impact the bubble clusters and bubble dispersion. These will be investigated in a future study.

\section*{Acknowledgements}

The authors would like to acknowledge T. Ziegenhein for many technical supports in the experimental method over the years. T.M. benefited from fruitful discussions on this topic with J. Fr\"{o}hlich, Y. Liao, P. Shi, S. Heitkam, X. Xu and Y. Tagawa.

\section*{Declaration of Interests}

The authors report no conflict of interest.

\section*{Data availability}
The data that support the findings of this study are available from the first author T.M. on request.\\

\appendix

\bibliographystyle{jfm}

\bibliography{BIT_exp_JFM}

\begin{thebibliography}{74}
\expandafter\ifx\csname natexlab\endcsname\relax\def\natexlab#1{#1}\fi
\def\au#1{#1} \def\ed#1{#1} \def\yr#1{#1}\def\at#1{#1}\def\jt#1{\textit{#1}}
  \def\bt#1{#1}\def\bvol#1{\textbf{#1}} \def\vol#1{#1} \def\pg#1{#1}
  \def\publ#1{#1}\def\arxiv#1{#1}\def\org#1{#1}\def\st#1{\textit{#1}}

\bibitem[Ahmed {\em et~al.\/}(2020)Ahmed, Izbassarov, Lu, Tryggvason, Muradoglu
  \& Tammisola]{2020_Ahmed}
{\sc \au{Ahmed, Z.}, \au{Izbassarov, D.}, \au{Lu, J.}, \au{Tryggvason, G.},
  \au{Muradoglu, M.} \& \au{Tammisola, O.}} \yr{2020}  \at{Effects of soluble
  surfactant on lateral migration of a bubble in a pressure driven channel
  flow}.  \jt{Int. J. Multiphase Flow}  \bvol{126},  \pg{103251}.

\bibitem[Atasi {\em et~al.\/}(2022)Atasi, Ravisankar, Legendre \&
  Zenit]{2022_Atasi}
{\sc \au{Atasi, O.}, \au{Ravisankar, M.}, \au{Legendre, D.} \& \au{Zenit, R.}}
  \yr{2022}  \at{The presence of surfactants controls the stability of bubble
  chains in carbonated drinks}.  \jt{arXiv preprint arXiv:2211.02253} .

\bibitem[Bachhuber \& Sanford(1974)]{1974_Bachhuber}
{\sc \au{Bachhuber, C.} \& \au{Sanford, C.}} \yr{1974}  \at{The rise of small
  bubbles in water}.  \jt{Journal of Applied Physics}  \bvol{45}~(6),
  \pg{2567--2569}.

\bibitem[Batchelor(1967)]{1967_Batchelor}
{\sc \au{Batchelor, G.~K.}} \yr{1967} {\em An introduction to fluid
  dynamics\/}.  \publ{Cambridge University Press}.

\bibitem[Bel~Fdhila \& Duineveld(1996)]{1996_Bel}
{\sc \au{Bel~Fdhila, R.} \& \au{Duineveld, P.~C.}} \yr{1996}  \at{The effect of
  surfactant on the rise of a spherical bubble at high reynolds and peclet
  numbers}.  \jt{Physics of Fluids}  \bvol{8}~(2),  \pg{310--321}.

\bibitem[Biferale {\em et~al.\/}(2012)Biferale, Perlekar, Sbragaglia \&
  Toschi]{2012_Biferale}
{\sc \au{Biferale, L}, \au{Perlekar, P}, \au{Sbragaglia, M} \& \au{Toschi, F}}
  \yr{2012}  \at{Convection in multiphase fluid flows using lattice boltzmann
  methods}.  \jt{Phys. Rev. Lett.}  \bvol{108}~(10),  \pg{104502}.

\bibitem[Br{\"o}der \& Sommerfeld(2007)]{2007_Broder}
{\sc \au{Br{\"o}der, D} \& \au{Sommerfeld, M}} \yr{2007}  \at{Planar shadow
  image velocimetry for the analysis of the hydrodynamics in bubbly flows}.
  \jt{Measurement Science and Technology}  \bvol{18}~(8),  \pg{2513}.

\bibitem[Cano-Lozano {\em et~al.\/}(2016)Cano-Lozano, Martinez-Bazan, Magnaudet
  \& Tchoufag]{2016_Cano}
{\sc \au{Cano-Lozano, J.~C.}, \au{Martinez-Bazan, C.}, \au{Magnaudet, J.} \&
  \au{Tchoufag, J.}} \yr{2016}  \at{Paths and wakes of deformable nearly
  spheroidal rising bubbles close to the transition to path instability}.
  \jt{Phys. Rev. Fluids}  \bvol{1}~(5),  \pg{053604}.

\bibitem[Carter \& Coletti(2017)]{2017_Carter}
{\sc \au{Carter, D.~W} \& \au{Coletti, F.}} \yr{2017}  \at{Scale-to-scale
  anisotropy in homogeneous turbulence}.  \jt{J. Fluid Mech.}  \bvol{827},
  \pg{250--284}.

\bibitem[du~Cluzeau {\em et~al.\/}(2022)du~Cluzeau, Bois, Leoni \&
  Toutant]{2022_Cluzeau}
{\sc \au{du~Cluzeau, A}, \au{Bois, G}, \au{Leoni, N} \& \au{Toutant, A}}
  \yr{2022}  \at{Analysis and modeling of bubble-induced agitation from direct
  numerical simulation of homogeneous bubbly flows}.  \jt{Phys. Rev. Fluids}
  \bvol{7}~(4),  \pg{044604}.

\bibitem[du~Cluzeau {\em et~al.\/}(2019)du~Cluzeau, Bois \&
  Toutant]{2019_Cluzeau}
{\sc \au{du~Cluzeau, A}, \au{Bois, G} \& \au{Toutant, A}} \yr{2019}
  \at{Analysis and modelling of reynolds stresses in turbulent bubbly up-flows
  from direct numerical simulations}.  \jt{J. Fluid Mech.}  \bvol{866},
  \pg{132--168}.

\bibitem[Cuenot {\em et~al.\/}(1997)Cuenot, Magnaudet \& Spennato]{1997_Cuenot}
{\sc \au{Cuenot, B}, \au{Magnaudet, J} \& \au{Spennato, B}} \yr{1997}  \at{The
  effects of slightly soluble surfactants on the flow around a spherical
  bubble}.  \jt{J. Fluid Mech.}  \bvol{339},  \pg{25--53}.

\bibitem[Drenckhan \& Saint-Jalmes(2015)]{2015_Drenckhan}
{\sc \au{Drenckhan, W.} \& \au{Saint-Jalmes, A.}} \yr{2015}  \at{The science of
  foaming}.  \jt{Adv. Colloid Interface Sci.}  \bvol{222},  \pg{228--259}.

\bibitem[Durst {\em et~al.\/}(1986)Durst, Sch{\"o}nung, Selanger \&
  Winter]{1986_Durst}
{\sc \au{Durst, F.}, \au{Sch{\"o}nung, B.}, \au{Selanger, K.} \& \au{Winter,
  M.}} \yr{1986}  \at{Bubble-driven liquid flows}.  \jt{J. Fluid Mech.}
  \bvol{170},  \pg{53--82}.

\bibitem[Eftekhari {\em et~al.\/}(2021)Eftekhari, Schwarzenberger, Heitkam,
  Javadi, Bashkatov, Ata \& Eckert]{2021_Eftekhari}
{\sc \au{Eftekhari, M.}, \au{Schwarzenberger, K.}, \au{Heitkam, S.},
  \au{Javadi, A.}, \au{Bashkatov, A.}, \au{Ata, S.} \& \au{Eckert, K.}}
  \yr{2021}  \at{Interfacial behavior of particle-laden bubbles under
  asymmetric shear flow}.  \jt{Langmuir}  \bvol{37}~(45),  \pg{13244--13254}.

\bibitem[Ellingsen \& Risso(2001)]{2001_Ellingsen}
{\sc \au{Ellingsen, K.} \& \au{Risso, F.}} \yr{2001}  \at{On the rise of an
  ellipsoidal bubble in water: oscillatory paths and liquid-induced velocity}.
  \jt{J. Fluid Mech.}  \bvol{440},  \pg{235--268}.

\bibitem[Frisch(1995)]{1995_Frisch}
{\sc \au{Frisch, U.}} \yr{1995} {\em Turbulence: the legacy of AN
  Kolmogorov\/}.  \publ{Cambridge university press}.

\bibitem[Frumkin \& Levich(1947)]{1947_Frumkin}
{\sc \au{Frumkin, A} \& \au{Levich, V.~G.}} \yr{1947}  \at{On surfactants and
  interfacial motion}.  \jt{Zh. Fiz. Khim.}  \bvol{21},  \pg{1183--1204}.

\bibitem[Hayashi \& Tomiyama(2018)]{2018_Hayashi}
{\sc \au{Hayashi, Kosuke} \& \au{Tomiyama, Akio}} \yr{2018}  \at{Effects of
  surfactant on lift coefficients of bubbles in linear shear flows}.  \jt{Int.
  J. Multiphase Flow}  \bvol{99},  \pg{86--93}.

\bibitem[Hessenkemper {\em et~al.\/}(2022)Hessenkemper, Starke, Atassi,
  Ziegenhein \& Lucas]{2022_Hessenkemper}
{\sc \au{Hessenkemper, H.}, \au{Starke, S.}, \au{Atassi, Y.}, \au{Ziegenhein,
  T.} \& \au{Lucas, D.}} \yr{2022}  \at{Bubble identification from images with
  machine learning methods}.  \jt{Int. J. Multiphase Flow}  \bvol{155},
  \pg{104169}.

\bibitem[Hessenkemper \& Ziegenhein(2018)]{2018_Hessenkemper}
{\sc \au{Hessenkemper, H} \& \au{Ziegenhein, T}} \yr{2018}  \at{Particle shadow
  velocimetry ({PSV}) in bubbly flows}.  \jt{Int. J. Multiphase Flow}
  \bvol{106},  \pg{268--279}.

\bibitem[Hessenkemper {\em et~al.\/}(2021{\natexlab{{\em a\/}}})Hessenkemper,
  Ziegenhein, Lucas \& Tomiyama]{2021_Hessenkemper}
{\sc \au{Hessenkemper, H}, \au{Ziegenhein, T}, \au{Lucas, D} \& \au{Tomiyama,
  A}} \yr{2021{\natexlab{{\em a\/}}}}  \at{Influence of surfactant
  contaminations on the lift force of ellipsoidal bubbles in water}.  \jt{Int.
  J. Multiphase Flow}  \bvol{145},  \pg{103833}.

\bibitem[Hessenkemper {\em et~al.\/}(2021{\natexlab{{\em b\/}}})Hessenkemper,
  Ziegenhein, Rzehak, Lucas \& Tomiyama]{2021a_Hessenkemper}
{\sc \au{Hessenkemper, H.}, \au{Ziegenhein, T.}, \au{Rzehak, R.}, \au{Lucas,
  D.} \& \au{Tomiyama, A.}} \yr{2021{\natexlab{{\em b\/}}}}  \at{Lift force
  coefficient of ellipsoidal single bubbles in water}.  \jt{Int. J. Multiphase
  Flow}  \bvol{138},  \pg{103587}.

\bibitem[Innocenti {\em et~al.\/}(2021)Innocenti, Jaccod, Popinet \&
  Chibbaro]{2021_Innocenti}
{\sc \au{Innocenti, A.}, \au{Jaccod, A.}, \au{Popinet, S.} \& \au{Chibbaro,
  S.}} \yr{2021}  \at{Direct numerical simulation of bubble-induced
  turbulence}.  \jt{J. Fluid Mech.}  \bvol{918}.

\bibitem[Lai \& Socolofsky(2019)]{2019_Lai}
{\sc \au{Lai, C. C.~K.} \& \au{Socolofsky, S.~A.}} \yr{2019}  \at{The turbulent
  kinetic energy budget in a bubble plume}.  \jt{J. Fluid Mech.}  \bvol{865},
  \pg{993–1041}.

\bibitem[Lance \& Bataille(1991)]{1991_Lance}
{\sc \au{Lance, M.} \& \au{Bataille, J.}} \yr{1991}  \at{Turbulence in the
  liquid phase of a uniform bubbly air-water flow}.  \jt{J. Fluid Mech.}
  \bvol{222},  \pg{95--118}.

\bibitem[Langevin(2014)]{2014_Langevin}
{\sc \au{Langevin, D.}} \yr{2014}  \at{Rheology of adsorbed surfactant
  monolayers at fluid surfaces}.  \jt{Annu. Rev. Fluid Mech.}  \bvol{46},
  \pg{47--65}.

\bibitem[Legendre {\em et~al.\/}(2009)Legendre, Lauga \&
  Magnaudet]{2009_Legendre}
{\sc \au{Legendre, D.}, \au{Lauga, E.} \& \au{Magnaudet, J.}} \yr{2009}
  \at{Influence of slip on the dynamics of two-dimensional wakes}.  \jt{J.
  Fluid Mech.}  \bvol{633},  \pg{437--447}.

\bibitem[Levich(1962)]{1962_Levich}
{\sc \au{Levich, V.~G.}} \yr{1962} {\em Physicochemical hydrodynamics\/}.
  \publ{Prentice-Hall Inc.}

\bibitem[Lohse(2018)]{2018_Lohse}
{\sc \au{Lohse, D.}} \yr{2018}  \at{Bubble puzzles: From fundamentals to
  applications}.  \jt{Phys. Rev. Fluids}  \bvol{3}~(11),  \pg{110504}.

\bibitem[Lohse(2022)]{2022_Lohse}
{\sc \au{Lohse, D.}} \yr{2022}  \at{Fundamental fluid dynamics challenges in
  inkjet printing}.  \jt{Annu. Rev. Fluid Mech.}  \bvol{54},  \pg{349--382}.

\bibitem[Loubi{\`e}re \& H{\'e}brard(2004)]{2004_Loubiere}
{\sc \au{Loubi{\`e}re, Karine} \& \au{H{\'e}brard, Gilles}} \yr{2004}
  \at{Influence of liquid surface tension (surfactants) on bubble formation at
  rigid and flexible orifices}.  \jt{Chem. Eng. Process}  \bvol{43}~(11),
  \pg{1361--1369}.

\bibitem[Lu {\em et~al.\/}(2017)Lu, Muradoglu \& Tryggvason]{2017_Lu}
{\sc \au{Lu, J.}, \au{Muradoglu, M.} \& \au{Tryggvason, G.}} \yr{2017}
  \at{Effect of insoluble surfactant on turbulent bubbly flows in vertical
  channels}.  \jt{Int. J. Multiphase Flow}  \bvol{95},  \pg{135--143}.

\bibitem[Lu \& Tryggvason(2013)]{2013_Lu}
{\sc \au{Lu, J.} \& \au{Tryggvason, G.}} \yr{2013}  \at{Dynamics of nearly
  spherical bubbles in a turbulent channel upflow}.  \jt{J. Fluid Mech.}
  \bvol{732},  \pg{166--189}.

\bibitem[Ma {\em et~al.\/}(2022)Ma, Hessenkemper, Lucas \& Bragg]{2022_Ma}
{\sc \au{Ma, T.}, \au{Hessenkemper, H.}, \au{Lucas, D.} \& \au{Bragg, A.~D.}}
  \yr{2022}  \at{An experimental study on the multiscale properties of
  turbulence in bubble-laden flows}.  \jt{J. Fluid Mech.}  \bvol{936},
  \pg{A42}.

\bibitem[Ma {\em et~al.\/}(2020)Ma, Lucas, Jakirli{\'c} \&
  Fr{\"o}hlich]{2020_Ma}
{\sc \au{Ma, T.}, \au{Lucas, D.}, \au{Jakirli{\'c}, S.} \& \au{Fr{\"o}hlich,
  J.}} \yr{2020}  \at{Progress in the second-moment closure for bubbly flow
  based on direct numerical simulation data}.  \jt{J. Fluid Mech.}  \bvol{883},
   \pg{A9}.

\bibitem[Ma {\em et~al.\/}(2021)Ma, Ott, Fr{\"o}hlich \& Bragg]{2021_Ma}
{\sc \au{Ma, T.}, \au{Ott, B.}, \au{Fr{\"o}hlich, J.} \& \au{Bragg, A.~D}}
  \yr{2021}  \at{Scale-dependent anisotropy, energy transfer and intermittency
  in bubble-laden turbulent flows}.  \jt{J. Fluid Mech.}  \bvol{927},
  \pg{A16}.

\bibitem[Maeda {\em et~al.\/}(2021)Maeda, Date, Sugiyama, Takagi \&
  Matsumoto]{2021_Maeda}
{\sc \au{Maeda, K.}, \au{Date, M.}, \au{Sugiyama, K.}, \au{Takagi, S.} \&
  \au{Matsumoto, Y.}} \yr{2021}  \at{Viscid--inviscid interactions of pairwise
  bubbles in a turbulent channel flow and their implications for bubble
  clustering}.  \jt{J. Fluid Mech.}  \bvol{919},  \pg{A30}.

\bibitem[Magnaudet \& Eames(2000)]{2000_Magnaudet}
{\sc \au{Magnaudet, J.} \& \au{Eames, I.}} \yr{2000}  \at{The motion of
  high-reynolds-number bubbles in inhomogeneous flows}.  \jt{Annu. Rev. Fluid
  Mech.}  \bvol{32}~(1),  \pg{659--708}.

\bibitem[Manikantan \& Squires(2020)]{2020_Manikantan}
{\sc \au{Manikantan, H.} \& \au{Squires, T.~M}} \yr{2020}  \at{Surfactant
  dynamics: hidden variables controlling fluid flows}.  \jt{J. Fluid Mech.}
  \bvol{892},  \pg{P1}.

\bibitem[Masuk {\em et~al.\/}(2021)Masuk, Salibindla \& Ni]{2021_Masuk}
{\sc \au{Masuk, A. U.~M.}, \au{Salibindla, A. K.~R.} \& \au{Ni, R.}} \yr{2021}
  \at{Simultaneous measurements of deforming hinze-scale bubbles with
  surrounding turbulence}.  \jt{J. Fluid Mech.}  \bvol{910},  \pg{A21}.

\bibitem[Mathai {\em et~al.\/}(2020)Mathai, Lohse \& Sun]{2020_Mathai}
{\sc \au{Mathai, V.}, \au{Lohse, D.} \& \au{Sun, C.}} \yr{2020}  \at{Bubbly and
  buoyant particle--laden turbulent flows}.  \jt{Annu. Rev. Condens. Matter
  Phys. 11}  \bvol{11},  \pg{529--559}.

\bibitem[Mclaughlin(1996)]{1996_Mclaughlin}
{\sc \au{Mclaughlin, John~B}} \yr{1996}  \at{Numerical simulation of bubble
  motion in water}.  \jt{Journal of colloid and interface science}
  \bvol{184}~(2),  \pg{614--625}.

\bibitem[Mendez-Diaz {\em et~al.\/}(2013)Mendez-Diaz, Serrano-Garc\'{i}a, Zenit
  \& Hern\'{a}ndez-Cordero]{2013_Mendez}
{\sc \au{Mendez-Diaz, S.}, \au{Serrano-Garc\'{i}a, J.~C.}, \au{Zenit, R.} \&
  \au{Hern\'{a}ndez-Cordero, J.~A.}} \yr{2013}  \at{Power spectral
  distributions of pseudo-turbulent bubbly flows}.  \jt{Phys. Fluids}
  \bvol{25}~(4).

\bibitem[Moore(1963)]{1963_Moore}
{\sc \au{Moore, D.~W.}} \yr{1963}  \at{The boundary layer on a spherical gas
  bubble}.  \jt{J. Fluid Mech.}  \bvol{16}~(2),  \pg{161--176}.

\bibitem[Mougin \& Magnaudet(2001)]{2001_Mougin}
{\sc \au{Mougin, G.} \& \au{Magnaudet, J.}} \yr{2001}  \at{Path instability of
  a rising bubble}.  \jt{Phys. Rev. Lett.}  \bvol{88}~(1),  \pg{014502}.

\bibitem[Mougin \& Magnaudet(2006)]{2006_Mougin}
{\sc \au{Mougin, G.} \& \au{Magnaudet, J.}} \yr{2006}  \at{Wake-induced forces
  and torques on a zigzagging/spiralling bubble}.  \jt{J. Fluid Mech.}
  \bvol{567},  \pg{185--194}.

\bibitem[N{\'e}el \& Deike(2021)]{2021_Neel}
{\sc \au{N{\'e}el, B} \& \au{Deike, L.}} \yr{2021}  \at{Collective bursting of
  free-surface bubbles, and the role of surface contamination}.  \jt{J. Fluid
  Mech.}  \bvol{917},  \pg{A46}.

\bibitem[Palaparthi {\em et~al.\/}(2006)Palaparthi, Papageorgiou \&
  Maldarelli]{2006_Palaparthi}
{\sc \au{Palaparthi, R.}, \au{Papageorgiou, D.~T.} \& \au{Maldarelli, C.}}
  \yr{2006}  \at{Theory and experiments on the stagnant cap regime in the
  motion of spherical surfactant-laden bubbles}.  \jt{J. Fluid Mech.}
  \bvol{559},  \pg{1--44}.

\bibitem[Pandey {\em et~al.\/}(2020)Pandey, Ramadugu \& Perlekar]{2020_Pandey}
{\sc \au{Pandey, V.}, \au{Ramadugu, R.} \& \au{Perlekar, P.}} \yr{2020}
  \at{Liquid velocity fluctuations and energy spectra in three-dimensional
  buoyancy-driven bubbly flows}.  \jt{J. Fluid Mech.}  \bvol{884},  \pg{R6}.

\bibitem[Pesci {\em et~al.\/}(2018)Pesci, Weiner, Marschall \&
  Bothe]{2018_Pesci}
{\sc \au{Pesci, C.}, \au{Weiner, A.}, \au{Marschall, H.} \& \au{Bothe, D.}}
  \yr{2018}  \at{Computational analysis of single rising bubbles influenced by
  soluble surfactant}.  \jt{J. Fluid Mech.}  \bvol{856},  \pg{709--763}.

\bibitem[Rensen {\em et~al.\/}(2005)Rensen, Luther \& Lohse]{2005_Rensen}
{\sc \au{Rensen, J.}, \au{Luther, S.} \& \au{Lohse, D.}} \yr{2005}  \at{The
  effect of bubbles on developed turbulence}.  \jt{J. Fluid Mech.}  \bvol{538},
   \pg{153--187}.

\bibitem[Riboux {\em et~al.\/}(2013)Riboux, Legendre \& Risso]{2013_Riboux}
{\sc \au{Riboux, G.}, \au{Legendre, D.} \& \au{Risso, F.}} \yr{2013}  \at{A
  model of bubble-induced turbulence based on large-scale wake interactions}.
  \jt{J. Fluid Mech.}  \bvol{719},  \pg{362--387}.

\bibitem[Riboux {\em et~al.\/}(2010)Riboux, Risso \& Legendre]{2010_Riboux}
{\sc \au{Riboux, G.}, \au{Risso, F.} \& \au{Legendre, D.}} \yr{2010}
  \at{Experimental characterization of the agitation generated by bubbles
  rising at high {R}eynolds number}.  \jt{J. Fluid Mech.}  \bvol{643},
  \pg{509--539}.

\bibitem[Risso(2018)]{2018_Risso}
{\sc \au{Risso, F.}} \yr{2018}  \at{Agitation, mixing, and transfers induced by
  bubbles}.  \jt{Annu. Rev. Fluid Mech.}  \bvol{50},  \pg{25–48}.

\bibitem[Roghair {\em et~al.\/}(2011)Roghair, Mercado, Van Sint~Annaland,
  Kuipers, Sun \& Lohse]{2011_Roghair}
{\sc \au{Roghair, I.}, \au{Mercado, J.~M.}, \au{Van Sint~Annaland, M.},
  \au{Kuipers, H.}, \au{Sun, C.} \& \au{Lohse, D.}} \yr{2011}  \at{Energy
  spectra and bubble velocity distributions in pseudo-turbulence: Numerical
  simulations vs. experiments}.  \jt{Int. J. Multiphase Flow}  \bvol{37},
  \pg{1093--1098}.

\bibitem[Santarelli {\em et~al.\/}(2016)Santarelli, Roussel \&
  Fr\"{o}hlich]{2016_Santarelli_b}
{\sc \au{Santarelli, C.}, \au{Roussel, J.} \& \au{Fr\"{o}hlich, J.}} \yr{2016}
  \at{Budget analysis of the turbulent kinetic energy for bubbly flow in a
  vertical channel}.  \jt{Chem. Eng. Sci.}  \bvol{141},  \pg{46--62}.

\bibitem[Schl{\"u}ter {\em et~al.\/}(2021)Schl{\"u}ter, Herres-Pawlis, Nieken,
  Tuttlies \& Bothe]{2021_Schlueter}
{\sc \au{Schl{\"u}ter, M.}, \au{Herres-Pawlis, S.}, \au{Nieken, U.},
  \au{Tuttlies, U.} \& \au{Bothe, D.}} \yr{2021}  \at{Small-scale phenomena in
  reactive bubbly flows: Experiments, numerical modeling, and applications}.
  \jt{Annu. Rev. Chem. Biomol. Eng.}  \bvol{12},  \pg{625--643}.

\bibitem[Shew \& Pinton(2006)]{2006_Shew}
{\sc \au{Shew, W.~L.} \& \au{Pinton, J.-F.}} \yr{2006}  \at{Dynamical model of
  bubble path instability}.  \jt{Phys. Rev. Lett.}  \bvol{97}~(14),
  \pg{144508}.

\bibitem[Soligo {\em et~al.\/}(2019)Soligo, Roccon \& Soldati]{2019_Soligo}
{\sc \au{Soligo, G.}, \au{Roccon, A.} \& \au{Soldati, A.}} \yr{2019}
  \at{Breakage, coalescence and size distribution of surfactant-laden droplets
  in turbulent flow}.  \jt{J. Fluid Mech.}  \bvol{881},  \pg{244--282}.

\bibitem[Sreenivasan \& Antonia(1997)]{1997_Sreenivasan}
{\sc \au{Sreenivasan, K.~R.} \& \au{Antonia, R.~A.}} \yr{1997}  \at{The
  phenomenology of small-scale turbulence}.  \jt{Annu. Rev. Fluid Mech.}
  \bvol{29}~(1),  \pg{435--472}.

\bibitem[Stone(1994)]{1994_Stone}
{\sc \au{Stone, H.~A.}} \yr{1994}  \at{Dynamics of drop deformation and breakup
  in viscous fluids}.  \jt{Annu. Rev. Fluid Mech.}  \bvol{26}~(1),
  \pg{65--102}.

\bibitem[Tagawa {\em et~al.\/}(2014)Tagawa, Takagi \& Matsumoto]{2014_Tagawa}
{\sc \au{Tagawa, Y.}, \au{Takagi, S.} \& \au{Matsumoto, Y.}} \yr{2014}
  \at{Surfactant effect on path instability of a rising bubble}.  \jt{J. Fluid
  Mech.}  \bvol{738},  \pg{124}.

\bibitem[Takagi \& Matsumoto(2011)]{2011_Takagi}
{\sc \au{Takagi, S.} \& \au{Matsumoto, Y.}} \yr{2011}  \at{Surfactant effects
  on bubble motion and bubbly flows}.  \jt{Annu. Rev. Fluid Mech.}  \bvol{43},
  \pg{615--636}.

\bibitem[Takagi {\em et~al.\/}(2009)Takagi, Ogasawara, Fukuta \&
  Matsumoto]{2009_Takagi}
{\sc \au{Takagi, S.}, \au{Ogasawara, T.}, \au{Fukuta, M.} \& \au{Matsumoto,
  Y.}} \yr{2009}  \at{Surfactant effect on the bubble motions and bubbly flow
  structures in a vertical channel}.  \jt{Fluid dynamics research}
  \bvol{41}~(6),  \pg{065003}.

\bibitem[Takagi {\em et~al.\/}(2008)Takagi, Ogasawara \&
  Matsumoto]{2008_Takagi}
{\sc \au{Takagi, S.}, \au{Ogasawara, T.} \& \au{Matsumoto, Y.}} \yr{2008}
  \at{The effects of surfactant on the multiscale structure of bubbly flows}.
  \jt{Phil. Trans. R. Soc. A}  \bvol{366}~(1873),  \pg{2117--2129}.

\bibitem[Tomiyama {\em et~al.\/}(2002)Tomiyama, Celata, Hosokawa \&
  Yoshida]{2002_Tomiyama}
{\sc \au{Tomiyama, A.}, \au{Celata, G.P.}, \au{Hosokawa, S.} \& \au{Yoshida,
  S.}} \yr{2002}  \at{Terminal velocity of single bubbles in surface tension
  force dominant regime}.  \jt{Int. J. Multiphase Flow}  \bvol{28}~(9),
  \pg{1497--1519}.

\bibitem[Townsend(1949)]{1949_Townsend}
{\sc \au{Townsend, A.}} \yr{1949}  \at{The fully developed wake of a circular
  cylinder}.  \jt{Aust. J. Chem.}  \bvol{2}~(4),  \pg{451--468}.

\bibitem[Veldhuis {\em et~al.\/}(2008)Veldhuis, Biesheuvel \&
  Van~Wijngaarden]{2008_Veldhuis}
{\sc \au{Veldhuis, C.}, \au{Biesheuvel, A.} \& \au{Van~Wijngaarden, L.}}
  \yr{2008}  \at{Shape oscillations on bubbles rising in clean and in tap
  water}.  \jt{Phys. Fluids}  \bvol{20}~(4),  \pg{040705}.

\bibitem[Verschoof {\em et~al.\/}(2016)Verschoof, Van Der~Veen, Sun \&
  Lohse]{2016_Verschoof}
{\sc \au{Verschoof, R.~A.}, \au{Van Der~Veen, R.~CA}, \au{Sun, C.} \&
  \au{Lohse, D.}} \yr{2016}  \at{Bubble drag reduction requires large bubbles}.
   \jt{Phys. Rev. Lett.}  \bvol{117}~(10),  \pg{104502}.

\bibitem[Ybert \& di~Meglio(1998)]{1998_Ybert}
{\sc \au{Ybert, C.} \& \au{di~Meglio, J.-M.}} \yr{1998}  \at{Ascending air
  bubbles in protein solutions}.  \jt{The European Physical Journal B-Condensed
  Matter and Complex Systems}  \bvol{4}~(3),  \pg{313--319}.

\bibitem[Ybert \& di~Meglio(2000)]{2000_Ybert}
{\sc \au{Ybert, C.} \& \au{di~Meglio, J.-M.}} \yr{2000}  \at{Ascending air
  bubbles in solutions of surface-active molecules: influence of desorption
  kinetics}.  \jt{The European Physical Journal E}  \bvol{3}~(2),
  \pg{143--148}.

\bibitem[Ziegenhein \& Lucas(2017)]{2017_Ziegenhein}
{\sc \au{Ziegenhein, T.} \& \au{Lucas, D.}} \yr{2017}  \at{Observations on
  bubble shapes in bubble columns under different flow conditions}.
  \jt{Experimental Thermal and Fluid Science}  \bvol{85},  \pg{248--256}.

\bibitem[Ziegenhein \& Lucas(2019)]{2019_Ziegenhein}
{\sc \au{Ziegenhein, T.} \& \au{Lucas, D.}} \yr{2019}  \at{The critical bubble
  diameter of the lift force in technical and environmental, buoyancy-driven
  bubbly flows}.  \jt{Int. J. Multiphase Flow}  \bvol{116},  \pg{26--38}.

\end{thebibliography}

\end{document}